\documentclass[12pt,a4paper]{article}
\usepackage{latexsym}
\usepackage{amsmath}
\usepackage{amsfonts}
\usepackage{amsbsy}
\usepackage{amssymb}
\usepackage{amscd}
\usepackage{graphicx}
\usepackage{mathrsfs}
\usepackage{epsfig}
\usepackage{bbm}
\usepackage{psfrag}

\def \C{{\mathbb C}}
\def \Z{\mathbb Z}
\def \N{\mathbb{N}}

\def \Diff{\text{Diff}}

\def \A{{\mathfrak A}}
\def \AL{\overline{\mathcal{A}}}
\def \Aut{\mathcal{A}\hspace{-1pt}{\it ut}(\mathcal{P})}

\def \Cyl{{\mathcal C}}
\def \Dyl{\mathcal D}
\def \Cat{\mathscr{C}}

\def \G{\mathcal{G}}
\def \Ga{\Gamma}

\def \M{\mathcal{M}}
\def \Mor{\text{Mor}}

\def \Or{\mathcal{O}r\mathcal{C}ut}
\def \P{\mathcal{P}}

\def \Susp{\text{\rm Susp}}

\def \T{{\mathcal T}}
\def \a{\alpha}

\def \H{\mathcal{H}}
\def \Hkin{\mathcal{H}_{\text{\rm kin}}}
\def \Hdiff{\mathcal{H}_{\text{\rm diff}}}
\def \Haut{\mathcal{H}_{\text{\rm Aut}}}

\def \d{\delta}
\def \id{\text{\rm id}}

\def \g{\gamma}
\def \Nyl{\mathcal{N}}
\def \r{\rho}
\def \s{\sigma}
\def \Sig{\Sigma}
\def \e{\varepsilon}
\def \o{\omega}
\def \Om{\Omega}
\def \t{\tau}

\def \SU{\text{Susp}(G)}

\def \n{\nu}
\def \m{\mu}

\def \PG{\mathcal{P}}

\def \tr{\text{ tr}}

\def \V{\mathcal V}

\newtheorem{Theorem}{Theorem}[section]
\newtheorem{Corollary}{Corollary}[section]
\newtheorem{Lemma}{Lemma}[section]
\newtheorem{Definition}{Definition}[section]

\renewcommand{\thefigure}{\thefigure.\arabic{figure}}
\numberwithin{equation}{section}
\numberwithin{figure}{section}

\begin{document}

\title{Automorphisms in Loop Quantum Gravity\\[20pt]}

\author{\it \small Benjamin Bahr, MPI f\"ur Gravitationsphysik, Albert-Einstein
Institut,\\ \it \small Am M\"uhlenberg 1, 14467 Golm, Germany\\[20pt]
\it \small Thomas Thiemann, MPI f\"ur Gravitationsphysik,
Albert-Einstein Institut, \\ \it \small Am M\"uhlenberg 1, 14467
Golm, Germany;\\ \it \small Perimeter Institute for Theoretical
Physics, \\ \it \small 31 Caroline St. N., Waterloo Ontario N2L 2Y5,
Canada}

\maketitle

\abstract{\noindent We investigate a certain distributional
extension of the group of spatial diffeomorphisms in Loop Quantum
Gravity. This extension, which is given by the automorphisms $\Aut$
of the path groupoid $\P$, was proposed by Velhinho and is inspired
by category theory. This group is much larger than the group of
piecewise analytic diffeomorphisms. In particular, we will show that
graphs with the same combinatorics but different generalized
knotting classes can be mapped into each other. We describe the
automorphism-invariant Hilbert space and comment on how a
combinatorial formulation of LQG might arise.}


\section{Introduction}

Loop Quantum Gravity is an attempt to quantize the theory of general
relativity (see \cite{ALLMT, ROVELLI, SMOLIN, INTRO} and references
therein). This is done by casting classical GR into a hamiltonian
formulation, in which it becomes a constrained theory with fields on
a Cauchy hypersurface $\Sig$. These fields are an
$\mathfrak{su}(2)$-connection $A_a^I(x)$ (the Ashtekar connection)
and its canonically conjugate $E_I^a(x)$, which is the analogue of
the electric field in $SU(N)$-Yang-Mills theory. Modulo physical
constants, which can be absorbed into the definition of $E_I^a(x)$,
the Poisson structure reads
\begin{eqnarray}
\big\{A_a^I(x),\,E_J^b(y)\big\}\;=\;\d_a^b\,\d^I_J\,\d(x,y).
\end{eqnarray}

\noindent The connection $A$ plays the r\^{o}le of the configuration
variable, and $E$ is the canonically conjugate momentum. The
classical configuration space is the space of smooth connections
$\mathcal{A}$.

By going over to the quantum regime, the space $\mathcal{A}$ is
extended to $\AL$, by adding also connections which are
``distributional'' in the sense that they can have support on
lower-dimensional subspaces of $\Sig$. The space $\AL$ can be
endowed with a topology that turns it into a compact Hausdorff
space, and a regular, normalized Borel measure $\m_{AL}$, the
\emph{Ashtekar-Isham-Lewandowski-measure}. The quantum theory is
then formulated in the kinematical Hilbert space
$\Hkin=L^2(\AL,d\m_{AL})$. The space $\mathcal{A}$ is topologically
dense in $\AL$, but measure-theoretically thin: $\mathcal{A}$ is
contained in a set of measure zero.\cite{INTRO}

This is analogous to the case of quantum field theory: Consider the
case of the Euclidean free scalar field. Then the path integral
formalism proposes to write the partition function $Z([J])$ as an
integral over all field configurations
\begin{eqnarray}
Z[J]\;=\;\,Z[0]\int\,\Dyl\phi\;e^{-\int
\left[\phi^*(\Box+m^2)\phi\,+\,i\phi J\right]}
\end{eqnarray}

\noindent In this case, the integration ranges over a set of field
configurations $\phi$, in which the smooth fields lie dense. The set
of these fields, however, are contained in a set of measure zero.\\

In the Hamiltonian formulation of GR employed in Loop Quantum
Gravity, the dynamics is contained in the constraint functions,
which are phase space functions that generate a Hamiltonian flow on
the constraint hypersurface. Two of the constraints, the Gauss- and
the Diffeomorphism- constraint, encode the invariance of the theory
under change of $SU(2)$-gauge and diffeomorphisms on $\Sig$.
Consequently, their Hamiltonian flows generate gauge-transformations
and diffeomorphisms on $\Sig$, respectively. Classically, the
corresponding group of smooth gauge transformations $\G$ and
diffeomorphisms $\Diff(\Sig)$ act on $\mathcal{A}$:
\begin{eqnarray}\label{Gl:ClassicalActionsOfGaugeAndDiffeos}
\a_g\;:\;\mathcal{A}\;&\longrightarrow&\;\mathcal{A},\qquad
g\in\G\\[5pt]\nonumber
\a_{\phi}\;:\;\mathcal{A}\;&\longrightarrow&\;\mathcal{A},\qquad
\phi\in\Diff(\Sig).
\end{eqnarray}

\noindent The action of the groups $\G$ and $\Diff(\Sig)$ can be
easily extended to the quantum configuration space $\AL$. But also
the groups themselves can be extended to groups of generalized gauge
transformations and generalized diffeomorphisms, $\overline{\G}$ and
$\overline{\Diff(\Sig)}$, to give

\begin{eqnarray}\label{Gl:QuantumActionsOfGaugeAndDiffeos}
\a_g\;:\;\AL\;&\longrightarrow&\;\AL,\qquad
g\in\overline\G\\[5pt]\nonumber
\a_{\phi}\;:\;\AL\;&\longrightarrow&\;\AL,\qquad
\phi\in\overline{\Diff(\Sig)}.
\end{eqnarray}

\noindent The group of generalized gauge transformation
$\overline\G$ is taken to be the group of all maps from $\Sig$ to
$SU(2)$, i.e.

\begin{eqnarray}
\overline\G\;=\;SU(2)^{\Sig}.
\end{eqnarray}

\noindent Although classical GR is not invariant under this group,
in the quantum theory this arises as a natural candidate for the
extension of the smooth gauge transformations.

There are reasons to believe that also the diffeomorphisms have to
be extended: First, the Hilbert space $\Hdiff$ of states invariant
under diffeomorphisms, is not separable and contains many degrees of
freedom (the so-called moduli) that are believed to be unphysical
\cite{INTRO, ROVELLIFAIBAIRN}. Also, there are other physical
reasons to believe that the group of diffeomorphisms is too small
\cite{KOSL}. For an extension of the diffeomorphisms $\Diff(\Sig)$
to $\overline{\Diff(\Sig)}$, several suggestions have been made.

In \cite{ROVELLIFAIBAIRN}, it was shown that already a slight
extension of the group of diffeomorphisms gives rise to a separable
diff-invariant Hilbert space. In \cite{ALSTATUS}, Ashtekar and
Lewandowski discussed $C^n$-diffeomorphisms on $\Sig$, which are
analytic except for lower-dimensional subsets of $\Sig$. The proof
of the uniqueness of the diffeomorphism-invariant state
$\omega_{AL}$ on the holonomy-flux-algebra for Loop Quantum Gravity
uses these for $n\geq 1$ \cite{LOST}. In \cite{KOSL}, the stratified
diffeomorphisms, introduced earlier by Fleisch\-hack have been
investigated. In \cite{ZAP2} the piecewise analytic diffeomorphisms
have been introduced, which are bijections on $\Sig$ that leave the
set of analytical graphs $\Ga$ invariant. In \cite{FLEISCHHACK2},
the graphomorphisms extended this concept to the smooth and other
categories. 

In \cite{BAEZCAT, VELH2}, it was displayed how the basic ingredients
of Loop Quantum Gravity can be formulated naturally as concepts of
category theory, i.e. as morphisms, functors and natural
transformations. In this language, the connections arise as functors
from the path groupoid $\P$ of $\Sig$ to the suspension of the gauge
group $\Susp(SU(2))$, and the generalized gauge transformations are
in one-to-one correspondence to the natural transformations of these
functors. Furthermore, the diffeomorphisms acting on $\Sig$ can be
naturally interpreted as elements in the automorphism group $\Aut$,
i.e. as invertible functors from $\P$ to itself. Velhinho pointed
out that in the light of category language, $\Aut$ arises as a
candidate for an extension of the diffeomorphisms $\Diff(\Sig)$, and
this extension appears to be natural, at least from the mathematical
point of view.

In this article, we will investigate the consequences of choosing
$\overline{\Diff(\Sig)}=\Aut$. The automorphisms $\phi\in\Aut$ are
invertible functors on the path groupoid $\P$, i.e. they permute
points in $\Sig$, and also the paths between them in a consistent
way. We will, however, encounter elements in $\Aut$ that can not be
interpreted as bijections of $\Sig$. By this, the elements in $\Aut$
will also be able to map graphs into each other that have the same
combinatorics, but lie in different generalized knotting classes. By
this, a combinatorial picture emerges, which is a desirable feature
for a quantum theory of gravity \cite{INTRO, ZAP1, TINA1}.\\

The emphasis of this article lies on two topics: First, we will
prove that the automorphisms $\Aut$ leave the
Ashtekar-Isham-Lewandowski measure $\m_{AL}$ invariant, and hence
have a well-defined unitary action on the kinematical Hilbert space
$\Hkin=L^2(\AL,d\m_{AL})$. Second, we will have a closer look at the
automorphisms and the orbits of its action on $\Hkin$, in order to
describe the automorphism-invariant Hilbert space.

We will start in chapter \ref{Sec:Basics} by reviewing the basic
concepts of Loop Quantum Gravity, with emphasis on the categorial
formulation, and for general gauge group $G$. We introduce the
concept of a (primitive) metagraph, which will be useful in the
investigation of the automorphisms $\Aut$. We will continue by
presenting two kinds of nontrivial automorphisms in section
\ref{Sec:ExamplesForAutomorphisms}, which will both be not induced
by a bijection on $\Sig$, but are most useful in what follows. In
particular, with the help of these automorphisms we will prove in
chapter \ref{Sec:UnitarityOfAutomorphisms} that the automorphisms
leave the Ashtekar-Isham-Lewandowski measure invariant, but also be
able to show that any two graphs (in fact, hyphs) with the same
combinatorics can be mapped into each other by an automorphism. It
is in particular this fact which suggests that by using the
automorphisms, a combinatorial picture emerges.

In section \ref{Sec:TheAutomorphismInvariantHilbertSpace} we will
investigate the orbits of vectors in $\Hkin$ under the action of
$\Aut$. For a certain choice of rigging map, we will define the
Hilbert space of vectors invariant under the action of $\Aut$. For
the case of abelian Loop Quantum Gravity, i.e. $G=U(1)$, an explicit
orthonormal basis will be given. for $G=SU(2)$, we will comment on
how to obtain such a basis.

In the appendix, we will, after briefly presenting notions from
category theory and combinatorial group theory, present a way of how
to write the exponentiated fluxes in category language, and present
a categorial version of the Weyl- and the holonomy-flux algebra.

\section{Basics}\label{Sec:Basics}

Loop Quantum Gravity rests on the observation that, instead of
knowing a $\mathfrak{g}$-connection $A_a$, it is equivalent to know,
for every possible path $p$, its parallel transport
$A(p)=\P\exp\int_p A_adx^a$ along $p$, which is an element of the
gauge group $G$. These parallel transports, or holonomies, form the
configuration space $\mathcal{A}$ of the classical theory, which is
extended to $\AL$, the set of generalized connections, in the
quantum theory. The close relation to lattice gauge theory is rooted
in this formalism.

\subsection{Curves, ways and paths}

In the following chapter we will review the basic notions to deal
with the set of generalized connections, as well as their
categorical formulation\footnote{We follow the notation of
\cite{INTRO}, where the concatenation of, say, two curves $c_1$, and
$c_2$ is denoted as $c_1\circ c_2$, rather than $c_2\circ c_1$,
which is usually employed in category theory texts.}. For a more
detailed mathematical treatment and the omitted proofs, see e.g.
\cite{INTRO, ALGRAPHS, FLEISCHHACK1}. Also, we will restrict
ourselves to the category of piecewise analytic paths, which is
usually employed in Loop Quantum Gravity. In all what follows,
$\Sig$ will denote an analytic, connected, closed manifold of
dimension $n>2$.

\begin{Definition}
Let $\Sig$ be an analytic, connected, closed manifold of dimension
$n>2$. A (piecewise analytic) curve in $\Sig$ is a map
$c:[0,1]\to\Sig$, such that there exist $0=t_0<t_1<\ldots<t_n=1$, so
that $c$ restricted to $[t_k,t_{k+1}]$ is an analytic embedding.
Denote the set of all curves as $\Cyl$. Write
\begin{itemize}
\item $s(c)\;:=\;c(0)$
\item $t(c)\;:=\;c(1)$
\item $r(c)\;:=\;c([0,1])\,\subset\,\Sig$.
\end{itemize}
\end{Definition}

\noindent Some curves can be concatenated, and all of them can be
inverted:

\begin{Definition}
Let $c_1,c_2\in\Cyl$ such that $t(c_1)=s(c_2)$. Then
\begin{eqnarray}
c_1\circ
c_2(t)\;:=\;\left\{\begin{array}{cl}c_1(2t)&\quad t\in[0,\frac{1}{2}]\\[5pt] c_2(1-2t)&\quad t\in[\frac{1}{2},1]\end{array}\right.
\end{eqnarray}

\noindent defines an element in $\Cyl$, which is called the product
of $c_1$ and $c_2$. Furthermore, for any $c\in\Cyl$:

\begin{eqnarray}
c^{-1}(t)\;:=\;c(1-t)
\end{eqnarray}

\noindent also defined an element in $c^{-1}\in\Cyl$ which is called
the inverse of $c$.
\end{Definition}

\noindent All this endows $\Cyl$ with some structure, but so far the
multiplication $\circ$ is not associative.\\

\begin{Definition}
Define an equivalence relation on curves: $c_1\sim c_2$ if there is
a piecewise analytic, monotonically increasing function
$\varphi:[0,1]\to[0,1]$ such that $c_1(t)=c_2(\varphi(t))$. So two
curves $c_1,\,c_2$ are equivalent if and only if there is a sequence
of curves $d_1,\ldots,d_n\in\Cyl$ such that $d_k\sim d_{k+1}$ in the
above sense and $d_1=c_1$, $d_n=c_2$.

The set of all equivalence classes $[c]$ is called the set of ways,
and is denoted by $\tilde\Cyl$. For $w=[c]\in\tilde\Cyl$,
$s(w):=s(c)$, $t(w):=t(c)$ and $r(w):=r(c)$ are well-defined.
\end{Definition}

\noindent The set $\tilde\Cyl$ carries the structure of a category
in the following way: The objects of $\tilde\Cyl$ are points in
$\Sig$: $|\tilde\Cyl|=\Sig$. For $x,y\in\Sig$ the set of morphisms
$\Mor_{\tilde\Cyl}(x,y)$ is given by the set of all ways $w$
starting at $x$ and ending at $y$, i.e. $s(w)=x$, $t(w)=y$.  The
concatenation $[c_1]\circ[c_2]:=[c_1\circ c_2]$ is well-defined and
associative, since ways do not depend on a parametrization. Also
$[c]^{-1}:=[c^{-1}]$ defines an involution on $\tilde\Cyl$. For each
$x\in\Sig$ the identity $\id_x$ is given by the equivalence class
$[c]$ of the constant curve $c(t)=x$.\\

\begin{Definition}
Let $x,y\in\Sig$. On the elements of $\Mor_{\tilde\Cyl}(x,y)$ define
an equivalence relation by: Let $w_1\approx w_2$ if there are ways
$v, v_1, v_2$ such that $w_1=v_1\circ v_2$ and $w_2=v_1\circ v\circ
v^{-1}\circ v_2$. This generates an equivalence relation $\approx$
on $\Mor_{\tilde\Cyl}(x,y)$. The set of equivalence classes $[[w]]$
are called paths. The set of paths is denoted by $\P$.

The category which is defined by $|\P|:=|\tilde\Cyl|$ and
$\Mor_{\P}(x,y):=\Mor_{\tilde\Cyl}(x,y)\big/\approx$, as well as
$[[w]]\circ[[v]]:=[[w\circ v]]$ is called the path groupoid, and
will also be denoted by $\P$.
\end{Definition}

\noindent Usually, $\P$ is defined by defining an equivalence
relation $\begin{array}{c}{\sim}\\[-10pt]{\approx}\end{array}$ on
$\Cyl$ by combining the two equivalence relations above, and
directly go from the curves $c$ to their equivalence classes
$p=[[[c]]]$. However, in this work it will turn out to be more
convenient at some points to work with the category $\tilde\Cyl$, so
we also define it here.

Also note that $s([[w]]):=s(w)$ and $t([[w]]):=t(w)$ can also be
defined in $\P$, but $r([[w]])=r(w)$ is not well-defined. One can
define
\begin{eqnarray}\label{Gl:DefinitionOfRangeForPaths}
r(p)\;:=\;\bigcap_{p=[[w]]}r(w),
\end{eqnarray}

\noindent but this has the property that $r(p\circ q)\neq r(p)\cup
r(q)$. But $r(p\circ q)= r(p)\cup r(q)$ is a desirable property if
one wants to speak about ``points $x\in\Sig$ lying on $p$''. It is
this particular fact that makes the set of automorphisms of $\P$
much larger than the set of automorphisms of $\tilde\Cyl$.

\begin{Definition}
Let $p$ be a path in $\Sig$. If there is a representative $c\in\Cyl$
with $p=[[[c]]]$, such that from $c(t)=c(t')$ it follows that either
$t=t'$ or $t=0,\,t'=1$ or $t=1,\,t'=0$, then $p$ is said to have no
self-intersections.
\end{Definition}

\noindent Note that for paths without self-intersections it is
easier to talk about points lying on that path. Let $c$ be a
representative of such a path, then $r(p):=r(c)$ is well-defined and
gives the same result as (\ref{Gl:DefinitionOfRangeForPaths}). We
will use this notion in some constructions later.

\begin{Definition}
A path $e\in\P$ is called an edge if there is a representative $c\in
Cyl$ with $e=[[[c]]]$, such that $c$ is analytic, and $e$ has no
self-intersections.
\end{Definition}

\noindent The edges are special elements in $\P$. They are the
equivalence classes of analytic curves that do not self-intersect
(unless they start and end at the same point). By definition, every
path $p$ can be decomposed into finitely many edges, which will be
crucial for the rest of the work.

Edges are the key to gain access to the analytic structure of Loop
Quantum Gravity.

\begin{Definition}
Let $\g=\{e_1,\ldots,e_E\}$ be a set of edges in $\P$, such that the
following holds: For each $e_k$ there is a representative
$e_k=[[[c_k]]]$ in $\Cyl$ such that the $c_k$ mutually intersect at
most in their beginning- or endpoints. Then $\g$ is called a graph.
Denote the set of all graphs by $\Ga$.
\end{Definition}

\begin{Lemma}
Given two graphs $\g_1,\,\g_2$, such that each edge in $\g_1$ is a
finite product of edges and their inverses of $\g_2$. Then one
writes $\g_1\leq\g_2$. This defines a partial ordering, in
particular for any two graphs $\g_1,\,\g_2$ there is a graph $\g_3$
such that $\g_1\leq\g_3$ and $\g_2\leq\g_3$.
\end{Lemma}

This Lemma has an important corollary: Given any finite number of
paths $\{p_1,\ldots,p_n\}$ in $\P$, there is always a graph $\g$
such that each $p_k$ is a product of edges (and their inverses) in
$\g$. For this to hold, the piecewise analyticity of the curves in
$\C$ is essential. In particular, the same does not hold if one
drops the condition of analyticity and works in the smooth category.
However, one can work with the so-called webs, or hyphs, which can
also be used in this context and which generalize the concept of
graphs. In the analytic category, the definition of hyphs is as
follows:

\begin{Definition}
Let $v=(p_1,\ldots,p_n)$ be a finite sequence of paths with the
following property:

 For each $k\in\{1,\ldots,n\}$ the path $p_k$ has a segment
which is free with respect to the $p_l,\,l\in\{1,\ldots,k-1\}$. This
means that, given a graph $\g$ such that each $p_k$ is a product of
edges in $\g$ and their inverses, then in the decomposition
$p_k=e_{k_1}\circ\ldots\circ e_{k_{n_k}}$ there is one edge
$e_{k_l}$ (or its inverse) which appears exactly once, and which
does not appear in the decompositions of all
$p_l,\,l\in\{1,\ldots,k-1\}$.

Denote the set of all hyphs by $\V$.
\end{Definition}

\noindent The set of hyphs is also partially ordered. Since all
graphs are also hyphs, this is trivial in the analytic category, but
stays true in other cases, such as the smooth category as well
\cite{FLEISCHHACK1}. Note that in the original definition of a hyph,
the additional condition that each $p_k$ has no self-intersections.
This is a much stronger condition, but we will need the weaker
definition presented here
later.\\

In Loop Quantum Gravity, GR is written in terms of a gauge theory,
i.e. the notion of a compact, connected, semisimple Lie Group $G$ is
employed. The following definition a priori depends on this gauge
group. For LQG, the cases $G=SU(2)$ and $G=U(1)$ are the most
important.

\begin{Definition}
Let $G$ be a semisimple, connected, compact Lie group. Let
furthermore $\m=\{p_1,\ldots,p_M\}$ be a set of paths in $\P$ that
are algebraically independent. This means that for each set
$\{g_1,\ldots,g_M\},\,g_k\in G$, there is a functor
$A:\P\to\Susp(G)$ such that
\begin{eqnarray}
A(p_k)\,=\,g_k.
\end{eqnarray}

\noindent Then $\m$ is called a metagraph. Denote the set of all
metagraphs by $\M$.
\end{Definition}

\noindent Note that this notion depends on $G$:  Consider for
instance two loops $l_1\neq l_2$ and $p=l_1\circ l_2\circ
l_1^{-1}\circ l_2^{-1}$. Then $\{p\}$ is a metagraph for $G=SU(2)$,
but is not for $G=U(1)$.

For the gauge groups in question, however, one can show that
$\Ga\,\subset\,\V\,\subset\,\M$. So, each graph is a hyph, and each
hyph is a metagraph, but not the other way round\footnote{We will
see, for instance, that for every path $p$ with $s(p)\neq t(p)$,
$\{p\}$ is a metagraph.}. Define a partial ordering on $\M$ by the
same rule as for graphs or hyphs: $\m_1\leq\m_2$ iff every path
$p\in\m_1$ can be composed by paths in $\m_2$, or their inverses.
Since for every finite set of paths $\{p_k\}$ one can find a graph
$\g$ so that each $p_k$ can be composed of edges in $\g$ and their
inverses, and each graph is again a metagraph, $\leq$ defines a
partial ordering on $\M$.

The metagraphs are a useful concept when investigating the
automorphisms $\Aut$, since $\Aut$ leaves $\M$ invariant, which can
be seen easily. In contrast, we will encounter explicit examples of
elements $\phi\in\Aut$ that do not leave $\Ga$ or $\V$ invariant.
So, the mathematical concept of metagraphs will be quite useful in
order to
investigate the action of $\Aut$ on $\P$.\\

\subsection{Notions from Loop Quantum
Gravity}\label{Sec:NotionsFromLQG}

We now briefly review how the concept of the path groupoid $\P$ is
used in Loop Quantum Gravity, in particular how to define the
quantum configuration space $\AL$. We will do this in terms staying
as close as possible to category language. For a brief introduction
to category notions, see appendix \ref{App:CategoryNotions}.

\begin{Definition}
Let $\P$ be the path groupoid of $\Sig$, and $G$ a compact,
connected, semisimple Lie group. Then a functor $A:\P\to\Susp(G)$ is
called a (generalized) connection\footnote{In the literature, this
space is also denoted as $Hom(\P, G)$, the set of groupoid
homomorphisms from $\P$ to $G$.}. This means that $A$ maps paths in
$\P$ to elements in $G$ such that
\begin{eqnarray}
A(p\circ q)\;&=&\;A(p)\cdot A(q)\\[5pt]\nonumber
A(p^{-1})\;&=&\;A(p)^{-1}.
\end{eqnarray}
\noindent The set of all connections is denoted by $\AL$.
\end{Definition}

\noindent It is clear that every smooth
$\mathfrak{su}(2)$-connection $A^I_a(x)$ gives rise to such a
functor, by mapping each path $p\in\Sig$ to the holonomy of $A^I_a$
along $p$:
\begin{eqnarray}
A(p)\;:=\;\P\exp\int_pA^I_a\frac{\t_I}{2}\,dx^a
\end{eqnarray}

\noindent In this sense the set of all smooth connections $\mathcal
A$ is a subset of $\AL$.

Under a gauge transformation $g\in \G=C^{\infty}(\Sig, SU(2))$, the
holonomy $A(p)$ along a path $p$ changes as
\begin{eqnarray}
A(p)\;\longrightarrow\;g_x\,A(p)\,g_y^{-1},
\end{eqnarray}

\noindent if $p$ starts at $x\in\Sig$ and ends at $y\in\Sig$, and
$g_x$ is the value of the function $g$ at $x\in\Sig$. This motivates
the following definition:

\begin{Definition}
A natural transformation $g$ on functors $A\in\AL$ is called a
(generalized) gauge transformation. The set of all such gauge
transformations is denoted by $\overline{\G}$.
\end{Definition}

\noindent Recall that functors $A_1$ and $A_2$ can be related by a
natural transformation, if there is for every object $x\in|\P|=\Sig$
a morphism $g_x:A_1(x)\to A_2(x)$ such that the following diagram
commutes for all $p\in\Mor_{\P}(x,y)$:
\begin{eqnarray*}
\begin{CD}
A_1(x) @>A_1(p)>> A_1(y) \\
@Vg_xVV       @VVg_yV\\
A_2(x) @>A_2(p)>> A_1(y)
\end{CD}
\end{eqnarray*}

\noindent Since $\Susp(G)$ has only one object $*$, $A(x)=*$ for all
$A$ and $x$. This amounts to say that for each $x\in\Sig$ there is
an element $g_x\in G$ such that
\begin{eqnarray}\label{Gl:GaugeTransformation}
A_1(p)\;=\;g_x\,A_2(p)\,g_y^{-1}.
\end{eqnarray}

\noindent This justifies the name gauge transformation, and shows
that the set $\overline\G\simeq G^{\Sig}$. Thus, given a functor $A$
and $\{g_x\}_{x\in\Sig}\in\overline\G$,
(\ref{Gl:GaugeTransformation}) can be seen as a definition of the
gauge-transformed functor
$\a_gA(p)\;:=\;g_{s(p)}\;A(p)\;g_{t(p)}^{-1}$. So the set
$\overline\G$ acts on $\AL$.

This immediately shows that $\overline\G=SU(2)^{\Sig}$, i.e. the set
of all maps from $\Sig$ to $SU(2)$, without any smoothness (or
continuity, measurability) condition. It is clear that this is a
tremendous extension to a symmetry group which is not a symmetry of
classical GR anymore. This shows that the quantum theory is in fact
invariant under larger groups, having to do with the fact that
space-time becomes discrete in some sense: Gauging can happen at
each point in space completely independent of each other. The same
can be done for the diffeomorphisms: Every, say analytical,
diffeomorphism $\phi$ acts in the path groupoid $\P$ in the
following way: points are mapped to points, and paths to paths:
\begin{eqnarray}
x\;&\longmapsto&\;\phi(x)\\[5pt]\nonumber
p\;&\longmapsto&\;\phi(p)
\end{eqnarray}

\noindent with $\phi([[[c]]]):=[[[\phi\circ c]]]$ for representative
curves $c\in\Cyl$. If $p$ starts at $x$ and ends at $y$, then
$\phi(p)$ starts at $\phi(x)$ and ends at $\phi(y)$. This means that
$\phi$ induces a functor on the path groupoid $\P$, which is
invertible since $\phi$ is as a map.

\begin{Definition}
Let $\phi:\P\to\P$ be an invertible functor. Then $\phi$ is called
an automorphism of $\P$. Denote the set of all automorphisms by
$\Aut$.
\end{Definition}

\noindent Note that $\Aut$ also acts on $\AL$, via
$\a_{\phi}A(p):=A(\phi(p))$. By this, $\Aut$ appears as an extension
of $\Diff(\Sig)$.\\


It is this set of automorphisms $\Aut$ that we will focus our
attention on for the rest of the article. The automorphisms extend
the analytic diffeomorphisms, and is the largest possible extension
\cite{VELH2}. We will comment on the actual size of $\Aut$ in
contrast to $\Diff(\Sig)$ later in this article. It should be noted
that each invertible functor $\zeta:\tilde\Cyl\to\tilde\Cyl$, i.e.
$\zeta\in\mathcal{A}\hspace{-1pt}{\it ut}(\tilde\Cyl)$ induces also
an element $\phi_{\zeta}\in\Aut$. In particular, the set
$\{\phi_{\zeta}|\zeta\in\mathcal{A}\hspace{-1pt}{\it
ut}(\tilde\Cyl)\}$ forms a subgroup of $\Aut$, which has been
investigated in the literature \cite{FLEISCHHACK2,ZAP1}, called
``piecewise analytic diffeomorphisms'', or ``graphomorphisms''.
However this is a proper subgroup: As we will see later, there are
many automorphisms in $\Aut$ which are no graphomorphisms.

The reason for this can be seen as follows: The automorphisms
permute points of $\Sig$, and also permutes the paths between
points, in a consistent way. Consistent means that if
$p\in\Mor_{\P}(x,y)$, then $\phi(p)\in\Mor_{\P}(\phi(x),\phi(y))$.
 However, a path itself consists (if it is without
self-intersection) of many points. So one could feel that for $z$
lying on $p$ (more precisely: $z\in r(p)$), then also $\phi(z)$
should lie on $\phi(p)$. Even more, since $p$ can be decomposed into
$p_1\circ p_2$, where $p_1$ is the part of $p$ which goes from $x$
to $z$, and $p_2$ is the remainder, which goes from $z$ to $y$.
Since $\phi(p)=\phi(p_1\circ p_2)=\phi(p_1)\circ \phi(p_2)$, one
might think that, since $\phi(p_1)$ ends at $\phi(z)$, and
$\phi(p_2)$ starts at $z$, $\phi(p)$ should pass through $\phi(z)$.
But this is not the case, as the following picture shows:

\begin{figure}[hbt!]\label{fig:AutomorphismsAreCool}
\begin{center}
    \psfrag{x}{$x$}
    \psfrag{y}{$y$}
    \psfrag{z}{$z$}
    \psfrag{p_1}{$p_1$}
    \psfrag{p_2}{$p_2$}
    \psfrag{p}{$p=p_1\circ p_2$}
    \psfrag{phix}{$\phi(x)$}
    \psfrag{phiy}{$\phi(y)$}
    \psfrag{phiz}{$\phi(z)$}
    \psfrag{phip}{$\phi(p)=\phi(p_1)\circ \phi(p_2)$}
    \psfrag{phip1}{$\phi(p_1)$}
    \psfrag{phip2}{$\phi(p_2)$}
    \includegraphics[scale=0.75]{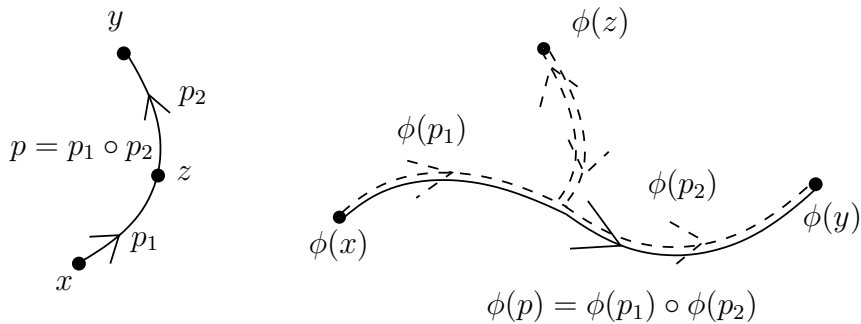}
    \end{center}
    \caption{\small Here $z$ lies on $r(p)$, but $\phi(z)$ does not lie on $r\big(\phi(p)\big)$.}
\end{figure}

\noindent Let $p$ be a path without self-intersections, which is
composed of $p=p_1\circ p_2$, i.e. $p$ passes through $z:=t(p_1)$.
The images of $p_1$ and $p_2$ under $\phi$ are given by the dashed
lines, the solid line is $\phi(p)$. $\phi(p_1)$ ends at $\phi(z)$
and $\phi(p_2)$ starts at it, but since $\phi(p_1)\circ \phi(p_2)$
contains a retracing, $\phi(p)=\phi(p_1)\circ\phi(p_2)$ does not
pass through $z$. We see that the fact that retracings cancel out in
$\P$, the ill-definedness of $r(p)$, and the existence of
automorphisms which are not induced by maps from $\Sig$ to $\Sig$,
are deeply interrelated to each other.

\subsection{Automorphisms and connections}\label{Sec:Metagraphs}

In this section we will investigate the metagraphs further, and
their relation to the automorphisms. We will extend the LQG-notions
for graphs \cite{ALGRAPHS} to metagraphs, and in particular show
that the topology on $\AL$ can also be defined in terms of
metagraphs. This will allow to prove i.e. continuity of the action
of automorphisms $\Aut$ on the set of connections $\AL$. So, fix a
compact, connected and semisimple Lie group $G$ for the rest of this
section. We will refer to $G$ as the gauge group.

The partial ordering defined on the set of metagraphs $\M$ has a
category theory background, which we will use in the following.

\begin{Definition}
Let $\P$ be the path groupoid of $\Sig$, and $\m\in\M$ a metagraph.
Then define $\P_{\m}$ to be the subgroupoid of $\P$ which is
generated by the elements in $\m$.
\end{Definition}

\noindent This groupoid contains the elements in $\m$, their
inverses, the identities $\id_{s(p_k)}$, $\id_{t(p_k)}$, and all
products that can be formed of them. Thus all $\P_{\m}$ are finitely
generated subgroupoids of $\P$\footnote{But not the other way round,
as the example for $G=U(1)$ above suggests.}. Note that by this
definition, also $\P_{\g}$ for $\g\in\Ga$ and $\P_{v}$ for $v\in\V$
are declared.

We immediately conclude:

\begin{Corollary}
For two metagraphs $\m_1,\m_2\in\M$, we have that $\m_1\leq\m_2$ if
and only if $\P_{\m_1}$ is a subgroupoid of $\P_{\m_2}$.
\end{Corollary}

\noindent The same holds for graphs and hyphs. In fact, a topology
on the set $\AL$ of generalized connections $A$, or equivalently
functors $A:\P\to\Susp(G)$ is defined by convergence on the finitely
generated subgroupoids $\P_{\g}$ for al $\g\in\Ga$. In
\cite{FLEISCHHACK1} it was shown that the same topology is defined
if one replaces $\Ga$ by $\V$. In fact, defining the topology on
$\AL$ by using the metagraphs leads to the same result, as we will
briefly indicate in the following.

\begin{Definition}
Let $\P$ be the path groupoid, and $\AL$ the set of all morphisms
$A:\P\to\Susp(G)$. For a metagraph $\m$ denote the set of all
morphisms from $\PG_{\m}$ to $\SU$ by $\AL_{\m}$. Define the
projection

\begin{eqnarray}
&&\pi_{\m}\;:\;\AL\;\longmapsto\;\AL_{\m}\\[5pt]\nonumber
&&(\pi_{\m}A)(p_k)\;:=\;A(p_k)
\end{eqnarray}

\noindent For $\m_1\leq\m_2$ define the projections
$\pi_{\m_1\m_2}\;:\AL_{\m_2}\to\AL_{\m_1}$ via

\begin{eqnarray}
\big(\pi_{\m_1\m_2}A\big)(p)\;:=\;A(q_1)\cdot\ldots\cdot A(q_n)
\end{eqnarray}

\noindent if $p\in\m_1$ can be written by $p=q_1\circ\ldots\circ
q_n$, the $q_k$ being elements in $\m_2$ or their inverses.
\end{Definition}

\noindent It is easy to show that
$\pi_{\m_1,\m_2}\circ\pi_{\m_2}\;=\;\pi_{\m_1}$ for $\m_1\leq\m_2$.
By the definition of metagraphs, each $\AL_{\m}$,
$\m=\{p_1,\ldots,p_M\}$ comes with a natural bijection
$A\longmapsto(A(p_1),\ldots,A(p_M))$. Pulling the topology of $G^M$
back to $\AL_{\m}$ makes all the $\AL_{\m}$ into compact Hausdorff
spaces.

\begin{Definition}
Define the topology on $\AL$ to be the weakest topology such that
all the projections $\pi_{\m}$ are continuous.
\end{Definition}

\noindent One can show that with this topology $\AL$ becomes a
compact Hausdorff space. The proof goes entirely along the same
lines as in the case for graphs or hyphs \cite{INTRO, FLEISCHHACK2},
and rests crucially on the compactness of $G$. We will not repeat
the proof here.

In fact, this topology coincides with the topology which is defined
by the condition that all $\pi_{\g}$ for $\g\in\Ga$ are continuous.
%
%
%
%

\begin{Lemma} Let $\T_1$ be the weakest topology
such that for all $\m\in\M$ the map $\pi_{\m}$ is continuous. Let
$\T_2$ be the weakest topology such that for all graphs $\g\in\Ga$
the map $\pi_{\g}$ is continuous. Then the identity map between the
two topological space
\begin{eqnarray}
\id\;:\;(\AL,\T_1)\;\longrightarrow\;(\AL,\T_2)
\end{eqnarray}
\noindent is a homeomorphism.
\end{Lemma}

\noindent\textbf{Proof:} Since $\pi_{\g}^{-1}(U)$ are a basis for
$\T_2$, it is sufficient to show that $\id^{-1}(\pi_{\g}^{-1}(U))$
are open in $\T_1$. But this is clear, since every graph is also a
metagraph. Thus, $\id$ is a continuous bijection between compact
Hausdorff spaces, hence also a homeomorphism.\\

Thus, defining the topology on $\AL$ by means of metagraphs is
completely analogous to defining it by graphs. However, since the
invertible functors $\phi\in\Aut$ leave the set of metagraphs
invariant, unlike the set of graphs, in this formulation it is much
easier to prove that the automorphisms act continuously on $\AL$.

\begin{Lemma}
Let $\phi:\P\to\P$ be an invertible functor, i.e. $\phi\in\Aut$.
Then the map
\begin{eqnarray}
&&\a_{\phi}\;:\;\AL\;\longrightarrow\;\AL\\[5pt]\nonumber
&&(\a_{\phi}A)(p)\;:=\;A(\phi(p))
\end{eqnarray}

\noindent is a homeomorphism.
\end{Lemma}

\noindent\textbf{Proof:} Let $\m=\{p_1,\ldots,p_M\}$ be a metagraph.
In the following, we deliberately use the homeomorphism
$\AL_{\m}\simeq G^M$. For $A\in\AL$ we have
\begin{eqnarray}
\pi_{\m}(A)\;&=&\;\Big(A(p_1),\ldots,A(p_M)\Big)\\[5pt]\nonumber
&=&\;\Big(\a_{\phi^{-1}}A\big(\phi(p_1)\big),\ldots,\a_{\phi^{-1}}A\big(\phi(p_M)\big)\Big)\\[5pt]\nonumber
&=&\;\pi_{\phi(\m)}\big(\a_{\phi^{-1}}A\big),
\end{eqnarray}

\noindent i.e. we get
\begin{eqnarray}
\pi_{\m}\circ\a_{\phi}\;=\;\pi_{\phi(\m)}.
\end{eqnarray}

\noindent From this it follows that, for each open $U\in
\AL_{\m}\simeq\AL_{\phi(\m)}$, one has
\begin{eqnarray}
(\pi_{\phi(\m)})^{-1}(U)\;=\;(\a_{\phi})^{-1}\Big((\pi_{\m})^{-1}(U)\Big).
\end{eqnarray}

\noindent But since the $(\pi_{\m})^{-1}(U)$ form a basis of the
topology on $\AL$, preimages of open sets under $\a_{\phi}$ are
open, hence $\a_{\phi}$ is continuous. Since this is true for all
automorphisms $\phi\in\Aut$ and each $\phi$ is invertible, all
automorphisms $\a_{\phi}:\AL\to\AL$ are homeomorphisms. This concludes the proof.\\


\noindent There is also a normalized regular Borel measure $\m_{AL}$
on $\AL$, which is called the Asthekar-Isham-Lewandowski measure. It
is uniquely determined by the condition that its push-forward
$(\pi_{\g})_*\m_{AL}=\m_H$ is the normalized Haar measure on
$\AL_{\g}\simeq G^{|\g|}$. One could think that the metagraphs
$\m\in\M$ could now used in a similar way to define a measure on
$\AL$, which then would automatically be $\m_{AL}$. However, the
concept of metagraphs is slightly too broad for this: There is in
general no measure $\n$ on $\AL$ such that $(\pi_{\m})_*\n\;=\;\m_H$
on $\AL_{\m}$. This can be seen as follows: Choose $G=SU(2)$ and
$l\in\P$ be a nontrivial loop, which is also an edge, i.e.
$l\in\Mor_{\P}(x,x)$ for some $x\in\Sig$. Then $\m_1:=\{l\}$ as well
as $\m_2:=\{l^2\}$ are metagraphs, since one can take square roots
in $SU(2)$. Now $(\pi_{\m_1})_*{\m_{AL}}=\m_H$ on $SU(2)$, but
$(\pi_{\m_2})_*{\m_{AL}}=\m_H$ would imply
\begin{eqnarray}
\int_{SU(2)}d\m_H(h)\;F(h)\;=\;\int_{SU(2)}d\m_H(h)\;F(h^2)
\end{eqnarray}

\noindent for all continuous functions $F$ on $SU(2)$. But this is
wrong! So, in order to define $\m_{AL}$ on $\AL$ without referring
to graphs of hyphs, we need the following notion:

\begin{Definition}
Let $\m\in\M$ be a metagraph such that the push-forward of the
Ashtekar-Isham-Lewandowksi measure $\m_{AL}$ by $\pi_{\m}$ is
\begin{eqnarray}
(\pi_{\m})_*\m_{AL}\;=\;\m_H
\end{eqnarray}
\noindent the normalized Haar measure $\m_H$ on $\AL_{\m}\simeq
G^M$. Then $\m$ is called a primitive metagraph.
\end{Definition}

\noindent This notion is the first which is not defined entirely in
terms of category theory, but makes use of the analytic structure on
$\Sig$, through the use of the Ashtekar-Isham-Lewandowski measure,
which is defined in terms of graphs (or hyphs). Thus, it is not at
all clear whether the automorphisms $\Aut$ preserve the set of
primitive metagraphs. Note that this is equivalent to
$(\a_{\phi})^*\m_{AL}=\m_{AL}$, i.e. the question whether the
automorphisms leave $\m_{AL}$ invariant, or the question whether the
induced operators on $\Hkin=L^2(\AL,d\m_{AL})$, given by $(\hat
U(\phi)\psi)(A):=\psi(\a_{\phi}A)$ are all unitary.

We will provide a proof for this assertion in section
\ref{Sec:UnitarityOfAutomorphisms}. For this proof some explicit
examples of automorphisms will play a crucial r\^{o}le,
which will be presented in the following section.\\


\section{Examples for
automorphisms}\label{Sec:ExamplesForAutomorphisms}

It is obvious that every analytic diffeomorphism $\phi:\Sig\to\Sig$
induces an automorphism $\phi\in\Aut$. In this sense, $\Diff(\Sig)$
is a subgroup of $\Aut$. But also all invertible functors on
$\tilde\Cyl$, which are in one-to-one correspondence with maps from
$\Sig$ to $\Sig$ mapping hyphs to hyphs, descend to automorphisms on
$\P$ by $\phi[[w]]:=[[\phi(w)]]$.

But, there are many automorphisms $\phi\in\Aut$ that are not induced
by a map $\Sig\to\Sig$. The reason for this is deeply connected to
the groupoid structure of $\P$. By declaring retracings to be
equivalent to the identity $[[\d\circ\d^{-1}]]=[[\id_{s(\d)}]]$, one
disconnects the paths from the points: If $p$ is a path that passes
through a point $x$, i.e. $x\in r(p)$, the path $p\circ q$ does not
necessarily also have this property. Thus, there is no good notion
for a point $x$ lying on a path $p$. This is the reason why there
are automorphisms $\phi\in\Aut$ that have actions on points
(objects) and paths (morphisms) which are not compatible with each
other. In the following, we will give two extreme examples for this:

The first example are the natural transformations of the identity,
which arbitrarily permute the points in $\Sig$, while leaving the
paths essentially invariant. These will be crucial in the proof that
the automorphisms act unitarily on the kinematical Hilbert space
$\Hkin$.

The second example for non-trivial automorphisms will be the
edge-interchangers, which interchange two edges with identical
beginning- and endpoints, but leave all objects (points $x\in\Sig$)
invariant, as well as all paths intersecting the two given edges at
most in finitely many points. These will be most helpful in
determining the size of the orbits of vectors in $\Hkin$ under the
action of $\Aut$, in order to compute the automorphism-invariant
Hilbert space $\Haut$.


\subsection{Natural transformations of the
identity}\label{Sec:NTDID}

\noindent Recall from category theory that, given two functors $F,
G:\Cyl\;\to\;\Dyl$, the two are called to be natural transformations
from each other, if for each object $X$ in $\Cyl$ there is a
morphism $g_X:F(X)\to G(X)$ such that the following diagram
commutes:
\begin{eqnarray*}
\begin{CD}
F(X) @>F(f)>> F(Y) \\
@Vg_XVV       @VVg_YV\\
G(X) @>G(f)>> G(Y)
\end{CD}
\end{eqnarray*}

\noindent In the context of functors $A:\P\to\Susp(SU(2))$, two such
functors (generalized connections) $A_1,\,A_2$ are natural
transformations of each other if and only if the one is a gauge
transformed of the other. But also automorphisms can be natural
transformations of each other. In particular, two automorphisms
$\phi_1,\phi_2\in\Aut$ are natural transformations of each other, if
and only if there is a bijection $b:\Sig\to\Sig$ and for each
$x\in\Sig$ a path $p_x\in\Mor(x, b(x))$, such that for every path
$p\in\Mor(x, y)$:
\begin{eqnarray}
\phi_2(p)\;=\;p_{\phi_1(x)}^{-1}\;\circ\;\phi_1(p)\;\circ\;p_{\phi_1(y)}
\end{eqnarray}

\noindent Note that this requires $b(x)=\phi_2\circ\phi_1^{-1}(x)$.
Given $\phi_1$, $b$ and $\{p_x\}_{x\in\Sig}$, this can also be seen
as a definition of the transformed functor $\phi_2$. One special
case occurs for $\phi_1=\id$ being the identity functor.

\begin{Definition} Let $b:\Sig\to\Sig$ be a bijection and, for every
$x\in\Sig$ a path $p_x\in\Mor(x, b(x))$ be given. The functor
$\phi_{b,p}$ defined by
\begin{eqnarray}
\phi_{b,p}(x)\;&:=&\;b(x)\\[5pt]
\phi_{b,p}(p)\;&:=&\;p_{s(p)}^{-1}\;\circ\;p\;\circ\;p_{t(p)}
\end{eqnarray}

\noindent is called a \emph{natural transformation of the identity}.
\end{Definition}

\noindent Given two bijections $b_1,\, b_2$ and $p^1_x\in\Mor(x,
b_1(x))$ and $p^2_x\in\Mor(x, b_2(x))$, then one has
\begin{eqnarray}
\phi_{b_1,p^2}\;\circ\;\phi_{b_1,p^1}\;=\;\phi_{b,p}
\end{eqnarray}

\noindent with $b:=b_2\circ b_1$ and $p_x\in\Mor(x, b_2(b_1(x)))$
given by $p_x:=p^1_x\circ p^2_{b_1(x)}$. In particular, by choosing
$b_2=b_1^{-1}$ and $p_x^2=\big(p^1_{b_1(x)}\big)^{-1}$, one sees
that every such functor is invertible, hence an automorphism.

\begin{Corollary}
The natural transformations of the identity form a subgroup $\Nyl$
of $\Aut$.
\end{Corollary}

\noindent The natural transformations of the identity will be of
particular importance later on. In particular the group structure
will play a prominent role in the proof that all automorphisms act
unitarily on $\Hkin$.\\


\subsection{Edge-interchanger}\label{Sec:Edge-Interchanger}

The following example of an automorphism will prove to be most
important in order to compute the automorphism-invariant Hilbert
space $\Haut$. It will be an example of a functor which acts
trivially on $\Sig$, but modifies the morphisms. In particular, it
will interchange two edges (or paths without self.intersections))
$e_1,\,e_2$ with the same starting- and ending point. On the other
hand, it will leave every edge which intersects $e_1,\,e_2$ in at
most finitely many points invariant. It will therefore be termed
``edge-interchanger''.\\

Let $e_1,\, e_2$ be two paths in $\Mor(x, y)$ for $x\neq y$ without
self-intersection, and such that they do not mutually intersect,
apart from their beginning- and endpoints. First, choose
representative curves $c_1,\,c_2\in\Cyl$, i.e.
$[[[c_1]]]=e_1,\,[[[c_2]]]=e_2$, which contain no retracings, i.e.
the maps $c_1,\,c_2:[0,1]\to\Sig$ are injective. Furthermore, choose
for any $t\in(0,1)$ a path $p_t\in\Mor_{\P}(c_1(t),\,c_2(t))$, such
that all paths $p_t$ have a representative that intersects
$r(c_1),\,r(c_2)$ only at the respective beginning- and endpoints.
Define $p_0=\id_{s(e_1)}$ and $p_1=\id_{t(e_1)}$ to be the constant
paths. For $t_1,t_2\in[0,1]$ denote by $c_1^{t_1,t_2}$ the curve

\begin{eqnarray}
c_1^{t_1,t_2}\;:\;[0,1]\;\ni\;t\;\longmapsto\;c_1(t_1+t(t_2-t_1)),
\end{eqnarray}

\noindent and a similar definition of $c_2^{t_1,t_2}$. Note that
this definition also makes sense for $t_1>t_2$, in particular
$c_1^{t_1,t_2}=(c_1^{t_2,t_1})^{-1}$.\\

With this data, we now build a functor $\tilde\phi:\tilde\Cyl\to\P$
as follows:

\begin{Lemma}\label{Lem:DefinitionOfEdgeInterchanger}
Let $w\in\Mor_{\tilde\Cyl}(x,y)$ be a way in $\Sig$. Then divide $w$
according to the edges $e_1,\,e_2$:
\begin{eqnarray}
w\;=\;w_1\circ\,w_2\,\circ\ldots\circ\,w_n.
\end{eqnarray}

\noindent where the $w_k$ falls into either of the following
categories:
\begin{itemize}
\item $r(w_k)\;\cap\;r(e_{1,2})\;=\;r(w_k)\quad\Leftrightarrow\quad
w_k\;=\;[c_1^{t_1,t_2}]$ or $[c_2^{t_1,t_2}]$
\item $r(w_k)\;\cap\;r(e_{1,2})\;\subset\;\{s(w_k),t(w_k)\}$.
\end{itemize}

\noindent Then the following assignment
\begin{eqnarray}\label{Gl:DefinitionOfEdgeInterchangerOnNotSegments}
\tilde\phi(w)\;:=\;\tilde\phi(w_1)\;\circ\,\ldots\,\circ\;\tilde\phi(w_n)\;\in\;\Mor_{\P}(x,y)
\end{eqnarray}

\noindent with
\begin{eqnarray}\nonumber
w_k\,=\,[c_1^{t_1,t_2}]\;&\quad&\;\Rightarrow\;\quad\;\tilde\phi(w_k)\;:=\;p_{t_1}\;\circ\;[[[c_2^{t_1,t_2}]]]\;\circ\;p_{t_2}^{-1}\\[5pt]\nonumber
w_k\,=\,[c_2^{t_1,t_2}]\;&\quad&\;\Rightarrow\;\quad\;\tilde\phi(w_k)\;:=\;p_{t_1}^{-1}\;\circ\;[[[c_1^{t_1,t_2}]]]\;\circ\;p_{t_2}\\[5pt]\label{Gl:DefinitionOfEdgeInterchangerOnSegments}
r(w_k)\;\cap\;r(e_{1,2})\;\subset\;\{s(w_k),t(w_k)\}\;&\quad&\;\Rightarrow\;\quad\;\tilde\phi(w_k)\;:=\;[[w_k]].
\end{eqnarray}

\noindent defines a functor $\tilde\phi\;:\;\tilde\Cyl\;\to\;\P$.
\end{Lemma}

\noindent\textbf{Proof:} What has to be shown first is that the
above assignment is well-defined, i.e. does not depend on the manner
the way $w\in\Mor_{\tilde\Cyl}(x,y)$ is decomposed w.r.t. the edges
$e_1,\,e_2$. First we note that, given two decompositions
$w_1\circ\ldots\circ w_n$ and $w'_1\circ\ldots\circ w'_m$ of $w$,
there is a decomposition $w_1''\circ\ldots\circ w''_N$ of $w$ such
that each $w_k,\,w_l'$ is a product of the $w''_r$. Thus, if we can
show that $\tilde\phi(w)$ defined by one decomposition of $w$ does
not change if we decompose the decomposition further, we are done.\\

So let $w=w_1\circ\ldots\circ w_n$ and $w=w_1'\circ\ldots\circ w'_M$
be two decompositions of $w$ w.r.t. the edges $e_1$ and $e_2$, such
that each $w_k$ is a product of the $w_l'$. We need to show that
\begin{eqnarray}
\tilde\phi(w_1)\circ\ldots\circ\tilde\phi(w_n)\;=\;\tilde\phi(w_1')\circ\ldots\circ\tilde\phi(w'_m)
\end{eqnarray}

\noindent where the $\tilde\phi(w_k),\,\tilde\phi(w_l')$ are defined
according to (\ref{Gl:DefinitionOfEdgeInterchangerOnSegments}). Let
now $w_k=[c_1^{t_1,t_2}]$, and $w_k=w'_{l}\circ\ldots\circ
w'_{l+l'}$. Then there are points $t_1=t_l<t_{l+1}<\ldots
<t_{l+l'+1}=t_2$ such that
$w'_{l+j}\,=\,[c_1^{t_{l+j},t_{l+j+1}}]$.\footnote{This only holds
if $t_1<t_2$. If $t_1>t_2$ then the corresponding points have the
property that $t_1=t_l>t_{l+1}>\ldots >t_{l+l'+1}=t_2$. The proof is
then analogous.} Then we have

\begin{eqnarray}\nonumber
\tilde\phi(w'_l)\circ\ldots\circ\tilde\phi(w'_{l+l'})\;&=&\;p_{t_l}\circ[[[c_2^{t_l,t_{l+1}}]]]\circ
p_{t_l+1}^{-1}\,\circ\,p_{t_l+1}\circ[[[c_2^{t_{l+1},t_{l+2}}]]]\circ
p_{t_{l+2}}^{-1}\circ\ldots\\[5pt]\nonumber
&&\ldots\;\circ p_{t_{l+l'}}^{-1}\circ
p_{t_{l+l'}}\circ[[[c_2^{t_{l+1},t_{l+2}}]]]\circ
p_{t_{l+l'+1}}^{-1}\\[5pt]
&=&\;p_{t_l}\circ[[[c_2^{t_l,t_{l+1}}]]]\circ\ldots\circ[[[c_2^{t_{l+1},t_{l+2}}]]]\circ
p_{t_{l+l'+1}}^{-1}\\[5pt]\nonumber
&=&\;p_{t_1}\circ[[[c_2^{t_l,t_{l+1}}\circ\ldots\circ
c_2^{t_{l+1},t_{l+2}}]]]\circ p_{t_2}^{-1}\\[5pt]\nonumber
&=&\;p_{t_1}\circ[[[c_2^{t_1,t_{2}}]]]\circ
p_{t_2}^{-1}\\[5pt]\nonumber
&=&\;\tilde{\phi}(w_k).
\end{eqnarray}

\noindent Analogous relations hold in the case that
$w_k=[[[c_2^{t_1,t_2}]]]$.

\begin{figure}[hbt!]\label{fig:EdgeInterchanger}
\begin{center}
    \psfrag{x}{$x$}
    \psfrag{y}{$y$}
    \psfrag{w1}{$w_1$}
    \psfrag{w2}{$w_2$}
    \psfrag{w3}{$w_3$}
    \psfrag{w4}{$w_4$}
    \psfrag{w5}{$w_5$}
    \psfrag{w6}{$w_6$}
    \psfrag{p}{$p$}
    \psfrag{phip}{$\phi(p)$}
    \psfrag{e1}{$e_1$}
    \psfrag{e2}{$e_2$}
    \includegraphics[scale=0.75]{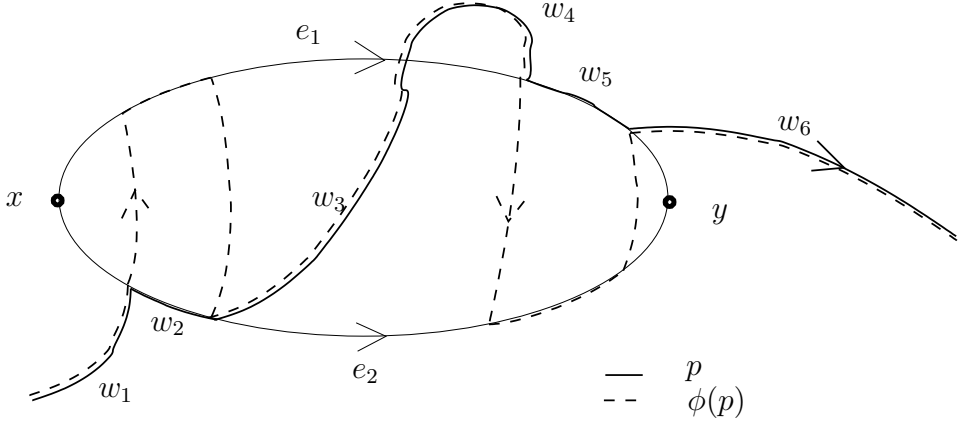}
    \end{center}
    \caption{\small The action of an edge-interchanger. The solid line shows some path $p$, the dashed line shows $\phi(p)$.}
\end{figure}

Let now $w_k$ be such that
$r(w_k)\;\cap\;r(e_{1,2})\;\subset\;\{s(w_k),t(w_k)\}$, and
$w_k=w'_l\circ w'_{l+l'}$. Then obviously also
$r(w'_{l+j})\;\cap\;r(e_{1,2})\;\subset\;\{s(w'_{l+j}),t(w'_{l+j})\}$
for all $j\in\{1,\ldots,l'\}$. By
(\ref{Gl:DefinitionOfEdgeInterchangerOnSegments}) we immediately see
that
\begin{eqnarray}
\tilde\phi(w'_l)\circ\ldots\circ\tilde\phi(w'_{l+l'})\;=\;\tilde\phi(w_k).
\end{eqnarray}

\noindent Since these are the only cases occurring, we conclude that
$\tilde\phi$ is a well-defined map of morphisms in
$\Mor_{\tilde\Cyl}(z,z')$ to morphisms in $\Mor_{\P}(z,z')$ for all $z,z'\in\Sig$. \\

\noindent The properties of a functor remain to be shown. But given
two morphisms $v,w$ in $\tilde\Cyl$ such that $t(v)=s(w)$, and
decompositions $v=v_1\circ\ldots \circ v_n$ and
$w=w_1\circ\ldots\circ w_m$ w.r.t. $e_1$ and $e_2$, then $v\circ
w\;=\;v_1\circ\ldots \circ v_n\circ w_1\circ\ldots\circ w_m$ is a
decomposition of $v\circ w$ w.r.t. $e_1$ and $e_2$. By
(\ref{Gl:DefinitionOfEdgeInterchangerOnNotSegments}) we see that

\begin{eqnarray}
\tilde\phi(v\circ w)\;=\;\tilde\phi(v)\,\circ\,\tilde\phi(w).
\end{eqnarray}

\noindent Since $\phi$ leaves beginning- and endpoints of paths
invariant, we conclude that the map induces a functor
$\tilde\phi:\tilde\Cyl\to\P$ that
acts trivially on the objects $|\tilde\Cyl|=|\P|=\Sig$.\\

\begin{Lemma}
For $v,w$ morphisms in $\tilde\Cyl$ with $[[v]]=[[w]]$, one has
$\tilde\phi(v)=\tilde\phi(w)$. Thus, $\tilde\phi$ descends to a
functor $\phi:\P\to\P$. Furthermore, one has $\phi^2=\id_{\P}$, i.e.
$\phi\in\Aut$.
\end{Lemma}

\noindent\textbf{Proof:} If $[[v]]=[[w]]$, then by definition one
can reach $w$ by starting with $v$ and deleting or inserting ways
$u\circ u^{-1}$. Hence we need to show
$\tilde\phi(w^{-1})\,=\,\tilde\phi(w)^{-1}$. Since for ways $w,\,v$
in $\tilde\Cyl$ one has $(v\circ w)^{-1}=w^{-1}\circ v^{-1}$, we
have only to show $\tilde\phi(w^{-1})\,=\,\tilde\phi(w)^{-1}$ for
$w$ being of one of the types
(\ref{Gl:DefinitionOfEdgeInterchangerOnSegments}). But for
$w=[c_1^{t_1,t_2}]$ this follows from the fact that

\begin{eqnarray}
[c_1^{t_1,t_2}]^{-1}\;=\;[(c_1^{t_1,t_2})^{-1}]\;=\;[c_1^{t_2,t_1}],
\end{eqnarray}

\noindent with an analogous relation for $c_2^{t_1,t_2}$. For $w$
touching the edges $e_1$ and $e_2$ at most at its beginning- and
endpoint, the assertion is trivial. We thus get
$\tilde\phi(v)=\tilde\phi(w)$ for $[[w]]=[[v]]$. Thus, $\tilde\phi$
descends to a functor $\phi:\P\to\P$.\\

It remains to show that $\phi^2=\id_{\P}$. Take a path $p$ in $\P$.
Choose a representative $w$ in $\tilde\Cyl$, i.e. $p=[[w]]$. Then
$\phi(p)=\tilde\phi(w)$. Decompose $w$ w.r.t the edges $e_1,\,e_2$:
\begin{eqnarray}
w\;=\;w_1\circ\ldots\circ w_n
\end{eqnarray}

\noindent where $w_k$ are of the type
(\ref{Gl:DefinitionOfEdgeInterchangerOnSegments}). Assume $w_k$ to
meet the edges $e_1$ and $e_2$ at most at their endpoints. Then

\begin{eqnarray}
\phi^2([[w_k]])\;=\;\phi(\tilde\phi(w_k))\;=\;[[w_k]].
\end{eqnarray}

\noindent Let now $w_k=[c_1^{t_1,t_2}]$. Then

\begin{eqnarray}
\tilde\phi(w_k)\;=\;p_{t_1}\;\circ\;[c_2^{t_1,t_2}]\;\circ\;p_{t_2}^{-1}
\end{eqnarray}

\noindent By construction, the $p_t$ are such that they have a
representative in $\tilde\Cyl$ which touches the edges $e_1,\,e_2$
only at their beginning- and endpoints $s(p_t), \,t(p_t)$. It
follows that $\phi(p_t)=p_t$ for all $t\in[0,1]$. So we have

\begin{eqnarray}\nonumber
\phi^2([[w_k]])\;&=&\;\phi\left(p_{t_1}\;\circ\;[[[c_2^{t_1,t_2}]]]\;\circ\;p_{t_2}^{-1}\right)\\[5pt]
&=&\;p_{t_1}\;\circ\;p_{t_1}^{-1}\;\circ\;[[[c_1^{t_1,t_2}]]]\;\circ\;p_{t_2}\;\circ\;p_{t_2}^{-1}\\[5pt]\nonumber
&=&\;[[[c_1^{t_1,t_2}]]]\;=\;[[w_k]].
\end{eqnarray}

\noindent Analogously for $w_k=[c_2^{t_1,t_2}]$. We conclude that $\phi^2=\id_{\P}$. So $\phi\in\Aut$.\\

\noindent From the definition is immediate that $\phi(e_1)=e_2$ and
vice versa. This justifies the name ``edge-interchanger'', although
not only edges can be interchanged, but every two paths without
self-intersection that do not self-intersect. Note that $\phi$ is an
automorphism that acts trivially on the points in $\Sig$, but
nontrivially on the paths. Every path, however, that is composed of
edges that meet $e_1,\,e_2$ in at most finitely many points is left
invariant under $\phi$. Note further that $\phi$ depends on a chosen
parametrization of $e_1,\,e_2$, as well as a choice
$\{p_t\}_{t\in(0,1)}$. This shows that there are many such
automorphisms that interchange $e_1$ and $e_2$.\\


\section{Unitarity of
automorphisms}\label{Sec:UnitarityOfAutomorphisms}

In section \ref{Sec:Metagraphs} we have seen that the set of
metagraphs is suitable for addressing topological questions on
$\AL$. However, for the measure theory on $\AL$, the concept of
metagraphs is too broad: Not every metagraph $\m\in\M$ has the
property that
\begin{eqnarray}
(\pi_{\m})_*\m_{AL}\;=\;\m_H.
\end{eqnarray}


\noindent The metagraphs that have this property are called
\emph{primitive metagraphs}. But since this definition rests on
analytical terms, it is a priori not clear whether the automorphisms
preserve the set of primitive metagraphs. The proof for this will be
delivered in this section. As an immediate consequence, the
operators on $\Hkin$ induced by elements $\Aut$ will all be unitary.

For the abelian gauge group $G=U(1)$ this is quite easy to see, and
we will give the short proof for this, before we will turn to the
general case and prove the assertion for general gauge group $G$.


\subsection{The kinematical Hilbert space}

In this section we review the kinematical Hilbert space $\Hkin$ for
general gauge groups $G$, and discuss the action of the
automorphisms on it.

\begin{Definition}
Let $\m=\{p_1,\ldots,p_E\}$ be a metagraph. Then a function
$f:\AL\to\C$ is called \emph{cylindrical over }$\m$, iff there is a
function $F:G^E\to\C$ such that
\begin{eqnarray}
f(A)\;=\;F\Big(A(p_1),\ldots,A(p_E)\Big).
\end{eqnarray}

\noindent If $F$ is continuous (differentiable, smooth) then $f$ is
called a continuous (differentiable, smooth) cylindrical function
over $\m$. The set of all smooth cylindrical functions over $\m$ is
denoted by $Cyl(\m)$. The set of all smooth cylindrical functions is
denoted by $Cyl$.
\end{Definition}

\noindent Obviously, if $f\in Cyl(\m)$, then $f\in Cyl(\m')$ for all
metagraphs $\m'$ such that $\m\leq\m'$. Thus, every cylindrical
function is cylindrical over a graph. The advantage of the
cylindrical functions lies in the following lemma.
\begin{Lemma}
The set of cylindrical functions $Cyl$ is a dense subspace in the
set of all continuous functions $C(\AL)$, and hence also in
$\Hkin=L^2(\AL,d\m_{AL})$.
\end{Lemma}

\noindent In the analytical category, the concept of graphs provides
an orthonormal basis of $\Hkin$.

\begin{Lemma}
Let $G$ be a semisimple, compact and connected gauge group. Let
$\g=\{e_1,\ldots,e_E\}$ be a minimal graph, i.e. there is no other
graph $\tilde\g\leq\g$. Let furthermore $\{\pi_k\}_{k=1}^E$ be a
sequence of irreducible representations of $G$, and
$m_k,n_k\in\{1,\ldots,\dim\pi_k\}$ be numbers. Then the function
\begin{eqnarray}\label{Gl:NetworkFunctions}
T_{\g,\vec\pi,\vec n, \vec
m}(A)\;:=\;\prod_{k=1}^E\sqrt{\dim\pi_k}\;\pi_k(A(e_k))_{n_km_k}
\end{eqnarray}

\noindent is cylindrical over $\g$. The set of all these functions
provide an orthonormal basis for $\Hkin$.
\end{Lemma}

\noindent In the case of $G=SU(2)$, the corresponding functions
(\ref{Gl:NetworkFunctions}) are called \emph{spin-network functions}
(SNF), and are denoted by $T_{\g,\vec j,\vec n,\vec m}$, where each
of the $j_k$ is a half-integer corresponding to an irreducible
representation of $SU(2)$, and $m_k,n_k\in\{-j,-j+1,\ldots,j\}$. In
the case of $G=U(1)$, where all representations are one-dimensional,
the functions $T_{\g,\vec n}$ are called \emph{charge-network
functions}, and the $n_k\in\Z$ denote irreducible representations of
$U(1)$.

\begin{Lemma}\label{Lem:ChargeNetworksStayChargeNetworksUnderAutomorphisms}
Let $\phi\in\Aut$ be an automorphism of the path groupoid, and
$T_{\g,\vec n}$ a charge-network function. Then the function
\begin{eqnarray}
\left(\hat U(\phi)T_{\g,\vec n}\right)(A)\;:=\;T_{\g,\vec
n}(\a_{\phi}A)
\end{eqnarray}

\noindent is again a charge-network function.
\end{Lemma}

\noindent\textbf{Proof:} If $\g$ is a minimal graph, then $\phi(\g)$
is not necessarily one, only a metagraph. But there is a minimal
graph $\g'=\{e'_1,\ldots,e'_{E'}\}$ such that $\phi(\g)\leq\g'$. In
particular
\begin{eqnarray}
\phi(e_k)\;=\;e'_{l_k^1},\circ\ldots,\circ e'_{l_k^{m_k}}
\end{eqnarray}

\noindent for each $k\in\{1,\ldots,E\}$, and the $e'_{l}$ are edges
(or their inverses) in $\g'$. With
\begin{eqnarray}
T_{\g,\vec n}(A)\;=\;\prod_{k=1}^E\,A(e_k)^{n_k}
\end{eqnarray}

\noindent we get
\begin{eqnarray}
\left(\hat U(\phi)T_{\g,\vec n}\right)(A)\;&=&\;T_{\g,\vec
n}(\a_{\phi}A)\\[5pt]\nonumber
&=&\;\prod_{k=1}^E\,A(e'_{l_k^1},\circ\ldots,\circ
e'_{l_k^{m_k}})^{n_k}.
\end{eqnarray}

\noindent But using the functorial properties of $A$, we get a
product of $A(e'_{l_k^m})^{n_k}$, which we can - due to the
abelianess of $U(1)$ - group together to obtain
\begin{eqnarray}
T_{\g,\vec
n}(\a_{\phi}A)\;=\;\prod_{l=1}^{E'}\,A(e'_l)^{m_l}\;=\;T_{\g',\vec
m}(A).
\end{eqnarray}

\noindent So the automorphism map charge-network functions into
charge-network functions.

\begin{Corollary}
Let $\phi\in\Aut$ be an automorphism and $\Hkin$ be the kinematical
Hilbert space for the gauge group $G=U(1)$. Then the operator
\begin{eqnarray}\label{Gl:DefinitionOfPullbackOfAutomorphisms}
\left(\hat U(\phi)f\right)\;:=\;f(\a_{\phi}A)
\end{eqnarray}

\noindent for any function $f\in\Hkin$ is unitary.
\end{Corollary}

\noindent\textbf{Proof:} By lemma
\ref{Lem:ChargeNetworksStayChargeNetworksUnderAutomorphisms} the map
(\ref{Gl:DefinitionOfPullbackOfAutomorphisms}) is defined on the
charge-network functions $T_{\g,\vec n}$, and map this set into
itself. Furthermore, the vector $\Om$, corresponding to the constant
function $1$ on $\AL$, which can be written as $T_{\g,\vec 0}$ for
any graph $\g$, is trivially left invariant by the $\phi\in\Aut$.
Let now $\g$ be a minimal graph and $f\in Cyl(\g)$ a smooth
cylindrical function over $\g$. Then
\begin{eqnarray}
f\;=\;\sum_{\vec n\in\Z^{E}}\;c_{\vec n}\;T_{\g,\vec n}.
\end{eqnarray}

\noindent One has, by definition of $\m_{AL}$ and the orthonormality
of the charge-network functions:
\begin{eqnarray}
\int_{\AL}\;d\m_{AL}(A)\;f(A)\;=\;c_{\vec 0}.
\end{eqnarray}

\noindent On the other hand:
\begin{eqnarray}
\int_{\AL}d\m_{AL}(A)\;f(\a_{\phi}A)\;=\;\sum_{\vec n\in\Z}c_{\vec
n}\;\int_{\AL}d\m_{AL}(A)\;T_{\g,\vec n}(\a_{\phi}A)
\end{eqnarray}

\noindent But if $T_{\g,\vec n}\neq T_{\g,\vec 0}$, then also $\hat
U(\phi)T_{\g,\vec n}\neq T_{\g',\vec 0}$, since no not constant
function on $\AL$ can be mapped to the constant function on $\AL$.
Thus, since the integral over $\AL$ of all not constant
charge-network functions is zero, we get
\begin{eqnarray}
\sum_{\vec n\in\Z}c_{\vec n}\;\int_{\AL}d\m_{AL}(A)\;T_{\g,\vec
n}(\a_{\phi}A)\;=\;c_{\vec 0}\;=\;\int_{\AL}\;d\m_{AL}(A)\;f(A)
\end{eqnarray}

\noindent So we have
\begin{eqnarray}\label{Gl:IntegralOverFunctionsDoesNotChange}
\int_{\AL}d\m_{AL}(A)\;f(\a_{\phi}A)\;=\;\int_{\AL}\;d\m_{AL}(A)\;f(A)
\end{eqnarray}

\noindent for all functions $f\in Cyl$. But this is a set dense in
the continuous functions w.r.t the supremum norm on $C(\AL)$, and
since $\a_{\phi}:\AL\to\AL$ is continuous, we conclude that
(\ref{Gl:IntegralOverFunctionsDoesNotChange}) holds for all
continuous functions $f\in C(\AL)$. Hence
\begin{eqnarray}
(\a_{\phi})^*\m_{AL}\;=\;\m_{AL},
\end{eqnarray}

\noindent i.e. the integration measure os preserved under the action
of $\phi\in\Aut$. Consequently, $\hat U(\phi)$ defined by
(\ref{Gl:DefinitionOfPullbackOfAutomorphisms}) is a unitary map on
$\Hkin=L^2(\AL,d\m_{AL})$.\\

It should be noted that the key for establishing this short proof is
the fact that the orthonormal basis elements $T_{\g,\vec n}$ are
mapped into themselves by the action of $\Aut$. This rested on the
fact that $G=U(1)$ is abelian, and is not true for nonabelian gauge
groups, such as $G=SU(2)$. In these cases, the automorphisms also
act unitarily on the corresponding $\Hkin$, however the proof is
much more involved, as we will see in the following section.


\subsection{Unitarity of Automorphisms}

In the last section, we have seen that in the case of $G=U(1)$ it is
quite straightforward to show that the action of the automorphisms
$\Aut$ on $\Hkin$ is unitary. However, this proof rested crucially
on the abelianess of $U(1)$. In the following, we will prove
unitarity for gauge group $G=SU(2)$.\\

\noindent For showing unitarity for the automorphisms $\Aut$ in the
case $G=SU(2)$, we will make explicit use of the natural
transformations of the identity $\phi_{b,p}\in\Nyl$ defined in
section \ref{Sec:NTDID}

In this section we will prove that all natural transformations of
the identity $\phi_{b,p}\in\Nyl$ act unitarily on $\Hkin$. The basic
idea of the proof is as follows: Since $\Nyl$ is a group, every
$\phi_{b,p}$ can be written as a sequence of other elements
$\phi_{b,p}=\phi_{b_N,p_N}\circ\ldots\circ\phi_{b_1,p_1}$, each of
which are not too different from the identity. For these we will be
able to prove that they act unitarily, because they do not
change a graph much, hence can be controlled.\\

\begin{Lemma}\label{Lem:ShortNaturalIdentityIsUnitary} Let $\g=\{e_1,\ldots,e_E\}$ be a graph. Let furthermore $v\in V(\g)$ be a vertex in the
graph, and $e_{k_1}\circ \ldots\circ e_{k_n}$ be a path in $\g$ with
the following properties:
\begin{itemize}
\item All the $e_{k_l}$ are edges in $\g$.
\item All the $e_{k_l}$ are distinct.
\item $s(e_{k_1})\;\neq\;t(e_{k_n})$.
\end{itemize}

\noindent Then the natural transformation of the identity
$\phi_{b,p}$ defined by
\begin{eqnarray*}
b(s(e_{k_1}))\;&:=&\;t(e_{k_n}),\qquad
b(t(e_{k_n}))\;:=\;s(e_{k_1}),\qquad b(x)\;:=\;x\text{ \rm else}\\[5pt]
p_{t(e_{k_n})}&\;:=\;&\big(e_{k_1}\circ \ldots\circ
e_{k_n}\big)^{-1},\qquad p_{s(e_{k_1})}\;:=\;e_{k_1}\circ
\ldots\circ e_{k_n},\qquad p_x\;:=\;\id_x\text{ \rm else}
\end{eqnarray*}

\noindent acts unitarily on $\H_{\g}$.
\end{Lemma}

\noindent\textbf{Proof:} First we note that $\phi_{b,p}$ maps
$\H_{\g}$ to itself. Since the definition of $\phi_{b,p}$ is such
that $(\phi_{b,p})^2=\id_{\P}$ is the identity functor, so does its
inverse. Thus, every edge $e_k$ is mapped to a combination of edges:
\begin{eqnarray}
\phi_{b,p}(e_k)\;=\;\vartheta_k(e_1,\ldots,e_E)
\end{eqnarray}

\noindent where the $\vartheta_k(e_1,\ldots, e_E)$ is a \emph{word}
of the $\{e_1,\ldots,e_E\}$, i.e. a product of edges and their
inverses. Thus, $\phi_{b,p}$ induces an endomorphism $\phi$ on the
free group $F_E$ in the $E$ letters $\{e_1,\ldots e_E\}$ (which is
defined by $\phi(e_k)\;=\;\vartheta_k(e_1,\ldots,e_E)$). But since
its inverse also does, $\phi$ is invertible (it is in fact its own
inverse), and hence defines an automorphism on $F_E$. By corollary
\ref{Cor:CorollaryToNielson}, we have

\begin{eqnarray}
\int_{SU(2)^E}d\m_{H}&&(h_1,\ldots,
h_E)\;F(h_1,\ldots,h_E)\\[5pt]\nonumber
&&\;=\;\int_{SU(2)^E}d\m_{H}(h_1,\ldots,
h_E)\;F\Big(\vartheta_1(h_1,\ldots,h_E),\ldots,\vartheta_E(h_1,\ldots,h_E)\Big),
\end{eqnarray}

\noindent and hence

\begin{eqnarray}\label{Gl:ShortNTDIDactUnitarily}
\int_{\AL}d\m_{AL}(A)\,f(A)\;=\;\int_{\AL}d\m_{AL}(A)\,\a_{\phi_{b,p}}f(A).
\end{eqnarray}

\noindent Since $f$ is cylindrical over $\g$, the immediate
conclusion from (\ref{Gl:ShortNTDIDactUnitarily}) is that
$\phi_{b,p}$ acts unitarily on $\H_{\g}$.\\

For $f$ being cylindrical over a graph $\g$, and $\phi_{b,p}\in\Nyl$
be a natural transformation of the identity functor, $\phi_{b,p}$
will generically not satisfy the conditions of lemma
\ref{Lem:ShortNaturalIdentityIsUnitary}, but it can be composed of a
sequence of elements in $\Nyl$ that do, which will be the key part
of the following lemma.

\begin{Lemma}\label{Lem:LongNaturalIdentitiyIsUnitary}
Let $f$ be a function cylindrical over a graph $\g$, and
$\phi_{b,p}$ a natural transformation of the identity. Then
\begin{eqnarray}
\int_{\AL}d\m_{AL}(A)\,f(A)\;=\;\int_{\AL}d\m_{AL}(A)\,\a_{\phi_{b,p}}f(A).
\end{eqnarray}
\end{Lemma}

\noindent\textbf{Proof:} Let $e_1,\ldots, e_E$ be the edges of the
graph $\g$, and $v_1,\ldots v_V$ be the vertices. Without loss of
generality, we can decompose the edges which are loops, such that
their beginning- and endpoints do not coincide. Consider the paths
$p_{v_1},\ldots,p_{v_V}$ which enter the definition of $\phi_{b,p}$.
Now decompose the $e_k$ and the $p_{v_l}$ into smaller edges
$e'_1,\ldots,e'_{E'}$ that form a graph $\g'$, such that all the
$e_k$ and the $p_{v_l}$ can be composed from edges in $\g'$ (and
their inverses). Now decompose the edges $e'_k$ of $\g'$ further
into edges $e_k''$ (forming a graph $\g''$), such that the following
holds: Each path $p_{v_l}$ can be decomposed into a product of $N$
paths $\tilde{e}_k$

\begin{eqnarray}\label{Gl:DecompositionOfPathsOfnTdid}
p_{v_l}\;=\;\tilde{e}_{1}^l\;\circ\;\tilde{e}_2^l\;\circ\;\ldots\;\tilde{e}_N^l
\end{eqnarray}

\noindent such that

\begin{itemize}
\item Each $\tilde{e}^l_m$,
$l\in\{1,\ldots,V\},\,m\in\{1,\ldots,N\}$ is a sequence of distinct
edges $e''_k$ (or their inverses) in $\g''$
\item For each $m\in\{1,\ldots,N\}$ the $2V$ points $s(\tilde{e}^l_m)$
and $t(\tilde{e}^l_m)$ for all $l\in\{1,\ldots,V\}$ are distinct.
%
\end{itemize}

\noindent Note that this decomposition is possible, even if some
$p_{v_l}$ are the constant paths: In this case, the paths in
(\ref{Gl:DecompositionOfPathsOfnTdid}) cancel each other out such
that each
$\tilde{e}_{1}^l\;\circ\;\tilde{e}_2^l\;\circ\;\ldots\;\tilde{e}_m^l$
is nontrivial, while
$\tilde{e}_{1}^l\;\circ\;\tilde{e}_2^l\;\circ\;\ldots\;\tilde{e}_N^l$
is a constant path again.\\

Then we define a natural transformation of the identity
$\phi_{b_m^l,p_m^l}$ by the following data:
\begin{eqnarray*}
b^l_m\big(s(\tilde{e}_m^l)\big)\;&=&\;t(\tilde{e}_m^l),\qquad
b^l_m\big(t(\tilde{e}_m^l)\big)\;=\;s(\tilde{e}_m^l),\qquad
b_m^l(x)=x\text{ else}\\[5pt]
(p_m^l)_{s(\tilde{e}_m^l)}\;&=&\;\tilde{e}_m^l,\qquad
(p_m^l)_{t(\tilde{e}_m^l)}\;=\;(\tilde{e}_m^l)^{-1}
\end{eqnarray*}

\noindent This defines an element in $\Nyl$, as one can readily see.
Furthermore, we can see that, for one fixed $m\in\{1,\ldots,N\}$,
the $\phi_{b_m^l,p_m^l}$ for all $l\in\{1,\ldots,V\}$ commute, due
to the condition that all the $\tilde{e}_m^l$ have distinct
beginning- and endpoints. Now define

\begin{eqnarray}\label{Gl:DefinitionOfCertainnTdidInProofOfUnitarity}
\phi_{\tilde b,\tilde p}\;:=\;\;\phi_{b_N^V,p_N^V}\;&\circ\;&\phi_{b_N^{V-1},p_N^{V-1}}\;\circ\ldots\circ\;\phi_{b_N^1,p_N^1}\;\circ\;\phi_{b_{N-1}^V,p_{N-1}^V}\;\circ\ldots\\[5pt]\nonumber
&\circ\;&\phi_{b_{N-1}^1,p_{N-1}^1}\;\circ\;\phi_{b_{N-2}^V,p_{N-2}^V}\;\circ\;\ldots\;\circ\;\phi_{b_1^1,p_1^1}.
\end{eqnarray}

\noindent Now let $e_k$ be an edge in $\g$, with starting point
$v_{l'}$ and ending point $v_{l''}$. Then for every
$m\in\{1,\ldots,N\}$ we have
\begin{eqnarray}\label{Gl:BeschisseneGleichung}
\phi_{b_m^V,p_m^V}\;\circ\;\ldots\;&\circ\;&\phi_{b_m^1,p_m^1}\big((\tilde{e}_{m-1}^{l'})^{-1}\circ\ldots\circ(\tilde{e}_1^{l'})^{-1}\circ
e_k\circ\tilde{e}_1^{l''}\circ\ldots\circ
\tilde{e}_{m-1}^{l''}\big)\\[5pt]\nonumber
&&\;=\;(\tilde{e}_{m}^{l'})^{-1}\circ\ldots\circ(\tilde{e}_1^{l'})^{-1}\circ
e_k\circ\tilde{e}_1^{l''}\circ\ldots\circ \tilde{e}_{m}^{l''}.
\end{eqnarray}

\noindent In order to show this, we first note that, since all the
$\phi_{b_m^l,p_m^l}$ commute for all $l\in\{1,\ldots,V\}$ by
construction, we can move $\phi_{b_m^{l'},p_m^{l'}}$ and
$\phi_{b_m^{l''},p_m^{l''}}$ to the left in
(\ref{Gl:BeschisseneGleichung}). Since all the starting- and
endpoints of the $\tilde{e}_m^l$ are different from
$s(\tilde{e}_m^{l'})$ and $s(\tilde{e}_m^{l''})$ by construction,
the action of the $\phi_{b_m^l,p_m^l}$ on
$(\tilde{e}_{m-1}^{l'})^{-1}\circ\ldots\circ(\tilde{e}_1^{l'})^{-1}\circ
e_k\circ\tilde{e}_1^{l''}\circ\ldots\circ \tilde{e}_{m-1}^{l''}$ for
$l\neq l',l''$ is trivial. Thus we get

\begin{eqnarray}
\phi_{b_m^V,p_m^V}\;\circ\;\ldots\;&\circ\;&\phi_{b_m^1,p_m^1}\big((\tilde{e}_{m-1}^{l'})^{-1}\circ\ldots\circ(\tilde{e}_1^{l'})^{-1}\circ
e_k\circ\tilde{e}_1^{l''}\circ\ldots\circ
\tilde{e}_{m-1}^{l''}\big)\\[5pt]\nonumber
&&\;=\;\phi_{b_m^{l'},p_m^{l'}}\;\circ\;\phi_{b_m^{l''},p_m^{l''}}\big((\tilde{e}_{m-1}^{l'})^{-1}\circ\ldots\circ(\tilde{e}_1^{l'})^{-1}\circ
e_k\circ\tilde{e}_1^{l''}\circ\ldots\circ
\tilde{e}_{m-1}^{l''}\big).
\end{eqnarray}

\noindent But starting- and endpoint of $\tilde{e}_m^{l'}$ and
$\tilde{e}_m^{l''}$ are all distinct, and hence, by definition of
the $\phi_{b_m^l,p_m^l}$, we get.

\begin{eqnarray}
\phi_{b_m^{l'},p_m^{l'}}\;&\circ&\;\phi_{b_m^{l''},p_m^{l''}}\big((\tilde{e}_{m-1}^{l'})^{-1}\circ\ldots\circ(\tilde{e}_1^{l'})^{-1}\circ
e_k\circ\tilde{e}_1^{l''}\circ\ldots\circ
\tilde{e}_{m-1}^{l''}\big)\\[5pt]\nonumber
&&\;=\;(\tilde{e}_{m}^{l'})^{-1}\circ\ldots\circ(\tilde{e}_1^{l'})^{-1}\circ
e_k\circ\tilde{e}_1^{l''}\circ\ldots\circ \tilde{e}_{m}^{l''}.
\end{eqnarray}

\noindent An immediate consequence from
(\ref{Gl:BeschisseneGleichung}), together with
(\ref{Gl:DefinitionOfCertainnTdidInProofOfUnitarity}), is that
\begin{eqnarray}
\phi_{b,p}(e_k)\;=\;\phi_{\tilde b,\tilde p}(e_k)
\end{eqnarray}

\noindent for all edges $e_k$ in the graph $\g$. But $\phi_{\tilde
b,\tilde p}$ is, by
(\ref{Gl:DefinitionOfCertainnTdidInProofOfUnitarity}) a product of
natural transformations of the identity $\phi_{b_m^l,p_m^l}$, all of
which satisfy the conditions of Lemma
\ref{Lem:ShortNaturalIdentityIsUnitary}. Thus, we get
\begin{eqnarray}
\int_{\AL}d\m_{AL}(A)\;\a_{\phi_{b,p}}f(A)\;=\;\int_{\AL}d\m_{AL}(A)\;\a_{\phi_{\tilde
b,\tilde p}}f(A)\;=\;\int_{\AL}d\m_{AL}(A)\;f(A)
\end{eqnarray}

\noindent for all functions $f$ cylindrical over $\g$. This finishes
the proof.\\

\begin{Corollary}
All natural transformations of the identity $\phi_{b,p}\in\Nyl$ act
unitarily on $\Hkin$.
\end{Corollary}

\noindent\textbf{Proof:} From lemma
\ref{Lem:LongNaturalIdentitiyIsUnitary} we get that

\begin{eqnarray}
\int_{\AL}d\m_{AL}(A)\;\a_{\phi_{b,p}}f(A)\;=\;\int_{\AL}d\m_{AL}(A)\;f(A)
\end{eqnarray}

\noindent holds for all functions $f\in Cyl$. Thus,
$\a_{\phi_{b,p}}$ induces an operator on $\Hkin$ which is isometric
on a dense subspace. Thus, it is also unitary on all of $\Hkin$. So,
all natural transformations of the identity $\phi_{b,p}\in\Nyl$ act
unitarily on $\Hkin$.\\

\noindent With this, we can now make the next step in the proof that
all automorphisms act unitarily on $\Hkin$.

\begin{Lemma}\label{Lem:AutomorphismsActUnitarilyOnOneEdges}
Let $e$ be an edge which is no loop, i.e. $s(e)\neq t(e)$. Let $f\in
Cyl(e)$ be a function cylindrical over $e$. Then, for any
automorphism $\phi\in\Aut$, one has
\begin{eqnarray}
\int_{\AL}d\m_{AL}(A)\;\a_{\phi}
f(A)\;=\;\int_{\AL}d\m_{AL}(A)\;f(A)
\end{eqnarray}
\end{Lemma}

\noindent\textbf{Proof:} Let $p=\phi(e)$. Since $s(e)\neq t(e)$,
also $s(p)\neq t(p)$. Now decompose $p$ into edges
$p=e_1\circ\ldots\circ e_E$. If necessary, decompose the edges $e_k$
further, until $s(e_1),\, t(e_1)$ and $t(e_E)$ are all distinct.
Then define the natural transformation of the identity $\phi_{b,p}$
by
\begin{eqnarray}
b(t(e_1))\;:=\;t(e_E),\qquad b(t(e_E))\;:=\;t(e_1),\qquad
b(x)=x\text{ \rm
else}\\[5pt]\nonumber
p_{t(e_1)}\;:=\;e_2\circ\ldots\circ e_E,\qquad
p_{t(e_E)}\;:=\;e_E^{-1}\circ\ldots\circ e_2^{-1},\qquad
p_x=\id_x\text{ \rm else}.
\end{eqnarray}

\noindent Since $f$ is cylindrical over the edge $e$, there is a
function $F$ such that

\begin{eqnarray}
\int_{\AL}d\m_{AL}(A)\;f(A)\;=\;\int_{SU(2)}d\m_H(h)\;F(h).
\end{eqnarray}

\noindent But obviously $\a_{\phi_{b,p}}\circ\a_{\phi}f$ is
cylindrical over the edge $e_1$. Since
$\phi_{b,p}(\phi(e))\;=\;e_1$, we have

\begin{eqnarray}
\int_{\AL}d\m_{AL}(A)\;\a_{\phi_{b,p}\circ\phi}f(A)\;=\;\int_{SU(2)}d\m_H(h)\;F(h),
\end{eqnarray}

\noindent so

\begin{eqnarray}
\int_{\AL}d\m_{AL}(A)\;f(A)
\;=\;\int_{\AL}d\m_{AL}(A)\;\a_{\phi_{b,p}}\circ\a_{\phi}f(A).
\end{eqnarray}

\noindent But since all natural transformations of the identity
$\phi_{b,p}\in\Nyl$ leave the Ashtekar-Isham-Lewandowski measure
invariant by lemma (\ref{Lem:LongNaturalIdentitiyIsUnitary}), we
have
\begin{eqnarray}
\int_{\AL}d\m_{AL}(A)\;f(A)
\;=\;\int_{\AL}d\m_{AL}(A)\;\a_{\phi}f(A),
\end{eqnarray}

\noindent which was the claim. \\

\noindent Lemma \ref{Lem:AutomorphismsActUnitarilyOnOneEdges} shows
that for $f,\,g$ being cylindrical over an edge $e$, one has
\begin{eqnarray}\label{Gl:AutomorphismsAreUnitaryOnEdges}
\langle\,\hat U(\phi) f\,|\,\hat U(\phi)\,g\,\rangle\;=\;\langle\,
f\,|\, g\,\rangle.
\end{eqnarray}

\noindent In fact, in the proof of lemma
\ref{Lem:AutomorphismsActUnitarilyOnOneEdges} we have shown that
every path $p$ with $s(p)\neq t(p)$ can be mapped to an edge by an
automorphism (in fact, by an element of $\Nyl$), hence
(\ref{Gl:AutomorphismsAreUnitaryOnEdges}) holds also for functions
$f, g\in Cyl(p)$. In the following we will look at arbitrary
automorphisms $\phi\in\Aut$, and use their properties as maps from
$\AL$ to $\AL$, as well as (\ref{Gl:AutomorphismsAreUnitaryOnEdges})
in order to show that they act unitarily on $\Hkin$.

\begin{Lemma}\label{Lem:I(phi)IsInTheWeakClosureOfC(A)}
Let $\phi\in\Aut$ and $\hat U(\phi)$ be the induced operator on
$\Hkin$, and $\hat U(\phi)^{\dag}$ be its adjoint. Define $\hat
I_{\phi}:=\hat U(\phi)^{\dag}\hat U(\phi)$, then there is a sequence
of continuous functions $f_n\in C(\AL)$ such that
\begin{eqnarray}
\lim_{n\to\infty}\hat a_{f_n}\;=\;\hat I_{\phi}
\end{eqnarray}

\noindent converges in the weak operator topology, where for a
continuous function $f\in C(\AL)$ the operator $\hat a_f$ denotes
multiplication with $f$ on $\Hkin$.
\end{Lemma}

%

%
%

\noindent\textbf{Proof:} Let $f\in C(\AL)$ be a continuous function
on $\AL$. Then $\a_{\phi}f$ is also continuous. Denote with $\bar f$
the complex conjugate function, then $\a_{\phi}\bar
f=\overline{\a_{\phi}f}$. Since $\hat U(\phi)\,\hat a_f\,\hat
U(\phi)^{-1}\;=\;\hat a_{\a_{\phi}f}$, we have

\begin{eqnarray}
\hat U(\phi)\,\hat a_{\bar f}\,\hat U(\phi)^{-1}\;&=&\;\hat
a_{\a_{\phi}\bar f}\;=\;\hat
a_{\overline{\a_{\phi}f}}\;=\;\left(\hat
a_{\a_{\phi}f}\right)^{\dag}\\[5pt]\nonumber
\;&=&\;\left(\hat U(\phi)\,\hat a_f\,\hat
U(\phi)^{-1}\right)^{\dag}\;=\;\left(\hat
U(\phi)^{\dag}\right)^{-1}\,\hat a_{f}\,\hat U(\phi)^{\dag}.
\end{eqnarray}

\noindent Multiplying with the respective operators $\hat
U(\phi)^{\dag},\;\hat U(\phi)$ from the right and the left, one gets

\begin{eqnarray}
\hat U(\phi)^{\dag}\hat U(\phi)\;\hat a_f\;=\;\hat a_f\,\hat
U(\phi)^{\dag}\hat U(\phi).
\end{eqnarray}

\noindent So the operator $\hat I_{\phi}$ commutes with all elements
$\hat a_f$, for $f\in C(\AL)$. Let $\Om$ be the cyclic and
separating\footnote{For an introduction to the $C^*$-algebraic
notions, see e.g. \cite{INTRO, BRATT}} vector for $C(\AL)$
(corresponding to the function $f(A)\equiv 1$, then, since $\Om$ is
also cyclic for the cylindrical functions $Cyl$, there is a sequence
of cylindrical functions $f_n\in Cyl$ such that
\begin{eqnarray}
\lim_{n\to\infty}\hat a_{f_n}\Om\;=\;\hat I_{\phi}\Om
\end{eqnarray}

\noindent in the topology of $\Hkin$. Thus, for any two continuous
functions $g,h\in C(\AL)$, one has

\begin{eqnarray}
\langle g\,|\,\hat I_{\phi}\,h\rangle\;&=&\;\langle g\,|\,\hat
I_{\phi}\,\hat a_h\,\Om\rangle\;=\;\langle g\,|\,\hat a_h\,\hat
I_{\phi}\,\Om\rangle\\[5pt]\nonumber
&\;=&\;\lim_{n\to\infty}\,\langle g\,|\,\hat a_h\,\hat
a_{f_n}\,\Om\rangle\;=\;\lim_{n\to\infty}\,\langle g\,|\hat
a_{f_n}\,h\rangle
\end{eqnarray}

\noindent since all the $\hat a_h$ are continuous operators on
$\Hkin$ and are mutually commuting. Since the
$|g\rangle,\;|h\rangle$ are a dense subset of
$\Hkin$, the statement follows.\\

Lemma (\ref{Lem:I(phi)IsInTheWeakClosureOfC(A)}) is in fact true for
all invertible maps $\phi:\AL\to\AL$. That the corresponding
sequence of $f_n$ converges to the constant function on $\AL$
however, is a consequence of Lemma
\ref{Lem:AutomorphismsActUnitarilyOnOneEdges}.

\begin{Lemma}
Let $\phi\in\Aut$ and $f_n$ be a sequence in $Cyl$ such that $\hat
a_{f_n}$ converges weakly to $\hat I_{\phi}=\hat U(\phi)^{\dag}\hat
U(\phi)$. Then, for every edge $e$ with $s(e)\neq t(e)$ there is an
$n_0\in \N$ such that $f_n$ depend trivially on all edges which are
not $e$ for all $n>n_0$.
\end{Lemma}

\noindent\textbf{Proof:} Let $f$ be a nontrivial function
cylindrical over $e$. Then by lemma
\ref{Lem:AutomorphismsActUnitarilyOnOneEdges}

\begin{eqnarray}
\langle f\,|\,\hat I_{\phi}\,f\rangle\;=\;\langle\hat
U(\phi)f\,|\,\hat U(\phi)f\rangle\;=\;\langle
f\,|\,f\rangle\;=\;\|f\|^2\,>\,0
\end{eqnarray}

\noindent it follows with lemma
\ref{Lem:I(phi)IsInTheWeakClosureOfC(A)} that $\langle f\,|\,\hat
a_{f_n}\,f\rangle\,>\,0$ for some $n>n_0$. But if $f_n$ would depend
nontrivially on any edge different from $e$, then so would the
product $f_nf$, but not $f$, so $\langle f\,|\,\hat
a_{f_n}\,f\rangle$ would be zero. Hence, $f_n$ cannot depend on any
edge not being $e$ for $n> n_0$.

\begin{Corollary}\label{Cor:OperatorsConvergeToTheConstantFunction}
There is an $n_0\in\N$ such that $f_n$ is the constant function $1$
on $\AL$ for every $n>n_0$.
\end{Corollary}

\noindent\textbf{Proof:} Choose two different edges $e_1$, $e_2$
which have differing beginning- and endpoints. Then $f_n$ depend
trivially on every edge not equal $e_1$ for $n>n_1$ and trivially on
every edge not equal to $e_2$ for $n>n_2$. But since $e_1\neq e_2$,
$f_n$ depend trivially on every edge for $n>n_0:=\text{\rm
max}(n_1,n_2)$. But the only cylindrical functions depending
trivially on every edge are the constant functions. But since $\hat
U(\phi)\Om=\Om$, we have
\begin{eqnarray}
\langle \Om\,|\,\hat
I_{\phi}\,\Om\rangle\;=\;\langle\Om\,|\,\Om\rangle\;=\;1,
\end{eqnarray}

\noindent so $\hat a_{f_n}$ must be the constant function $1$ for
all $n>n_0$.\\

\noindent We conclude this section by stating the main result.

\begin{Theorem}
For any automorphism $\phi\in\Aut$ the corresponding operator $\hat
U(\phi)$ on $\Hkin$ is unitary.
\end{Theorem}

\noindent\textbf{Proof:} From corollary
\ref{Cor:OperatorsConvergeToTheConstantFunction} it follows that the
$\hat a_{f_n}$ converge to the identity operator. But with lemma
\ref{Lem:I(phi)IsInTheWeakClosureOfC(A)} we conclude that $\hat
I_{\phi}$ must be the identity operator. So $\hat U(\phi)^{\dag}\hat
U(\phi)\;=\;\mathbbm{1}$, and multiplying by $\hat U(\phi)^{-1}$
from the right gives the desired result.


\section{The automorphism-invariant Hilbert
space}\label{Sec:TheAutomorphismInvariantHilbertSpace}

We now review the definition of the $\Diff(\Sig)$-invariant Hilbert
space for the case  $G=SU(2)$ and $\Diff(\Sig)$ being the group of
analytical diffeomorphisms on $\Sig$ \cite{ALLMT, ALSTATUS}.

Since there is no known nontrivial topology or Borel measure on
$\Diff(\Sig)$, the group of (analytical) diffeomorphisms, defining
the rigging map naively via the group averaging:
\begin{eqnarray}
\eta[\psi](\chi)\;=\;\int_{\Diff(\Sig)}d\m(\phi)\,\langle
\psi\,|\hat U(\phi)\chi\rangle
\end{eqnarray}

\noindent is not possible. But there are other ways of defining a
antilinear map
\begin{eqnarray}
\eta\;:\;D\,\longrightarrow\,D_{\Diff}^*
\end{eqnarray}

\noindent from a dense subspace $D\subset\Hkin$ invariant under
$\Diff(\Sig)$ into the linear functionals on $D$ that are invariant
under $\Diff(\Sig)$. This is usually done as follows:

For any $\phi\in\Diff(\Sig)$ and any graph $\g$, also $\phi(\g)$ is
a graph. It follows that, in the case of $G=SU(2)$, which we
consider here, spin network functions $T_{\g,\vec j,\vec n,\vec m}$
are mapped into spin network functions. So for two SNFs one has that
$\langle T_{\g,\vec j\vec n,\vec m}\,|\,\hat U(\phi)T_{\g',\vec
j'\vec n',\vec m'}\rangle$ is either $0$ or $1$, depending on
$\phi\in\Diff(\Sig)$. Then $\eta$ is defined by

\begin{eqnarray}\nonumber
\eta[T_{\g,\vec j,\vec n\vec m}](T_{\g',\vec j'\vec n',\vec
m'})\;:=\;\sum_{\varphi\in\Diff(\Sig)\big/\Diff_{\g}}\;F(|GS_{\g}|)\;\sum_{\tilde\varphi\in
GS_\g}\;\big\langle\hat U(\varphi\circ\tilde\varphi)\,T_{\g,\vec
j,\vec n\vec m}\,\big|\,T_{\g',\vec j'\vec n',\vec
m'}\big\rangle\\[5pt]\label{Gl:DefinitionOfDiffInvariantRiggingMap}
\end{eqnarray}

\noindent where $\Diff_{\g}$ is the set of diffeomorphisms which
leave $\P_{\g}$ invariant. $GS_{\g}$ is the group of graph
symmetries, i.e. the quotient of $\Diff_{\g}$ and the subgroup of
$\Diff(\Sig)$ which leaves $\g$ invariant. This is a finite group.
$F$ is a function, which can be chosen arbitrarily. This defines an
antilinear map from the span of the SNFs to the
$\Diff(\Sig)$-invariant linear functionals, which serves as rigging
map, and defines a $\Diff(\Sig)$-invariant inner product via
\begin{eqnarray}
\langle\eta[\psi]\,|\,\eta[\chi]\rangle_{\Diff}\;:=\;\eta[\psi](\chi)
\end{eqnarray}

\noindent In the case of $\Diff(\Sig)$, an orthogonal basis for
$\Hdiff$ is given by the set of equivalence classes of SNFs under
the action of $\Diff(\Sig)$. The normalization of these vectors is a
nontrivial issue \cite{HANNO}, and is governed by the function $F$
in
(\ref{Gl:DefinitionOfDiffInvariantRiggingMap}).\\

If we now replace the diffeomorphisms $\Diff(\Sig)$ by the
automorphisms $\Aut$, we can try to use the same techniques to
define an automorphism-invariant inner product
$\langle\cdot|\cdot\rangle_{\rm Aut}$ and an automorphism-invariant
Hilbert space $\Haut$.

\subsection{Graph combinatorics}

\noindent The automorphisms $\Aut$ act unitarily on the Hilbert
space $\Hkin$, as was demonstrated in the last section. Furthermore,
we have seen that there are many automorphisms $\phi\in\Aut$ that do
not correspond simply to a piecewise analytic map $\Sig\to\Sig$. In
particular, given any bijection $b:\Sig\to\Sig$, any choice of paths
$p_x\in\Mor(x,b(x))$ defines a functor $\phi_{b,p}\in\Nyl$. On the
other hand, each edge-interchanger acts trivially on the objects in
$\Sig$, but changes morphisms. Thus the automorphisms allow for a
lot of freedom how to change metagraphs $\m\in\M$. In this section
we will present a lemma that shows how
large the orbit of a graph is under the action of $\Aut$.\\ 
%
%

\begin{Lemma}\label{Lem:TwoCombinatoriallyEqualHyphsCanBeMappedInotEachOther}
Given any two hyphs $v_1=(p_1,\ldots,p_H)$ and
$v_2=(q_1,\ldots,q_H)$. Assume the two are combinatorially the same,
i.e. there is a bijection $b$ between the vertices $b:V(v_1)\to
V(v_2)$ and a bijection $c:\{p_1,\ldots,p_H\}\to\{q_1,\ldots,q_H\}$
such that
\begin{eqnarray}
s(c(p_k))\;=\;b(s(p_k)),\qquad t(c(p_k))\;=\;b(t(p_k)).
\end{eqnarray}

\noindent Then there is an automorphism $\phi\in\Aut$ such that
$\phi(v_1)=v_2$.\footnote{Note that this is slightly weaker than
demanding that every automorphism between the subgroupoids
$\P_{v_1}\to\P_{v_2}$ can be extended to an automorphism on $\P$.}
\end{Lemma}

\noindent\textbf{Proof:} We will explicitly construct this
automorphism as a product of finitely many natural transformations
of the identity and edge-interchangers. First we show that there is
a sequence of edge-interchangers such that $v_1$ can be mapped to a
graph $\g$, which has the same vertices as $v_1$, as well as the
same combinatorics as $v_1$.

Choose a graph $\g'=\{e'_1,\ldots,e'_{E'}\}$ such that $v_1\leq
\g'$. Then, by definition, there is an edge $e'_l$ in $\g'$ such
that $e'_l$ meets $\{p_1,\ldots,p_{H-1}\}$ at most in its endpoints.
\begin{eqnarray}
p_H\;=\;\tilde p_1\circ e'_l\circ \tilde p_2
\end{eqnarray}

\noindent Choose an analytic curve $c$ (i.e. [[[c]]] is an edge)
from $s(p_H)$ to $t(p_H)$ that does not meet any of the $e'_k$,
apart from the points $s(p_H)$ and $t(p_H)$. Then choose two curves
$c_1$ and $c_2$ with the following properties:

\begin{itemize}
\item $s(c_1)=c(\frac{1}{4})$ and $t(c_1)=s(e'_l)$
\item $s(c_2)=c(\frac{3}{4})$ and $t(c_2)=t(e'_l)$
\item The curves $c_1$, $c_2$ are injective and $r(c_1)$, $r(c_2)$
do not intersect, as well as they do not intersect with any of the
$r(c)$, $r(e'_k)$, apart from their beginning- and endpoints.
\end{itemize}

\noindent For any curve $c$ and $0\leq t_1,t_2\leq 1$ denote by
$c^{t_1,t_2}$ the curve
\begin{eqnarray}
c^{t_1,t_2}(t):=c(t_1+t(t_2-t_1)).
\end{eqnarray}

\noindent Then by  construction the following paths
\begin{eqnarray}\label{Gl:DefinitonOfInterchangedEdges}
e_1\;&:=&\;[[[c_1^{\frac{1}{2},1}]]]\;\circ\;e'_l\;\circ\;[[[c_2^{\frac{1}{2},1}]]]^{-1}\\[5pt]\nonumber
e_2\;&:=&\;[[[c_1^{0,\frac{1}{2}}]]]^{-1}\;\circ\;[[[c^{\frac{1}{4},\frac{3}{4}}]]]\;\circ\;[[[c_2^{0,\frac{1}{2}}]]]
\end{eqnarray}

\noindent are paths without self-intersection that mutually
intersect only at their beginning- and endpoints. Furthermore, they
intersect with the other edges $e'_k,\,k\neq l$ only at most in
$s(e'_l),\, t(e'_l)$. Thus, by construction, the same is true for
the paths $p_1,\ldots,p_{H-1}$, since they are built from the
$e'_k,\,k\neq l$.\\


Now choose representative curves $d_1$ and $d_2$ for $e_1$ and $e_2$
such that

\begin{itemize}
\item $d_1\left(\frac{1}{4}\right)=s(e'_l)$ and
$d_1\left(\frac{3}{4}\right)=t(e'_l)$
\item $d_2\left(\frac{1}{4}\right)=c\left(\frac{1}{4}\right)$ and
$d_1\left(\frac{3}{4}\right)=c\left(\frac{3}{4}\right)$.
\end{itemize}

\noindent Finally choose for every
$t\in(0,1)\backslash\left\{\frac{1}{4},\frac{3}{4}\right\}$ a path
$p_t$ that goes from $d_1(t)$ to $d_2(t)$, and which does not meet
$e_1,\,e_2$ apart from their respective beginning- and endpoint.
Furthermore, define
\begin{eqnarray}\label{Gl:DefinitionOfIntermittingPaths}
p_{\frac{1}{4}}\;&:=&\;(\tilde
p_1)^{-1}\,\circ\,[[[c^{0,\frac{1}{4}}]]]\\[5pt]\nonumber
p_{\frac{3}{4}}\;&:=&\;\tilde
p_2\,\circ\,[[[c^{\frac{3}{4},1}]]]^{-1}.
\end{eqnarray}

\noindent Now we have chosen two paths without self-intersections
$e_1,\,e_2$ which meet only in their beginning- and endpoint, a
parametrization $d_1,d_2$ for each of the paths, and for each
$t\in(0,1)$ a path $p_t\in\Mor(d_1(t),d_2(t))$ that does not meet
$e_1,e_2$, apart from its beginning- and endpoint. This data defines
an automorphism $\phi\in\Aut$ by lemma
\ref{Lem:DefinitionOfEdgeInterchanger}.\\

\noindent Since $e_1,\,e_2$ have been chosen to be such that they
meet the $p_1,\ldots,p_{H-1}$ only in at most finitely many points,
we conclude that $\phi(p_k)=p_k$ for $k=1,\ldots,H-1$. By the
definition of $\phi$ and (\ref{Gl:DefinitonOfInterchangedEdges}),
(\ref{Gl:DefinitionOfIntermittingPaths}), we have
\begin{eqnarray}\nonumber
\phi(p_H)\;&=&\;\phi(\tilde
p_1)\,\circ\,\phi(e'_l)\,\circ\,\phi(\tilde p_2)\\[5pt]\nonumber
&=&\;\tilde p_1\,\circ\,\big(p_{\frac{1}{4}}\circ
[[[c^{\frac{1}{4},\frac{3}{4}}]]]\circ
p_{\frac{3}{4}}^{-1}\big)\,\circ\,\tilde p_2\\[5pt]
&=&\;[[[c^{0,\frac{1}{4}}]]]\,\circ\,[[[c^{\frac{1}{4},\frac{3}{4}}]]]\,\circ\,[[[c^{\frac{3}{4},1}]]]\\[5pt]\nonumber
&=&\;[[[c]]].
\end{eqnarray}

\noindent So by this construction, all paths $p_k$ have been left
invariant, except for $p_H$ which has been mapped to an edge
$e_H:=[[[c]]]$, which meets the $p_k,\,k<H$ at most in its
beginning- and endpoint. For this construction it was crucial that
$p_H$ had a free segment w.r.t all the other $p_k,\,k<H$, but also
w.r.t. $e_H$, by construction. This means that
$(e_H,p_1,\ldots,p_{H-1})$ is also a hyph, i.e. each of the $p_k$
has a free segment w.r.t all $p_l,\,l<k$, but also w.r.t $e_H$. So
the same construction can be carried out once again to obtain an
automorphism $\phi'\in\Aut$ that leaves $e_H$ as well as all
$p_k,\,k>H-1$ invariant, but maps $p_{H-1}$ to an edge $e_{H-1}$
which meets $e_H,p_k,\,k<H-1$ at most in its beginning-and endpoint.
This gives a hyph $(e_{H-1},e_H,p_1,\ldots,p_{H-2})$. By repeating
this process, we arrive at a hyph $(e_1,\ldots,e_H)$, where all the
$e_k$ are edges and meet at most in their beginning- and endpoint.
So $\g:=\{e_1,\ldots,e_H\}$ is a graph. Thus, we have constructed a
finite sequence of automorphisms that map the hyph
$(p_1,\ldots,p_H)$ into the graph $\g$.\\

Now let $\g''=\{e''_1,\ldots,e''_H\}$ be any graph which has the
same combinatorics as $\g=\{e_1,\ldots,e_H\}$, but the edges of
which do not intersect with the edges of $\g$.
$V(\g)=\{v_1,\ldots,v_V\}$ and $V(\g'')=\{v''_1,\ldots,v''_V\}$, and
\begin{eqnarray}
s(e_k)=v_a,\,t(e_k)=v_b\;\qquad\Leftrightarrow\;\qquad
s(e''_k)=v''_a,\,t(e''_k)=v''_b.
\end{eqnarray}

Then construct an automorphism mapping one to the other by the
following method: For each $v_l\in V(\g)$ choose a path
$p_{v_l}\in\Mor(v_l,\,v''_l)$ such that all $p_{v_l}$ are without
self-intersections, do not intersect each other, and intersect the
$e_k,e''_k$ only in the vertices $v_l,v''_l$. Then define a natural
transformation of the identity $\phi_{b,p}\in\Nyl$ by the following
data:
\begin{eqnarray}
b(v_l)\;&:=&\;v''_l,\quad b(v'')\;:=\;v_l,\quad b(x)\;=\;x\;\text{
\rm
else}\\[5pt]\nonumber
p_{v_l}\;&:=&\;p_{v_l},\quad p_{v''_l}\;:=\;p_{v_l}^{-1},\quad
p_x\;=\;\id_x\,\text{ \rm else}
\end{eqnarray}

\noindent Then each of the $p'_k:=\phi_{b,p}(e_k)$ is a path without
self-intersections, and all the $p'_k$ have free segments w.r.t each
other (in particular the $e_k$). So $(p'_1,\ldots,p'_H)$ is a hyph
with vertices $v''_1,\ldots,v''_V$. On the other hand, $\g''$ is a
graph, the edges $e''_k$ of which intersect all the $p'_k$ at most
in the vertices $v''_l$. Since for each path $p'_k$, the edge $e'_k$
starts and ends at the same points as $p'_k$ by construction, there
is a sequence of $H$ edge-interchangers that maps each $p'_k$ to
$e''_k$.\\

So we have started with a hyph $v_1=(p_1,\ldots,p_H)$, have mapped
$v_1$ to $\g=(e_1,\ldots,e_H)$ by a sequence of $H$
edge-interchangers, have mapped $\g$ by a natural transformation of
the identity $\phi_{b,p}$ to the hyph $v_1'=(p'_1,\ldots,p'_H)$, and
have mapped $v'_1$ by another sequence of $H$ edge-interchangers to
any other graph $\g''$, of which we only demanded that it has the
same combinatorics as $v_1$, and that it has empty intersection with
$\g$. By the same construction, we can show that we can map any two
graphs that have the same combinatorics but empty intersection into
each other by a sequence of automorphisms. But this shows that one
can map any two hyphs $v_1,\,v_2$ having the same combinatorics into
each other by a sequence of edge-interchangers and natural
transformations of the identity. Thus, the lemma is proven.

\subsection{Orbits of the Automorphisms}

In order to define the automorphism-invariant Hilbert space $\Haut$,
it is instructive to get an idea of the orbits of vectors
$\psi\in\Hkin$ under the action of $\Aut$. First we will investigate
these orbits in the following for arbitrary gauge groups $G$, in
order to arrive at some general statements. Then, we will specialize
to $G=U(1)$ and $G=SU(2)$ in order to say something about the
corresponding spaces $\Haut$.\\

\begin{Lemma}\label{Lem:NTDIDLeaveGaugeInvariantFunctionsInvariant}
Let $f\in Cyl$ be a gauge-invariant cylindrical function. Then for
any natural transformation of the identity $\phi_{b,p}\in\Nyl$ one
has
\begin{eqnarray}
\hat U(\phi_{b,p})f\;=\;f
\end{eqnarray}
\end{Lemma}

\noindent\textbf{Proof:} Let $f$ be cylindrical over the graph
$\g=\{e_1,\ldots,e_E\}$. For each $v\in V(\g)$ and each $a\in\AL$
denote $g^A_v:=A(p_{v})\in G$. With
\begin{eqnarray}
f(A)\;=\;F(A(e_1),\ldots,A(e_E))
\end{eqnarray}

\noindent we get
\begin{eqnarray}\nonumber
\big(\hat
U(\phi_{b,p})f\big)(A)\;&=&\;f\big(\a_{\phi_{b,p}}A)\\[5pt]\nonumber
&=&\;F\Big(A(p_{s(e_1)}^{-1}\circ e_1\circ
p_{t(e_1)}),\,\ldots,A(p_{s(e_E)}^{-1}\circ e_E\circ
p_{t(e_E)})\Big)\\[5pt]
&=&\;F\Big((g_{s(e_1)}^A)^{-1}
\,A(e_1)\,g_{t(e_1)},\ldots,(g_{s(e_E)}^A)^{-1}
\,A(e_E)\,g_{t(e_E)}\Big)\\[5pt]\nonumber
&=&\;F\Big(A(e_1),\ldots,A(e_E)\Big)\\[5pt]\nonumber
&=&\;f(A).
\end{eqnarray}

\noindent Lemma \ref{Lem:NTDIDLeaveGaugeInvariantFunctionsInvariant}
shows one important fact: Although $\phi_{b,p}$ can change graphs
quite arbitrarily, the corresponding functions on that graph remain
unchanged. The reason for this is that a function cylindrically over
a metagraph $\m\in\M$ does generally not carry all of the
information in order to reconstruct $\m$ from its dependence $A\to
f(A)$.

\begin{figure}[hbt!]\label{fig:Fig1}
\begin{center}
    \psfrag{p1}{$p_1$}
    \psfrag{p2}{$p_2$}
    \psfrag{p3}{$p_3$}
    \psfrag{p4}{$p_4$}
    \includegraphics[scale=0.75]{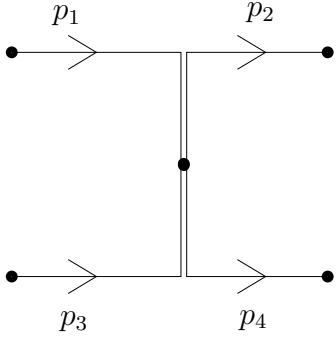}
    \end{center}
    \caption{\small A metagraph $\m$ (in fact a hyph) consisting of four paths and five vertices.}
\end{figure}

Consider the following example: Given a metagraph
$\m=\{p_1,\ldots,p_4\}$, as in picture \ref{fig:Fig1}.

\noindent Note that the paths $p_1\circ p_2$ and $p_3\circ p_4$ each
contain a retracing. Now consider the function cylindrical over the
metagraph $\m$ by

\begin{eqnarray}\label{Gl:AFunctionCylindricalOverAHyph}
f(A)\;=\;F\big(A(p_1),\,A(p_2),\,A(p_3),\,A(p_4)\big)
\end{eqnarray}

\noindent with $F(h_1,h_2,h_3,h_4)=\tilde F(h_1h_2,\,h_3h_4)$ for
some smooth function $\tilde F$ on $G^2$. Note that $f$ does not
depend on the parallel transports along all the $p_1,\ldots,p_4$,
but only on the ones along $p_1\circ p_2$ and $p_3\circ p_4$. So in
particular, $f$ does not depend at all on the parallel transports
along the retracings. Consequently, $f$ is cylindrical over the
graph $\g$, which consists of the following two edges:

\begin{figure}[hbt!]\label{fig:Fig2}
\begin{center}
    \psfrag{p1p1}{$p_1\circ p_2$}
    \psfrag{p3p4}{$p_3\circ p_4$}
    \includegraphics[scale=0.75]{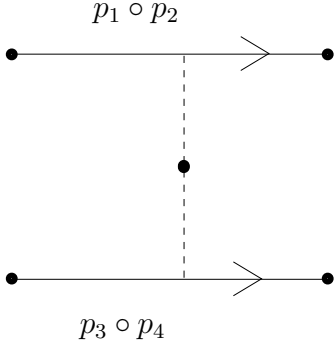}
    \end{center}
    \caption{\small The function $f$ given by (\ref{Gl:AFunctionCylindricalOverAHyph}) is also cylindrical over the graph $\tilde\g\leq\m$, which only consists of two edges and four vertices}
\end{figure}

\noindent So we see that the dependence of the function $f$ is not
on all of $\m$, but only of a certain subgraph. This is of
particular importance, since the following graph $\g$:

\begin{figure}[hbt!]\label{fig:Fig3}
\begin{center}
    \psfrag{e1}{$e_1$}
    \psfrag{e2}{$e_2$}
    \psfrag{e3}{$e_3$}
    \psfrag{e4}{$e_4$}
    \includegraphics[scale=0.75]{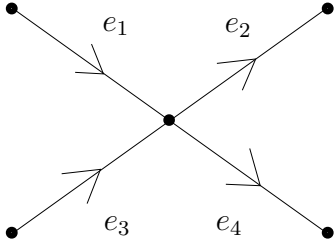}
    \end{center}
    \caption{\small The metagraph $\m$ can be mapped to this graph $\g$ by an automorphism}
\end{figure}

\noindent has the same combinatorics as $\m$. So, by lemma
\ref{Lem:TwoCombinatoriallyEqualHyphsCanBeMappedInotEachOther} there
is an automorphism $\phi\in\Aut$ mapping one to the other, i.e.
$\phi(e_k)=p_k$. Thus, a function $f$ cylindrical over $\g$ is
mapped into $\a_{\phi}f$, which is cylindrical over $\m$. But if $f$
depends only on the parallel transports along $e_1\circ e_2$ and
$e_3\circ e_4$, then $\a_{\phi}$ is also cylindrical over
$\tilde\g\leq\m$, as we have just seen. So, although $\phi$ respects
the number of vertices and paths in a metagraph, $\hat U(\phi)$ can
map a function which is cylindrical over a metagraph to a function
cylindrical over another metagraph, which does not have the same
combinatorics, such as $\g$ and $\tilde\g$ in the example above.

Although this seem paradoxical at first, it is in fact quite
natural: Consider the function $f$, which is cylindrical over the
graph $\g$, as in figure (\ref{fig:Fig3}), and which depends only on
the parallel transports along $e_1\circ e_2$ and $e_3\circ e_3$.
Then the function does not ``know'' that it is cylindrical over a
graph with four edges, in particular it does not know anything about
the middle vertex $t(e_1)$. For instance, it is automatically
gauge-invariant w.r.t. gauging at this vertex. This does, of course,
not happen to an arbitrary function $g$ cylindrical over $\g$, whose
dependence on the parallel transports along all four edges
$e_1,\ldots,e_4$ is nontrivial. \\

We see that the cylindrical functions can carry much less
information than the graph that they are cylindrical over. In
particular, the functions only carry information about on how many
parallel transports they depend, and which of these start or end at
the same points. It is exactly this information that is preserved by
the automorphisms.

This is of particular importance for the gauge-invariant functions,
since these carry only information about the first fundamental class
of the graph, but not about the graph topology itself, which is
summarized by the following lemma.

\begin{Lemma}\label{Lem:EveryGaugeInvariantFunctionCanBeMadeIntoAFlower}
Let $f\in Cyl$ be a gauge-invariant cylindrical function over a
graph $\g$ with $E$ edges and $V$ vertices. Then there is a
$E-V+1$-flower graph $\tilde \g$ (a graph with one vertex and
$E-V+1$ edges all starting and ending at that vertex) and an
automorphism $\phi\in\Aut$ such that $\hat U(\phi)f$ is a
gauge-invariant function cylindrical over $\tilde \g$.
\end{Lemma}

\begin{figure}[hbt!]\label{fig:Fig3}
\begin{center}
    \psfrag{e1}{$e_1$}
    \psfrag{e2}{$e_2$}
    \psfrag{e3}{$e_3$}
    \psfrag{...}{\ldots}
    \psfrag{eE}{$e_E$}
    \includegraphics[scale=0.75]{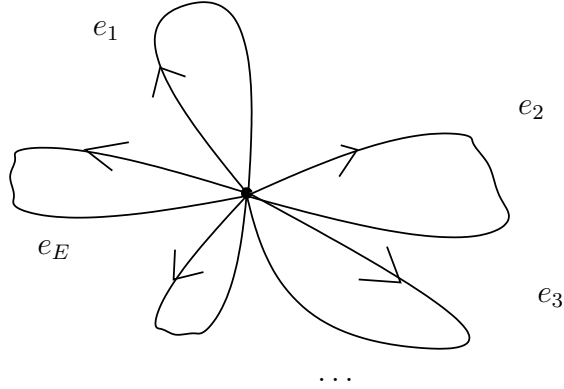}
    \end{center}
    \caption{\small An $E$-flower graph.}
\end{figure}

\noindent\textbf{Proof:} Choose a maximal tree $\t$ in the graph
$\g$. Choose a vertex $x\in V(\t)$ and for each edge $e_l\in
E(\g)\backslash E(\t)$ not belonging to the tree denote the unique
path from $x$ to $s(e_l)$ lying in $\t$ as $p_l^1$, and from $v$ to
$t(e_l)$ as $p_l^2$. Then define the $E-V+1$ paths
\begin{eqnarray}
p_l\;:=\;p_l^1\circ e_l\circ (p_l^2)^{-1}.
\end{eqnarray}

\noindent Then $v=(p_1,\ldots,p_{E-V+1})$ is a hyph, since each path
$p_l$ contains $e_l$, which is a free segment for all the other
paths. Note that, due to gauge-invariance, the function $f$ only
depends on the parallel transports along the paths $p_l$ (see, e.g.
\cite{FREI}). In particular, $f$ is cylindrical over the hyph
$v:=(p_1,\ldots,p_{E-V+1})$: $f\in Cyl(v)$. But since all paths
$p_l$ in $v$ start and end in the vertex $x$, by lemma
\ref{Lem:TwoCombinatoriallyEqualHyphsCanBeMappedInotEachOther} there
is an automorphism mapping the hyph $v$ to an $E-V+1$-flower graph
$\tilde\g$. So $f$ gets mapped to a function $\a_{\phi}f$
being cylindrical over that flower graph.\\

 Lemma \ref{Lem:EveryGaugeInvariantFunctionCanBeMadeIntoAFlower}
shows the tremendous size that the orbits of a vector under the
action of $\Aut$ have. In the case of the abelian gauge group
$G=U(1)$, the size is so large that one can in fact compute the
orbits, in the sense that one can determine the set of equivalence
classes of vectors in $\Hkin$ under the action of $\Aut$.


\subsection{The automorphism-invariant Hilbert space for gauge group $G=U(1)$}

Cylindrical functions for gauge group $G=U(1)$ carry particularly
few information about the graph they are cylindrical over.

\begin{Lemma}\label{Lem:HowOneIsAbleToModifyFlowers}
Let $\g=\{e_1,\ldots,e_E\}$ be an $E$-flower graph. Consider a word
$\vartheta(e_2,\ldots,e_E)$ in the edges of $\g$, apart from $e_1$.
Since $\g$ is a flower graph, $\vartheta$ is a path in the groupoid
$\P_{\g}$, i.e. a path starting and ending at the one vertex in
$V(\g)$, and going through the edges $e_2,\ldots,e_E$ in an order
determined by the word $\vartheta$. Then there is an automorphism
$\phi\in\Aut$ with the following properties:
\begin{eqnarray}
\phi\;:\;e_1\,&\longmapsto&\,e_1\circ\vartheta(e_2,\ldots,e_E)\\[5pt]\nonumber
e_k\,&\longmapsto&\,e_k,\quad\text{\rm for }k\in\{2,\ldots,E\}
\end{eqnarray}
\end{Lemma}

\noindent\textbf{Proof:}  Note that
$(e_2,\ldots,e_E,e_1\circ\vartheta)$ is a hyph, since
$e_1\circ\vartheta(e_2,\ldots,e_E)$ has a free segment w.r.t.
$e_2,\ldots,e_E$ (which is $e_1$), and the remaining
$(e_2,\ldots,e_E)$ form a graph. Since $(e_1,\ldots,e_E)$ is also a
hyph, having the same combinatorics, the assertion follows directly
from lemma \ref{Lem:TwoCombinatoriallyEqualHyphsCanBeMappedInotEachOther}.\\

\begin{Lemma}\label{Lem:HowOneIsAbleToModifyChargeNetworks}
Let $\g$ be an $E$-flower graph, $\vec n\in\Z^{E}$ and $T_{\g,\vec
n}$ be a charge network. Let $(m_2,\ldots,m_E)\in\Z^{E-1}$, then
there is an automorphism $\phi\in\Aut$ such that
\begin{eqnarray}
\hat U(\phi)T_{\g,\vec n}\;=\;T_{\g,\vec n'}
\end{eqnarray}

\noindent with
\begin{eqnarray}\label{Gl:HowOneIsABleToChangeTheCharges}
n'_1\;&=&\;n_1\\[5pt]\nonumber
n'_k\;&=&\;n_k\,+\,n_1m_k\quad\text{ \rm for }k\in\{2,\ldots,E\}
\end{eqnarray}
\end{Lemma}

\noindent\textbf{Proof:} Choose the word
\begin{eqnarray}
\vartheta(e_2,\ldots,e_E)\;=\;e_2^{m_2}\circ\ldots\circ e_E^{m_E}.
\end{eqnarray}

\noindent Then, by lemma \ref{Lem:HowOneIsAbleToModifyFlowers},
there is an automorphism $\phi\in\Aut$ with
\begin{eqnarray}
\phi(e_1)=e_1\circ e_2^{m_2}\circ\ldots\circ e_E^{m_E},
\end{eqnarray}

\noindent and which leaves all other $e_k$ invariant. Thus we get
\begin{eqnarray}\nonumber
\Big(\a_{\phi}T_{\g,\vec
n}\Big)(A)\;&=&\;\prod_{k=1}^E\,A(\a_{\phi}e_k)^{n_k}\\[5pt]
&=&\;\left[A(e_1)\,\prod_{k=2}^EA(e_k)^{m_k}\right]^{n_1}\,\prod_{k=2}^EA(e_k)^{n_k}\\[5pt]\nonumber
&=&\;A(e_1)^{n_1}\,\prod_{k=2}^E\,A(e_k)^{n_k+n_1m_k}\\[5pt]\nonumber
&=&\;T_{\g,\vec n'}(A),
\end{eqnarray}

\noindent where $T_{\g,\vec n'}$ is given by
(\ref{Gl:HowOneIsABleToChangeTheCharges})\\

\begin{Lemma}\label{Lem:HowOneIsAbleToDetermineTheOrbit}
Let $\g$ be an $E$-flower graph, $\vec n\in\Z^E$, and $T_{\g,\vec
n}$ a charge-network function. Then there is an automorphism
$\phi\in\Aut$ such that
\begin{eqnarray}
\a_{\phi}T_{\g,\vec n}\;=\;T_{\g,(p,0,0,\ldots,0)},
\end{eqnarray}

\noindent where $p\in\Z$ is the greatest common divisor of the
$|n_1|,\ldots,|n_E|$.
\end{Lemma}

\noindent\textbf{Proof:} First assume that all $n_k\geq 0$. If this
is not the case, one can, by lemma
\ref{Lem:TwoCombinatoriallyEqualHyphsCanBeMappedInotEachOther}, find
an automorphism that reverses the direction of all edges $e_k$ such
that $n_k<0$:

\begin{eqnarray*}
\phi(e_k)&=&e_k^{-1}\qquad\text{\rm if }n_k<0\\[5pt]
\phi(e_k)\;&=&\;e_k\qquad\text{\rm else}
\end{eqnarray*}

\noindent We will consider a sequence of 'steps'
(\ref{Gl:HowOneIsABleToChangeTheCharges}), each of which changes the
charge distribution among the edges $e_1,\ldots,e_E$, and
corresponds to the action of an automorphism $\phi\in\Aut$. We will
construct the sequence such that, at the end, the charge
distribution $(n_1,\ldots,n_E)$ will have changed into
$(p,0,0,\ldots,0)$.\\

\noindent First, assume that not all of the $n_k$ are zero,
otherwise $T_{\g,\vec n}=\Omega$ is the vector corresponding to the
constant function $f(A)\equiv 1$, and we are done.

Choose one of the smallest $n_k$, which is without loss of
generality $n_1$. Then choose another $n_l$, $l\geq 1$ with $n_1\leq
n_l$. Then, define $m_l:=-1,\,m_k=0$ for $1\neq k\neq l$. Then, by
lemma \ref{Lem:HowOneIsAbleToModifyChargeNetworks} there is an
automorphism $\phi^{(1)}\in\Aut$ that maps $T_{\g,\vec n}$ to
$T_{\g,\vec n^{(1)}}$ with $n^{(1)}_k=n_k$, for $k\neq l$, and
$n_l^{(1)}=n_l-n_k\geq 0$. Continue with choosing one of the
smallest $n_k^{(1)}$ (by renumbering, this is again wlog
$n^{(1)}_1$), and choose one $n^{(1)}_l$ with $n^{(1)}_1\leq
n_l^{(1)}$, define $m^{(1)}_l=-1$, $m^{(1)}_k=0$ for $1\neq k\neq
l$. By lemma \ref{Lem:HowOneIsAbleToModifyChargeNetworks} find an
automorphism $\phi^{(2)}$ mapping $T_{\g,\vec n^{(2)}}$ with
$n_k^{(2)}=n_k^{(1)}$ for $k\neq l$ and
$n_l^{(2)}=n_l^{(1)}-n_1^{(1)}\geq 0$, choose the smallest nonzero
$n_k^{(3)}$, and continue this algorithm.

Since each of the $\vec n^{(r)}$ is a sequence of nonnegative
numbers, and in each step one of the numbers is reduced by some
$n_1^{(r-1)}>0$, the algorithm stops after finitely many steps, say
after $N\leq \sum_kn_k$ steps. The algorithm cannot be continued
further, if after choosing the smallest $n_1^{(N)}$, there is no
other nonzero $n_l^{(N)}$, i.e. if all other $n_k^{(N)}$ are zero.
Then, we have constructed a series of automorphisms mapping
$T_{\g,\vec n}$ to some $T_{\g,(n_1^{(N)},0,0,\ldots,0)}$:
\begin{eqnarray}
\hat U(\phi^{(N)})\hat U(\phi^{(N-1)})\cdots\hat
U(\phi^{(1)})\,T_{\g,\vec n}\;=\;T_{\g,(p,0,0,\ldots,0)}
\end{eqnarray}

\noindent with $p=n_1^{(N)}$. It is easy to see that in each step,
the greatest common divisor of the nonzero $n_k^{(r)}$ never
changes. Thus, since $p$ is its own greatest common divisor, it has
to be equal to the greatest common divisor of all the nonzero $n_k$.
This completes the proof.\\

\begin{Lemma}\label{Lem:CharegOnOneFlowerIsFixedUpToASign}
Let $\g$ be a one-flower graph. If for $n,m\in\Z$ there is an
automorphism $\phi\in\Aut$ such that $\hat U(\phi)T_{\g,
n}=T_{\g,m}$, then $n=\pm m$.
\end{Lemma}

\noindent Assume there is a $\phi\in\Aut$ such that $\hat
U(\phi)T_{\g, n}=T_{\g,m}$. Then $\phi(\g)$ is a metagraph
$\phi(\g)\leq \g$. Thus, $\phi(\g)\in\P_{\g}$. But
$\P_{\g}=\{\g^k\,|\,k\in\Z\}$. Thus, $\phi(\g)=\g^k$ for some
$k\in\Z$. Since $\hat U(\phi^{-1})T_{\g,m}=T_{\g,n}$, we also have
$\phi^{-1}(\g)\leq\g$, i.e. $\phi^{-1}(\g)=\g^l$ for some $l\in\Z$.
On the other hand,
\begin{eqnarray}
\g\;=\;\phi\Big(\phi^{-1}(\g)\Big)\;=\;\phi\Big(\g^l\Big)\;=\;\Big(\phi(\g)\Big)^l\;=\;\g^{kl}
\end{eqnarray}

\noindent by the functorial properties of $\phi$. Thus, $kl=1$, but
the only two pairs of integer numbers $k,l\in\Z$ with this property
are $k=l=1$ and $k=l=-1$. So either $\phi(\g)=\g$ or
$\phi(\g)=\g^{-1}$. The claim follows.\\


By lemma \ref{Lem:EveryGaugeInvariantFunctionCanBeMadeIntoAFlower},
every gauge-invariant charge-network function can be mapped by an
automorphism to one on a flower graph (on which all charge-networks
are automatically gauge-invariant). In lemma
\ref{Lem:HowOneIsAbleToDetermineTheOrbit} we have seen that for
gauge group $G=U(1)$, each charge-network function on a flower graph
can be mapped by an automorphism into $T_{\g,n}$ for some one-flower
graph $\g$ and some $n\in\Z$. This $n$ is unique up to a sign, as we
have seen in Lemma \ref{Lem:CharegOnOneFlowerIsFixedUpToASign}. We
have thus determined the complete set of orbits of charge-network
states under the action of $\Aut$.

\begin{Lemma}
Let $D$ be the linear span of all charge-network states in $\Hkin$.
Then the following map:
\begin{eqnarray}\label{Gl:AutomorphismInnerProductForU(1)}
\eta\;&:&\;D\;\longrightarrow\;D_{\Aut}^*\\[5pt]\nonumber
\eta[f]g\;&:=&\;\sum_{[\phi]\in\Aut/\sim} \langle f\,|\,\hat
U(\phi)g\rangle
\end{eqnarray}

\noindent defines an antilinear map from $D$ to the
automorphism-invariant linear functionals over $D$. Here
$\phi_1\sim\phi_2$ if $\langle f|\hat U(\phi_1)g\rangle=\langle
f|\hat U(\phi_2)g\rangle$.
\end{Lemma}

\noindent\textbf{Proof:} We have seen that for each charge-network
state $T_{\g,\vec n}$ there is a unique $n\geq 0$ such that
\begin{eqnarray}
\Big[T_{\g,\vec n}\Big]\;=\;\Big[\bigcirc_n\Big]
\end{eqnarray}

\noindent where $\bigcirc$ is a one-flower graph, $\bigcirc_n$
denotes the charge-network function on $\bigcirc$ given by the one
charge $n$, and $[\cdot]$ denotes the orbit of $\cdot$ under $\Aut$.
Denote this $n$ by $\o(\vec n)$, then we have
\begin{eqnarray}
\sum_{[\phi]\in\Aut/\sim} \langle T_{\g,\vec n}\,|\,\hat
U(\phi)T_{\g',\vec n'}\rangle\;=\;\d_{\o(\vec n), \o(\vec n')},
\end{eqnarray}

\noindent due to the fact that any $\phi\in\Aut$ maps a charge
network into another charge network, so $\langle T_{\g,\vec
n}\,|\,\hat U(\phi)T_{\g',\vec n'}\rangle$ is either $0$ or $1$.\\

With this rigging map, an inner product
$\langle\cdot|\cdot\rangle_{\rm Aut}$ can be defined on the set of
all finite linear combinations of charge-network functions, and we
thus see

\begin{Corollary}
The automorphism-invariant inner product on $\eta(D)$ given by the
rigging map (\ref{Gl:AutomorphismInnerProductForU(1)}) can be
completed to the automorphism-invariant Hilbert space
\begin{eqnarray}
\Haut\;=\;\left\{\sum_{k=0}^{\infty}c_n\,\Big[\bigcirc_n\Big]\;\Bigg|\;\sum_n|c_n|^2<\infty\right\}
\end{eqnarray}
\end{Corollary}

\noindent\textbf{Proof:} This is clear from the fact that the orbit
of every finite linear combination $f$ of charge networks is given
by
\begin{eqnarray}
\big[f\big]\;=\;\sum_{k=0}^Nc_n\,\Big[\bigcirc_n\Big]
\end{eqnarray}

\noindent and
\begin{eqnarray}
\left\langle\,\Big[\bigcirc_n\Big]\;\Big|\;\,\Big[\bigcirc_m\Big]\right\rangle_{\rm
Aut}\;=\;\d_{nm}.
\end{eqnarray}

\noindent As we have seen, the automorphism-invariant Hilbert space
can be computed directly, for a certain choice of rigging map
(\ref{Gl:AutomorphismInnerProductForU(1)}). For each charge-network
function $T_{\g,\vec n}$ on a graph $\g$, there is a natural number
$n$ and an automorphism $\phi$ such that $\a_{\phi}T_{\g,\vec
n}\;=\;\bigcirc_n$. This shows how tremendously large the orbits of
the automorphism group are in the case of $G=U(1)$. The reason for
this is the following: given any graph $\g$ with $E$ edges and a
charge-network function $T_{\g,\vec n}$, then
\begin{eqnarray}
T_{\g,\vec
n}(A)\;=\;e^{i\left(n_1\varphi_1+\cdots+n_E\varphi_E\right)}
\end{eqnarray}

\noindent where $e^{i\varphi_k}=A(e_k)$ is the holonomy of $A$ along
the edge $e_k$. Then, our previous results show that there is a
closed loop $l$ in $\g$, i.e. a path consisting of edges in $\g$ and
their inverses, which starts and ends at the same point, such that
each $e_k$ is traversed exactly $n_k$ times (counting going against
the orientation of $e_k$ as $-1$). Then

\begin{eqnarray}
T_{\g,\vec n}(A)\;=\;e^{i \varphi}
\end{eqnarray}

\noindent with $\varphi=n_1\varphi_1+\cdots+n_E\varphi_E$, and
$e^{i\varphi}=A(l)$ is the holonomy along $l$. So the charge-network
function $T_{\g, \vec n}$ is cylindrical over the metagraph $l$. If
$l=\tilde l^n$ for a simple loop $\tilde l$, then $T_{\g,\vec n}$ is
also cylindrical over $\tilde l$, and we have shown that there is an
automorphism $\phi$ mapping the loop $\tilde l$ to a (say) circle
$\bigcirc$, hence $T_{\g,\vec n}$ to $\bigcirc_n$.\\

The above consideration rests crucially on the abelianess of $U(1)$.
In particular, it does not matter in which order $l$ transverses the
paths in $\g$, just how many times. This will not be true for
non-abelian gauge groups, such as $G=SU(2)$, as we will see in the
following.\\


\subsection{The automorphism-invariant Hilbert space for gauge group $G=SU(2)$.}

In this section, we investigate the set of orbits of vectors in
$\Hkin$ under the action of $\Aut$, in the case of $G=SU(2)$.
Ultimately, the goal is to compute the set of linear functionals
invariant under $\Aut$ with some inner product, such as in the case
for $G=SU(2)$, for example
\begin{eqnarray}\label{Gl:AutomorphismInnerProductForSU(2)}
\eta\;&:&\;D\;\longrightarrow\;D_{\Aut}^*\\[5pt]\nonumber
\eta[f]g\;&:=&\;\sum_{[\phi]\in\Aut/\sim} \langle f\,|\,\hat
U(\phi)g\rangle
\end{eqnarray}

\noindent Again, $\phi_1\sim\phi_2$ if $\langle f|\hat
U(\phi_1)g\rangle=\langle f|\hat U(\phi_2)g\rangle$.\\

To compute the set of orbits of vectors in $\Hkin$ under the action
of $\Aut$ for the case of $G=SU(2)$ is more difficult, due to the
fact that for a spin network function $T_{\g,\vec j,\vec n,\vec m}$,
the transformed $\a_{\phi}T_{\g,\vec j,\vec n,\vec m}$ is not
necessarily a spin network function anymore. The same holds for any
gauge-invariant spin network function $T_{\g,\vec j,\vec I}$: If one
chooses, for each graph $\g$, an orthonormal basis of intertwiners
$\vec I$ at each vertex, then any such basis vector can be mapped to
a finite linear combination of basis vectors by an automorphism

From this difficulty arises the phenomenon that, for two vectors
$T_{\g,\vec,\vec I}$, $T_{\g',\vec j',\vec I'}$ the set of all
possible overlaps
\begin{eqnarray}
\Big\{\langle T_{\g,\vec,\vec I}\,|\,\a_{\phi}\,T_{\g',\vec j',\vec
I'}\rangle\;\Big|\;\phi\in\Aut\Big\}
\end{eqnarray}

\noindent is not just $\{0,1\}$, as in the case for piecewise
analytic diffeomorphisms, or for automorphisms and gauge group
$G=U(1)$. We consider a simple example:\\

We have already seen in lemma
\ref{Lem:EveryGaugeInvariantFunctionCanBeMadeIntoAFlower} that every
gauge-invariant function on a graph can be mapped by an automorphism
to a gauge-invariant function on a flower graph. Thus, it is
sufficient to consider the orbits of gauge-invariant functions on
flower graphs. Consider the\footnote{Technically, there are many
flower graphs, but all of them can be mapped into each other by the
automorphisms by Lemma
\ref{Lem:TwoCombinatoriallyEqualHyphsCanBeMappedInotEachOther}, so
in the following we speak of ``the'' $2$-flower graph.} $2$-flower
graph $\g=\{e_1,e_2\}$.

\begin{figure}[hbt!]\label{Fig:2-Flower}
\begin{center}
    \psfrag{e1}{$e_1$}
    \psfrag{e2}{$e_2$}
    \psfrag{j1}{$j_1$}
    \psfrag{j2}{$j_2$}
    \includegraphics[scale=0.75]{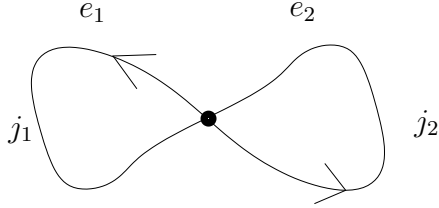}
    \end{center}
    \caption{\small The $2$-flower graph}
\end{figure}

\noindent For fixed $j_1,\,j_2\in\frac{1}{2}\N$, the intertwiner
space is $j_1+j_2-|j_1-j_2|+1$-dimensional. Assume
$j_1=j_2=\frac{1}{2}$. Then the intertwiner space is
two-dimensional. If we choose the normalized vector
\begin{eqnarray}
T_1(A)\;:=\;\tr_{\frac{1}{2}}\,\big(A(e_1)\big)\,\tr_{\frac{1}{2}}\,\big(A(e_2)\big)
\end{eqnarray}

\noindent to be one of the two orthonormal basis vectors in the
intertwiner space, and choose some normalized $T_2$ orthogonal to
it, we get an orthonormal base of the intertwiner space. On the
other hand, consider the one-flower graph (in fact, a Wilson loop)
$\bigcirc=\{e\}$ having only one edge $e$, and the gauge-invariant
function $T_0$ cylindric on $\bigcirc$ given by
\begin{eqnarray}
T_0(A)\;:=\;\tr_{\frac{1}{2}}\big(A(e)\big).
\end{eqnarray}

\noindent With the spin $j=\frac{1}{2}$ on the edge $e$, there is
only one gauge-invariant function on $\bigcirc$ (the
intertwiner-space is one-dimensional), which is given exactly by
$T_0$. By lemma
\ref{Lem:TwoCombinatoriallyEqualHyphsCanBeMappedInotEachOther}
however, we know that there is an automorphism $\phi$ such that
$\phi(e)\;=\;e_1\circ e_2$, so $\a_{\phi}T_0$ is cylindrical over
the two-flower graph $\g$, and in fact can be decomposed into
$T_1,\,T_2$. Since

\begin{eqnarray}
\a_{\phi}T_0(A)\;=\;\tr_{\frac{1}{2}}\big(A(e_1)\cdot A(e_2)\big),
\end{eqnarray}

\noindent a short calculation reveals that
\begin{eqnarray}\label{Gl:T_0AndT_1HaveAnAverlapOfSixtyDegrees}
\big\langle\,T_1\,\big|\,\a_{\phi}T_0\,\big\rangle\;=\;\frac{1}{2}.
\end{eqnarray}

\noindent So

\begin{eqnarray}
\a_{\phi}T_0\;=\;\cos\frac{\pi}{3}\,T_1\;+\;\sin\frac{\pi}{3}\,T_2.
\end{eqnarray}

\noindent We see that, since there are nontrivial ways to embed an
$E$-flower into an $E'$-flower (with $E<E'$) by ``wrapping loops'',
certain basis vectors on the $E$-flower will be mapped to nontrivial
combinations of basis vectors on the $E'$-flower. All these
nontrivial overlaps will show up in the automorphism-invariant inner
product (\ref{Gl:AutomorphismInnerProductForSU(2)}). To compute all
the contributions is now a combinatorial task. We refrain from doing
so here, but simply state a conjecture about the nature of the set
of orbits of vectors in $\Hkin$ under the action of $\Aut$.

In our example above, we have seen that the function $T_1$ and
$\a_{\phi}T_0$ constitute a basis of the intertwiner space for
functions on a $2$-flower graph with $j_1=j_2=\frac{1}{2}$. Note
that this is not an orthonormal basis, but by
(\ref{Gl:T_0AndT_1HaveAnAverlapOfSixtyDegrees}) have an angle of
$\cos\frac{\pi}{3}$ with respect to each other. Now consider again
the two-flower graph $\g=\{e_1,e_2\}$, but with
$j_1=\frac{1}{2},\,j_2=1$. Then again, the intertwiner space is
two-dimensional, and one normalized vector is
\begin{eqnarray}
T_3(A)\;:=\;\tr_{\frac{1}{2}}\big(A(e_1)\big)\,\tr_1\big(A(e_2)\big).
\end{eqnarray}

\noindent On the other hand, consider the automorphism
$\phi'\in\Aut$ mapping the closed loop $e$ to $e_1\circ (e_2)^2$.
One can then show that the projection of $\a_{\phi'}T_0$ to the
intertwiner space on the $2$-flower graph with
$j_1=\frac{1}{2},\,j_2=1$ is unequal $T_3$, but has non-vanishing
inner product with $T_3$.  Thus, this projection (denote it as $\Pi
(\a_{\phi'}T_0))$, together with $T_3$ form a basis for the
intertwiner space for $j_1=\frac{1}{2},\,j_1=1$. Again, this is no
orthonormal basis, since the two vectors are only linearly
independent, not orthogonal.

For each $E$-flower graph $\g=\{e_1,\ldots,e_E\}$ with spins
$j_1,\ldots,j_E$ there is a normalized vector in the intertwiner
space given by
\begin{eqnarray}
T_{E,\vec j}(A)\;:=\;\prod_{k=1}^E\,\tr_{j_k}\big(A(e_k)\big).
\end{eqnarray}

\noindent In a graphical notation similar to the one used in the
last chapter, we write
\begin{eqnarray}
\left[\begin{array}{cccc}\bigcirc_{j_1}& & & \\ & & &
\bigcirc_{j_2}\\& \bigcirc_{j_{E-1}} & & \\ \bigcirc_{j_E}& & \cdots
&
\end{array}\right]\;:=\;[T_{E,\vec j}]
\end{eqnarray}

\noindent Extending what we have just seen in the examples to higher
flowers with higher spins suggests that all other elements in the
intertwiner space can be composed by certain $\a_{\phi}T_{E',\vec
j'}$ with $E'<E$. This would mean that the orbits $[T_{E,\vec j}]$
would constitute a basis for the set of orbits of all smooth
cylindrical functions in $\Hkin$. Since these vectors have
nontrivial overlap, however, one would have to compute their
automorphism-invariant inner product by
(\ref{Gl:AutomorphismInnerProductForSU(2)}) and deduce an
orthonormal basis from them by the Gram-Schmidt-procedure. This
would provide a way to find an orthonormal basis for $\Haut$.\\

This would suggest that the linear combinations of the
form
\begin{eqnarray}
\psi\;=\;\sum_{E=0}^{N}\sum_{j_1,\ldots,j_E}\,c_{E, \vec
j}\;\left[\begin{array}{cccc}\bigcirc_{j_1}& & & \\ & & &
\bigcirc_{j_2}\\& \bigcirc_{j_{E-1}} & & \\ \bigcirc_{j_E}& & \cdots
&
\end{array}\right]
\end{eqnarray}

\noindent form a dense set in $\Haut$. If one could then derive a
good formula for the inner product between two such vectors, we
would be able to write down an orthonormal basis for $\Haut$. So
far, this has not been done due to the complicated combinatorics,
but we will address this point in a later publication.


\section{Summary and Outlook}

\subsection{Summary of the work}

In this publication, we have investigated the consequence of
extending the group of spatial diffeomorphisms $\Diff(\Sig)$ to path
groupoid automorphisms $\Aut$ in Loop Quantum Gravity. This
extension is inspired by category theory, and contains many elements
that cannot be interpreted as diffeomorphisms from $\Sig$ to itself.
This mimics the extension of smooth to generalized gauge
transformations: while the first consists of smooth maps from $\Sig$
to the gauge group $G$, the latter one consists of all such maps,
without continuity or even measurability assumption.

An automorphism is given by a permutation of the points in $\Sig$,
and a permutation of paths in $\Sig$ which are compatible with each
other in the sense that if the path $p$ starts at $x$ and ends at
$y$, then the transformed path $\phi(p)$ starts at $\phi(x)$ and
ends at $\phi(y)$. But by the groupoid structure of $\P$, the set of
piecewise analytic paths in $\Sig$, this does not necessarily
restrict what happens to points that lie ``in the middle'' of $p$.
So, the notion of a point lying on a path is not invariant under
automorphisms, which makes many nontrivial constructions possible.
In particular, some automorphisms simply cannot be interpreted as
maps from $\Sig$ to itself.

We have given some explicit examples for automorphisms that do not
arise as diffeomorphisms on $\Sig$. The first example was given by
the natural transformations of the identity, which are able to
arbitrarily permute the points in $\Sig$, but keep the paths
essentially the same (and which act, in particular, as identity on
the gauge-invariant part of the kinematical Hilbert space $\Hkin$).
The second example was given by the edge-interchangers, which left
all points invariant, but swapped two edges $e_1,e_2$ with the same
beginning- and endpoints. All paths that meet these two edges at
finitely many points are left invariant, however. In this sense,
these automorphisms have support only at two edges $e_1, e_2$, and
hence can be viewed as distributional.

We have used these two types of automorphisms in order to show that
every two graphs with the same combinatorics can be mapped into each
other by an automorphism $\phi\in\Aut$. This shows how little
information about the differential structure, and in fact even the
topology of $\Sig$ is encoded in the path groupoid $\P$.

We have delivered a proof that the action of $\Aut$ on $\AL$, the
set of (distributional) connections leave the
Ashtekar-Isham-Lewandowski measure $\m_{AL}$ invariant. This was
straightforward in the case of the abelian gauge group $G=U(1)$, but
nontrivial for arbitrary gauge groups.\\

In the last part of this work, we have investigated the induced
action of the automorphisms $\Aut$ on the kinematical Hilbert space
$\Hkin$. In particular, we have showed that due to the size of
automorphisms, the only information that is conserved by acting with
automorphisms on a cylindrical function, is the combinatorics of
paths, over which it is cylindrical. This information is highly
redundant in the description of function cylindrical over paths, and
it is the use of graphs to provide a way of finding a good
representative in the set of collections of paths over which a
function is cylindrical. Since the automorphisms do not leave the
set of graphs invariant, functions cylindrical over one graph can be
mapped to a function cylindrical over another graph, although the
graphs themselves are not mapped to each other.

A gauge-invariant function on a graph $\g$ does only depend on the
holonomies along a number of loops in $\g$, which correspond to the
first fundamental class $\pi_1(\g)$. Consequently, any
gauge-invariant function on a graph can be mapped to a
gauge-invariant function on some flower graph. This enabled us to
gain some control over the size of the orbits of vectors in $\Hkin$
under the action of $\Aut$. For some choice of rigging map, we
derived the automorphism-invariant Hilbert space $\Haut$ for the
gauge group $G=U(1)$. It was found that the space is
infinite-dimensional and separable, the generic element being of the
form
\begin{eqnarray}\label{Gl:ThisCouldBeAAutomorphismInvariantFunctionForU(1)}
\psi_{\rm Aut}\;=\;\sum_{k=0}^{\infty}c_n\,\Big[\bigcirc_n\Big],
\end{eqnarray}

\noindent with square-summable coefficients $\{c_n\}_{n\in\N}$, and
$\big[\bigcirc_n\big]$ being the equivalence class of one Wilson
loop with charge $n$.

For $G=SU(2)$, the combinatorics to work out the exact form of the
orbits is harder, due to the recoupling scheme of $SU(2)$. However,
we argued why we believe an element of the automorphism-invariant
Hilbert space would be of the form

\begin{eqnarray}\label{Gl:ThisCouldBeAAutomorphismInvariantFunctionForSU(2)}
\psi_{\rm Aut}\;=\;\sum_{E=0}^{\infty}\sum_{j_1,\ldots,j_E}\,c_{E,
\vec j}\;\left[\begin{array}{cccc}\bigcirc_{j_1}& & & \\ & & &
\bigcirc_{j_2}\\& \bigcirc_{j_{E-1}} & & \\ \bigcirc_{j_E}& & \cdots
&
\end{array}\right],
\end{eqnarray}

\noindent where the vectors are equivalence classes of $E$ separate
Wilson loops with spin charges $j_k$. These vectors will, however
not be orthonormal to each other. Rather, their inner product will
be determined by embedding-combinatorics of $E$-flowers into
$E'$-flowers for $E<E'$, and the corresponding recoupling scheme. We
will return to a detailed analysis of this space in a future
publication.

\subsection{Further directions}

By the form of the vectors
(\ref{Gl:ThisCouldBeAAutomorphismInvariantFunctionForU(1)}) and
(\ref{Gl:ThisCouldBeAAutomorphismInvariantFunctionForSU(2)}), we see
that the information about the degrees of freedom is completely
delocalized, as one would have expected from a
diffeomorphism-invariant theory. However, in order to get a good
physical intuition for the meaning of the states in $\Haut$, one
would have the following possibility: One can repeat the whole
analysis, but with matter degrees of freedom coupled to gravity.
Note that the automorphism-invariant content of a cylindrical
function for pure gravity is just given by the combinatorics of
paths, the holonomies of which it depends on, but not on how these
paths are embedded into space, i.e. if they intersect, or are
partially parallel. So the automorphism-invariant vectors do not
know about the vertices of a graph, for instance. Having matter
degrees of freedom coupled to gravity, this might change, due to the
following reason: Consider a, say, fermionic matter field coupled to
gravity. This field, when quantized along the lines of Loop Quantum
Gravity, would become a field sitting on the vertices of a graph,
and is transformed by some nontrivial representation of $SU(2)$. The
information about which field is situated at which vertex will still
be contained in the gauge-invariant sector of the theory. Shifting
these gauge-invariant cylindrical functions for matter plus gravity
around by automorphisms would result in graphs being mapped to other
metagraphs, that look like different graphs, since the paths can
intersect, or partially overlap. But that some matter is excited at
some vertex will not be changed by this, so one would be able to
distinguish the real vertices (where matter is excited) from the
ones that just appear as vertices, because the paths are embedded
into space in some peculiar way. This would provide a natural
mechanism of how matter could be used to localize gravitational
degrees of freedom, as has been advocated e.g. in \cite{ROVELLI1}.

Another important point is to see whether operators corresponding to
physical observables can be defined, i.e. like the volume operator.
This would not only allow for an interpretation of the
automorphism-invariant states in terms of physical quantities. Also,
if one could define a volume operator on the automorphism-invariant
Hilbert space, one would be able to define the master constraint
operator \cite{QSD8} on the automorphism-invariant Hilbert space.
Since the graph-changing version of the master constraint changes in
particular the first fundamental group of a graph, and the
automorphism-invariant vectors contain exactly this information,
this might provide a way to rephrase the quantum dynamics in some
combinatorial way in $\Haut$. This would be particularly
interesting, since one could make contact to \cite{TINA1}, where
such an operator already exists, leading to a combinatorial version
of the
dynamics.\\

One key step in this proof was theorem \ref{Thm:Nielsen}, a
classical result from combinatorial group theory by Nielsen ('36).
On the other hand, we also relied on the explicit construction of
the kinematical Hilbert space $\Hkin$ for this proof. From an
aesthetical point of view, however, this is unsatisfactory, since we
feel that the fundamental reason for the automorphisms to leave the
measure $\m_{AL}$ invariant is exactly Nielsen's theorem. It states
a deep connection between automorphisms of free groups and the
symmetries of the Haar measure. We are convinced that one can write
down a proof that all automorphisms preserve $\m_{AL}$, with only
relying on Nielsen's theorem, without referring to the analytical
structure of the manifold $\Sig$ at all. This would enable us to
transfer the proof to arbitrary groupoids $\P$, not necessarily the
path groupoid of a spatial manifold $\Sig$\footnote{At least as long
as these groupoids are sets, since in these cases definitions of
topology and measure make sense.}. With this one might be able to
quantize much more general theories of connections on groupoids, not
only on manifolds.\\

The path groupoid $\P$, as a category, is a useful concept when
investigation quantizations of Riemannian metrics on $3$-manifolds.
It was indicated \cite{BEAZSF, HYKGIR, MAAK} that for the
investigation of Lorentzian metrics on $4$-manifolds the notion of
$2$-category is an appropriate concept. It would be interesting to
investigate, in which sense the analysis presented here could be
repeated in such a framework, in order to see space-time
diffeomorphisms as automorphisms of $2$-categories.

\section*{Acknowledgements}

BB would like to thank Carla Cederbaum for time and patience, and
Christian Fleischhack for discussions about hyphs, groupoids and
analyticity. Also, thanks goes to Johannes Brunnemann, Andreas
D\"{o}ring, Cecilia Flori, Jerzy Lewandowski, Hendryk Pfeiffer and
Hanno Sahlmann for numerous discussions about automorphisms.


\appendix

\section{Elements from category theory}\label{App:CategoryNotions}


In this section, we will briefly review the basic notions of
category theory that are used in this article\footnote{Note that in
this appendix, we use the standard convention for the composition,
i.e. $f:X\to Y$ and $g:Y\to Z$ have a composition $g\circ f:X\to Z$.
In the rest of the article, however, we will use the notation
$f\circ g$ for the composition, in order to stay consistent with
large parts of LQG literature.}. Details and more about categories
can be found in \cite{CAT}.

\begin{Definition}
A category $\Cyl$ consists of a class of objects $X, Y, Z, \ldots$,
denoted by $|\Cyl|$, and, for each pair of objects $X,Y\in|\Cyl|$ a
class of morphisms $f,g,h,\ldots$, denoted by $\Mor_{\Cyl}(X,Y)$.
For these, the following rules hold:
\begin{itemize}
\item For each $f\in\Mor_{\Cyl}(X,Y)$ and $g\in\Mor_{\Cyl}(Y,Z)$
there is a morphism $g\circ f\in\Mor_{\Cyl}(X,Z)$ (the composite).
\item Composition is associative, i.e. $h\circ(g\circ f)=(h\circ
g)\circ f$.
\item For each object $X\in|\Cyl|$ there is a morphism
$\id_X\in\Mor_{\Cyl}(X,X)$ such that for all morphisms
$f\in\Mor_{\Cyl}(X,Y)$ and $g\in\Mor_{\Cyl}(Z, X)$ one has
\begin{eqnarray*}
f\circ \id_X\;=\;f,\qquad\qquad \id_X\circ g\;=\;g
\end{eqnarray*}
\end{itemize}

\noindent If $f\in\Mor_{\Cyl}(X,Y)$, then the \emph{source} and the
\emph{target} of $f$ are denoted by $s(f):=X$ and $t(f):=Y$.
\end{Definition}

\noindent There are plenty of examples for categories:

\begin{itemize}
\item The category $Set$, the objects of which are
sets, and the morphisms between two sets $X, Y$ are exactly all maps
between these sets.
\item For a manifold $\Sig$, the category $Hom(\Sig)$, the objects of
which are the points in $\Sig$, and the morphisms of which are
homotopy equivalence classes of curves between points.
\item For a manifold $\Sig$, the category $\mathcal{W}(\Sig)$, the objects
of which are points in $\Sig$, and the morphisms between two points
are all curves between these points, modulo reparametrization.
\item For a group $G$, the category $\Susp(G)$, called the
\emph{suspension of $G$}. This category has only one object, denoted
by $*$: $|\Susp(G)|=\{*\}$, and the morphisms from $*$ to itself are
in one-to-one correspondence with elements in $G$:
$\Mor_{\Susp(G)}(*,*)=G$.
\end{itemize}

\noindent A category $\Cyl$ in which every morphism has a right and
a left inverse is called a \emph{groupoid}. In the above examples,
$Set$ and $\mathcal W(\Sig)$ are no groupoids, but $Hom(\Sig)$ and
$\Susp(G)$ are.\\

It is customary in category theory to write statements as diagrams:
Consider a category $\Cyl$ and objects $X,Y,Z,W\in|\Cyl|$. Then the
following diagram
\begin{eqnarray*}
\begin{CD}
X @>f>> Y \\
@VgVV       @VVhV\\
Z @>k>> W
\end{CD}
\end{eqnarray*}

\noindent  is said to \emph{commute}, of $f\in\Mor_{\Cyl}(X,Y)$,
$g\in\Mor_{\Cyl}(X,Z)$, $h\in\Mor_{\Cyl}(Y,W)$ and
$k\in\Mor_{\Cyl}(Z,W)$, and $k\circ g\;=\;h\circ f$, as morphisms in
$\Mor_{\Cyl}(X, W)$. Working with commuting diagrams makes reasoning
in category theory fairly intuitive. Since the proofs in this paper
will be thoroughly analytical however, we will use commutative
diagrams only at some points, to point out some connections.\\

Consider two morphisms $f,g$ in $\mathcal{W}(\Sig)$ with $s(g)=t(f)$
can be composed, i.e. two curves (up to reparametrization) can be
concatenated and again give a curve modulo parametrization. It
should be noted that, due to the above definitions, the
concatenation is denoted by $g\circ f$. However, in the Loop Quantum
Gravity literature, this concatenation is usually denoted as $f\circ
g$. The reason is that with this convention the generalized
connections are functors from the path groupoid into the suspension
of the gauge group $\Susp(G)$, not its opposite category
$\Susp(G)^{op}$.

This is just a matter of convention, of course, but it should be
noted that, throughout this paper, the composition of morphisms
$f,g$ will usually be denoted as $f\circ g$, not as $g\circ f$.\\


\begin{Definition}
Let $\Cyl,\,\Dyl$ be two categories. A functor is an assignment $F:
|\Cyl|\to |\Dyl|$ and
$F:\Mor_{\Cyl}(X,Y)\,\to\,\Mor_{\Dyl}(F(X),F(Y))$, such that
\begin{itemize}
\item $F(g\circ f)\;=\;F(g)\circ F(f)$
\item $F(id_X)\;=\;id_{F(X)}$
\end{itemize}
\end{Definition}

\noindent So a functor assigns objects to objects and morphisms to
morphisms in a compatible way. As an example, consider the category
which has smooth manifolds as objects and smooth maps between them
as morphisms. Then the Cartan differential is a functor from this
category in itself. In particular, let $f:M\to N$ and $g:N\to O$ be
smooth maps, then the chain rule guarantees that
\begin{eqnarray}
d(g\circ f)\;=\;dg\,\circ\,df
\end{eqnarray}

\noindent as maps  $df:TM\to TN$, i.e. $dg:TN\to TO$.\\


\begin{Definition}
Let $F:\Cyl\to \Dyl$ and $G:\Cyl\to \Cyl$ be two functors. One calls
these two functors to be related by a natural transformation (or
being natural transformations from each other), if there is, for
each object $X\in|\Cyl|$ a morphism $g_X\in\Mor_{\Dyl}(F(X), G(X))$
such that the following diagram commutes:
\begin{eqnarray*}
\begin{CD}
F(X) @>F(f)>> F(Y) \\
@Vg_XVV       @VVg_YV\\
G(X) @>G(f)>> G(Y)
\end{CD}
\end{eqnarray*}
\end{Definition}

\noindent Note that, if $\Dyl$ is a groupoid (i.e. its morphisms can
be inverted), then given a functor $F$ and for each $X\in|\Cyl|$ a
morphism $g_X\in\Mor_{\Dyl}(F(X), Y)$, then
\begin{eqnarray*}
G(X)\;&:=&\;t(g_X)\qquad\text{\rm for all }X\in |\Cyl|\\[5pt]
G(f)\;&=&\;g_Y\circ F(f)\circ g_X^{-1}\qquad \text{\rm for all
}f\in\Mor_{\Cyl}(X,Y)
\end{eqnarray*}

\noindent\emph{defines} a functor $G$, which can be related to $F$
by a natural transformation. We will use this construction often for
defining functors which are natural transformations of other, given
functors.

\section{Elements from combinatorial group
theory}\label{App:CombinatorialGroupTheory}

\noindent Going over from piecewise analytic diffeomorphisms
$\Diff(\Sig)$ to the automorphisms $\Aut$ is a significant
enlargement of the gauge group. While the former preserves notions
of (generalized) knotting classes of a graph, the latter one only
keeps combinatorial information of the graph, i.e. which vertices
are attached to each other by paths and which are not. Consequently,
elements of combinatorial group theory enter the description, as
soon as automorphisms are considered. In this section we review some
basic notions from combinatorial group theory (details can be found
in \cite{COMB}) and conclude with a classical result of Nielson,
which shows a connection between automorphisms of free groups and
the symmetries of Haar measures. This will be one key point in
proving that the elements of $\Aut$
act unitarily on $\Hkin$. \\

\begin{Definition}
Let $E\in\N$, and $\{e_1,\ldots,e_E\}$ be $E$ abstract symbols,
called \emph{letters}. A (finite) \emph{word} in these letters is a
(finite) sequence in these letters, i.e. $e_2,\,e_4e_1$ or
$e_1e_1e_7e_1$. A word in the $2E$ letters
$\{e_1,\ldots,e_E,e_1^{-1},\ldots,e_E^{-1}\}$ is called
\emph{reduced}, if no $e_k$ occurs next to its inverse $e_k^{-1}$.
\end{Definition}

\noindent It is clear that every word can be put into its reduced
form, by successively eliminating $e_ke_k^{-1}$ or $e_k^{-1}e_k$
from it.

\begin{Definition}
Consider the set of all reduced words in the $2E$ letters
$\{e_1^{\pm1},\ldots,e^{\pm1}_E\}$, together with the ``empty
word'', denoted by $1$, and define multiplication between two words
as concatenating (and possibly reducing) them. The set of all these
reduced words form a group under this multiplication, which is
called the free group in $E$ letters, and is denoted by $F_E$.
\end{Definition}

\noindent Examples:
\begin{eqnarray}
\big(e_1e_2e_3\big)&\cdot&\big(e_3e_2e_1\big)\;=\;e_1e_2e_3e_3e_2e_1\\[5pt]
\big(e_1e_2e_3\big)&\cdot&\big(e_3^{-1}e_2^{-1}e_{17}\big)\;=\;e_1e_{17}
\end{eqnarray}

\noindent Similarly one can define the free group in zero
$F_0=\{1\}$ and in countable many parameters $F_{\omega}$. Free
groups play a prominent r\^{o}le in combinatorics, algorithm theory
and graph theory. Obviously, for $E<E'$, there is a natural
inclusion of $F_E$ as a subgroup $F_{E'}$. What is less intuitive is
that the free group in two parameters $F_2$ has a subgroup
isomorphic to $F_{\omega}$. As a consequence, every $F_E$ has a
subgroup isomorphic to $F_{E'}$ for $E<E'$. This makes these groups
more difficult than their linear counterparts, the $E$-dimensional
vector spaces.\\

\noindent A homomorphism $\phi$ from a free group $F_E$ into another
group is completely determined by its values on the basic letters,
$\phi(e_1),\ldots,\phi(e_E)$. An invertible group homomorphism from
$F_E$ to itself is called an automorphism, and is completely
determined by the $E$ words $\phi(e_k)=\vartheta_k(e_1,\ldots,e_E)$
which are the images of the $e_k$ under $\phi$. The automorphisms of
a free group are in fact well understood, which is shown by the
following theorem.

\begin{Theorem}\label{Thm:Nielsen} (Nielsen)
Let $\phi$ be an automorphism on $F_E$. Then $\phi$ can be written
as a finite product of ``elementary'' automorphisms
\begin{eqnarray}
\phi\;=\;\xi_n\circ\xi_{n-1}\circ \ldots\circ\xi_1
\end{eqnarray}

\noindent where every $\xi_r$ is one of the following:

\begin{itemize}
\item A permutation ($l\neq k\in\{1,\ldots,E\}$):
\begin{eqnarray}\label{Gl:ElementarySubstitution}
\xi_r(e_k)\;=\;e_l,\qquad\xi_r(e_l)\;=\;e_r,\qquad
\xi_r(e_m)\;=\;e_m\text{ \rm else}
\end{eqnarray}
\item An inversion ($k\in\{1,\ldots E\}$):
\begin{eqnarray}\label{Gl:ElementaryInversion}
\xi_r(e_k)\;=\;e_k^{-1},\qquad \xi_r(e_l)\;=\;\xi_r(e_l)\text{ \rm
else}
\end{eqnarray}
\item A shift ($k\neq l\in\{1,\ldots,E\})$:
\begin{eqnarray}\label{Gl:ElementaryShift}
\xi_r(e_k)\;=\;e_k e_l,\qquad\xi_r(e_l)\;=\;\xi_r(e_l),\qquad
\xi_r(e_m)\;=\;e_m\text{ \rm else}
\end{eqnarray}
\end{itemize}

\end{Theorem}

\noindent This important structural theorem has an immediate
consequence:

\begin{Corollary}\label{Cor:CorollaryToNielson}
Let $G$ be a compact Lie group with bi-invariant Haar measure
$d\m_H$. Let $F\in L^1\big(G^E,\,d\m_H^{\otimes E}\big)$, and $\phi$
be an automorphism on $F_E$. Denote
$\phi(e_k)=\vartheta_k(e_1,\ldots,e_E)$. Then
\begin{eqnarray}\label{Gl:AutomorphismsLeaveIntegralInvariant}
\int_{G^E}d\m^{\otimes E}_{H}&&(h_1,\ldots,
h_E)\;F(h_1,\ldots,h_E)\\[5pt]\nonumber
&&\;=\;\int_{G^E}d\m^{\otimes E}_{H}(h_1,\ldots,
h_E)\;F\Big(\vartheta_1(h_1,\ldots,h_E),\ldots,\vartheta_E(h_1,\ldots,h_E)\Big),
\end{eqnarray}
\end{Corollary}

\noindent\textbf{Proof:}  Nielson's theorem tells us that the
substitution
\begin{eqnarray}
\begin{array}{rcl}h_1&\;\longmapsto\,&\vartheta_1(h_1,\ldots,h_E)\\[5pt]\vdots& & \vdots\\[5pt]h_E&\;\longmapsto\,&\vartheta_E(h_1,\ldots,h_E)\end{array}
\end{eqnarray}

\noindent can be achieved by successively applying the
``elementary'' substitutions (\ref{Gl:ElementarySubstitution}),
(\ref{Gl:ElementaryInversion}), (\ref{Gl:ElementaryShift}) (with the
symbols $h_k$, instead of the $e_k$). But these are exactly the
substitutions under which the Haar measure $d\m^{\otimes
E}_H(h_1,\ldots,h_E)$ is invariant. The statement
(\ref{Gl:AutomorphismsLeaveIntegralInvariant}) follows.


\section{Categorical Weyl algebra of quantum
gravity}\label{App:CategorialWeylAlgebra}

\noindent The holonomies and their transformations, i.e.
gauge-transformations and diffeomorphisms, can be formulated in
terms of category theory, and this formulation suggests to enlarge
the diffeomorphism group $\Diff(\Sig)$ to the automorphisms of the
path groupoid $\Aut$, which act unitarily on the kinematical Hilbert
space $\H_{\text{kin}}$ of Loop Quantum Gravity.

The field algebra of Loop quantum gravity, however, consists of
holonomies as well as fluxes, and the symmetry groups act on both.
In this section, we will show that also the fluxes can be formulated
naturally in the languages of categories\footnote{Again, we use the
convention that the concatenation of two morphisms $p$ and $q$ is
denoted as $p\circ q$ if $t(p)=s(q)$, in order to stay consistent
with the notation in \cite{INTRO}}.

We will work with the exponentiated fluxes instead of the fluxes
themselves, which leads to the Weyl algebra of LQG, for technical
reasons. The fluxes themselves can be recovered from these
exponentiated versions easily.\\

\subsection{Categorial formulation of oriented surfaces}

We want to obtain the categorical formulation of an ''oriented
surface''. An oriented surface does in principle two things: First,
it cuts a path into several pieces (wherever a path intersects the
surface), and second, it assigns to each of these pieces two numbers
in $\{-1,\,0,\,1\}$. These numbers depend on whether the the
starting- or, respectively, the end point of the piece meets the
surface along or against the orientation of the surface $(\pm 1)$,
or whether the starting- or end part of the piece lies entirely away
from, or entirely within the surface (in which case it gets assigned
$0$). We will now turn this intuition into a categorical language.\\

\begin{Definition}
Let $\Cat$ be a category. Then define the category $\Or(\Cat)$ as
follows:
\begin{itemize}
\item The objects in both categories are the same:
$|\Or(\Cat)|\,=\,|\Cat|$.
\item Let $X,\,Y$ be objects in $\Or(\Cat)$. The morphisms $\text{Mor }(X,\,Y)$
in $\Or(\Cat)$ can be constructed as follows: Consider three finite
sequences with all the same length $N$: A sequence of paths
$p_1,\ldots,p_N$, as well as two sequences of natural numbers
$m_1,\ldots,m_N$ and $n_1,\ldots n_N$. We arrange these three
sequences as
\begin{eqnarray}
\left(\begin{array}{ccc}p_1,\,&\ldots,\,&p_N\\m_1,\,&\ldots,\,&m_N\\n_1,\,&\ldots,\,&n_N\end{array}\right)
\end{eqnarray}

 The paths have the properties that $t(p_k)=s(p_{k+1})$ for
$k=1,\ldots N-1$, as well as $s(p_1)=X$ and $t(p_N)=Y$. We define an
equivalence relation on these three sequences, by
\begin{eqnarray}
\left(\begin{array}{cccccc}p_1,\,&\ldots,\,&p_k,\,&p_{k+1},\,&\ldots,\,&p_N\\m_1,\,&\ldots,\,&m_k,\,&-n,\,&\ldots,\,&m_N\\n_1,\,&\ldots,\,&n,\,&n_{k+1},\,&\ldots,\,&n_N\end{array}\right)\;\sim\;
\left(\begin{array}{ccccc}p_1,\,&\ldots,\,&p_k\circ
p_{k+1},\,&\ldots,\,&p_N\\m_1,\,&\ldots,\,&m_k,\,&\ldots,\,&m_N\\n_1,\,&\ldots,\,&n_{k+1},\,&\ldots,\,&n_N\end{array}\right)
\end{eqnarray}

\noindent and

\begin{eqnarray}
\left(\begin{array}{ccccc}p_1,\,&\ldots,\,&\text{\rm
id},\,&\ldots,\,&p_N\\m_1,\,&\ldots,\,&m,\,&\ldots,\,&m_N\\n_1,\,&\ldots,\,&n,\,&\ldots,\,&n_N\end{array}\right)\;\sim\;
\left(\begin{array}{ccccc}p_1,\,&\ldots,\,&\text{\rm
id},\,&\ldots,\,&p_N\\m_1,\,&\ldots,\,&m+k,\,&\ldots,\,&m_N\\n_1,\,&\ldots,\,&n-k,\,&\ldots,\,&n_N\end{array}\right)
\end{eqnarray}

 for all $k\in\Z$. The morphisms from $X$ to $Y$ then
contain all equivalence classes of these sequences, which we denote
by

\begin{eqnarray}
\left[\begin{array}{ccc}p_1,\,&\ldots,\,&p_N\\m_1,\,&\ldots,\,&m_N\\n_1,\,&\ldots,\,&n_N\end{array}\right]\;\in\;\text{Mor}(X,\,Y)
\end{eqnarray}

Concatenation of morphisms is obtained by just concatenating the
sequences:

\begin{eqnarray}
\left[\begin{array}{ccc}p_1,\,&\ldots,\,&p_N\\m_1,\,&\ldots,\,&m_N\\n_1,\,&\ldots,\,&n_N\end{array}\right]\;\circ\;
\left[\begin{array}{ccc}q_1,\,&\ldots,\,&q_M\\r_1,\,&\ldots,\,&r_M\\s_1,\,&\ldots,\,&s_M\end{array}\right]\;:=\;
\left[\begin{array}{ccccc}p_1,\,&\ldots,\,&p_N,\,q_1,\,&\ldots,\,&q_M\\m_1,\,&\ldots,\,&m_N,\,r_1,\,&\ldots,\,&r_M\\n_1,\,&\ldots,\,&n_N,\,s_1,\,&\ldots,\,&s_M\end{array}\right]
\end{eqnarray}

\noindent while the identity functor in $\text{Mor }(X, X)$ is given
by

\begin{eqnarray}
\text{\rm
id}\,:=\;\left[\begin{array}{c}\id_X\\0\\0\end{array}\right]
\end{eqnarray}

\end{itemize}
\end{Definition}

\noindent It is straightforward to check that with these equivalence
relations, the category $\Or(\Cat)$ becomes a groupoid, if $\Cat$ is
one, if one defines the inverse of a morphism by

\begin{eqnarray}
\left[\begin{array}{ccc}p_1,\,&\ldots,\,&p_N\\m_1,\,&\ldots,\,&m_N\\n_1,\,&\ldots,\,&n_N\end{array}\right]^{-1}
\;:=\;\left[\begin{array}{ccc}p_N^{-1},\,&\ldots,\,&p_1^{-1}\\-n_N,\,&\ldots,\,&-n_1\\-m_N,\,&\ldots,\,&-m_1\end{array}\right]
\end{eqnarray}

\noindent There is a projection functor

\begin{eqnarray}
\pi:\;\Or(\Cat)\;\to\;\Cat
\end{eqnarray}

\noindent the action of which is given by the identity on the
objects, and by

\begin{eqnarray}
\pi\;:\;\left[\begin{array}{ccc}p_1,\,&\ldots,\,&p_N\\m_1,\,&\ldots,\,&m_N\\n_1,\,&\ldots,\,&n_N\end{array}\right]\;\longmapsto\;p_1\circ\cdots\circ
p_N.
\end{eqnarray}

\noindent One checks quickly that this action is well-defined.\\

In the following, we will consider the path groupoid $\PG$ and the
category $\Or(\PG)$. With the category $\Or(\PG)$ we have a notion
at hand to say what ''cutting an edge'' $p$ means in category
language.

\begin{Definition}
 Let $\PG$ be the path groupoid of a manifold $\Sig$. Then a
functor

\begin{eqnarray}
S\,:\,\PG\,\longrightarrow\,\Or(\PG)
\end{eqnarray}

\noindent is called a \emph{generalized oriented surface}, if the
following two conditions hold:

\begin{itemize}
\item The functor $\pi\circ S$ is the identity functor on $\PG$.
\item For each primitive metagraph $\m$ with morphisms $q_1,\ldots,\,q_n$ in $\PG$,
there is a primitive metagraph $\m'\geq \m$ with morphisms
$p_1,\ldots,\,p_m$,such that each $p_k$ is \emph{$S$-trivial}, which
means that
\begin{eqnarray}
S(p_k)\;=\;\left[\begin{array}{c}p_k\\m_k\\n_k\end{array}\right]
\end{eqnarray}

\noindent for some natural numbers $n_l,\,m_l$.
\end{itemize}

\end{Definition}

\noindent The interpretation of the first condition is obvious: To
each edge $p$, $S$ assigns a collection $p_1,\,\ldots,\, p_N$ such
that $p_1\circ\cdots\circ p_N=p$, i.e. it cuts the path into pieces.
Furthermore, to each piece there are assigned two natural numbers,
which determine the position of the ends of the piece with respect
to $S$. It is obvious that each oriented surface in $\Sig$, and also
each quasi-surface together with an orientation function in the
sense of \cite{FLEISCHHACK3} determines a generalized surface in the
above sense.

The second condition ensures that the surface does not cut a path in
a too ''wild'' way, and in particular cuts several different paths
consistently with each other. It is this condition that will ensure
that the Weyl operators constructed from the
generalized oriented surfaces will be unitarities.\\

\begin{Definition}\label{Def:Theta}
Let $\Sig$ be a manifold and $\PG$ its path groupoid. Let $G$ be a
compact Lie group and $\AL$ be the set of all functors from $\PG$ to
$\text{\rm Susp}(G)$. Let furthermore $S$ be a generalized oriented
surface and $d:\Sig\to G$ be a map.

 Then there is a
transformation of functors
\begin{eqnarray}
\Theta_{S,d}\;:\;\AL\;\longrightarrow\;\AL
\end{eqnarray}

\noindent which is given by the following rule: If

\begin{eqnarray}
S(p)\;=\;\left[\begin{array}{ccc}p_1,\,&\ldots,\,&p_N\\m_1,\,&\ldots,\,&m_N\\n_1,\,&\ldots,\,&n_N\end{array}\right]
\end{eqnarray}

\noindent then

\begin{eqnarray}\label{Gl:DefinitionOfWeylElement}
\Theta_{S,d}A\,(p)\;:&=&\;\overrightarrow{\prod_{k=1\ldots N}}
d(s(p_k))^{m_k}A(p_k)d(t(p_k))^{n_k}\\[5pt]\nonumber
:&=&\;d(s(p_1))^{m_1}A(p_1)d(t(p_1))^{n_1}\,\cdot\;\ldots\;\cdot\,d(s(p_N))^{m_N}A(p_N)d(t(p_N))^{n_N}.
\end{eqnarray}

\end{Definition}

\noindent It is easy to check that this action is well-defined, i.e.
does not depend on the representant of the equivalence class $S(p)$.
One can also readily see that this action coincides with the action
of the Weyl elements determined by oriented quasi-surfaces
\cite{FLEISCHHACK3}.

\begin{Lemma}
Let $S$ be a generalized oriented surface and $d:\Sig\to G$ be a
map. Then $\Theta_{S,d}$ given by definition \ref{Def:Theta} has the
following properties:

\begin{itemize}
\item [i)]$\a_{\phi}\circ
\Theta_{S,d}\circ\a_{\phi^{-1}}\;=\;\Theta_{\a_{\phi}S,\,d\circ\phi^{-1}}$
\item [ii)] $\Theta_{S, d}$ is continuous on $\AL$.
\item [iii)] $\Theta_{S, d}$ leaves the Ashtekar-Isham-Lewandowski measure
invariant, i.e. $\Theta_{S,d}^*\m_{AL}\;=\;\m_{AL}$.
\end{itemize}
\end{Lemma}

\noindent\textbf{Proof:}\\

\begin{itemize}
\item[\emph{i)}] Each automorphism $\phi\in\Aut$ acts naturally on the
set of generalized oriented surfaces, by
\begin{eqnarray}
S(p)\;=\;\left[\begin{array}{ccc}p_1,\,&\ldots,\,&p_N\\m_1,\,&\ldots,\,&m_N\\n_1,\,&\ldots,\,&n_N\end{array}\right]\quad\Rightarrow\quad
\a_{\phi}S\big(\phi(p)\big)\;=\;\left[\begin{array}{ccc}\phi(p_1),\,&\ldots,\,&\phi(p_N)\\m_1,\,&\ldots,\,&m_N\\n_1,\,&\ldots,\,&n_N\end{array}\right]
\end{eqnarray}

\noindent By direct computation we obtain
\begin{eqnarray}\nonumber
\a_{\phi}\circ\Theta_{S,d}\circ\a_{\phi}^{-1}A\;\big(\phi(p)\big)\;&=&\;\Theta_{S,
d}\circ\a_{\phi}^{-1}A\,(p)\\[5pt]
&=&\;\overline{\prod_{k=1,\ldots,N}}d(s(p_k))^{m_k}\,(\a_{\phi}^{-1}A)(p_k)\,d(t(p_k))^{n_k}\\[5pt]\nonumber
&=&\;\overline{\prod_{k=1,\ldots,N}}(d\circ\phi^{-1})(s(\phi(p_k)))^{m_k}\,A(\phi(p_k))\,(d\circ\phi^{-1})(t(\phi(p_k)))^{n_k}\\[5pt]\nonumber
&=&\;\Theta_{\a_{\phi}S,\,d\circ\phi^{-1}}A\;(\phi(p))
\end{eqnarray}

\noindent Since $\phi$ is an automorphism of $\PG$, it follows that
\begin{eqnarray}
\a_{\phi}\circ\Theta_{S,d}\circ\a_{\phi}^{-1}\;=\;\Theta_{\a_{\phi}S,\,d\circ\phi^{-1}}.
\end{eqnarray}

\item[\emph{ii)}] Since for each primitive metagraph $\m$ there is a finer
primitive metagraph $\m'\geq\m$ which consists only of $S$-trivial
morphisms, the set of such metagraphs is a partially ordered,
directed set. As a consequence, the set of all
\begin{eqnarray}
\pi_{\m'}^{-1}\big(U_1\,\times\,\ldots\,\times\,
U_{|E(\m)|}\big)\;\subset\;\AL
\end{eqnarray}

\noindent constitutes a basis for the topology on $\AL$, if the
$U_k\subset G$ are open.

\item[\emph{iii)}] Let $\m=\{p_1,\ldots,p_M\}$ be a primitive metagraph and $f\in
Cyl(\m)$ smooth and cylindrical over $\m$. Then
\begin{eqnarray}
\int_{\AL}d\m_{AL}(A)\;f(A)\;=\;\int_{G^E}d\m_H(h_1,\ldots,h_M)\;F(h_1,\ldots,h_M)
\end{eqnarray}

\noindent for some smooth function $F$ on $G^M$. Let
$\m'=\{q_1,\ldots,q_{M'}\}$ be a metagraph finer than $\m$ with
$S$-trivial paths $q_k$. For each $p_k\in\m$ we can then find paths
$q_1^k,\ldots,q^k_{n_k}\in\m'$, such that
\begin{eqnarray}\label{Gl:ActionOfSOnAPath}
S(p_k)\;=\left[\begin{array}{ccc}q_1^k,\,&\ldots,\,&q_{n_k}^k\\m^k_1,\,&\ldots,\,&m^k_{n_k}\\n^k_1,\,&\ldots,\,&n^k_{n_k}\end{array}\right].
\end{eqnarray}

\noindent Then by (\ref{Gl:DefinitionOfWeylElement}),
$\Theta_{S,d}f$ is cylindrical over $\m'$, and one has
\begin{eqnarray}\label{Gl:IntegrapOverThetaTransformedFunction}
&&\int_{\AL}d\m_{AL}(A)\;\Theta_{S,d}f(A)\;=\;\int_{G^{M'}}d\m_H(h_1,\ldots,h_{M'})\;\\[5pt]\nonumber
&&F\left(\overrightarrow{\prod_{l=1\ldots n_1}}
d(s(q^1_l))^{m^1_l}\,h^1_l\,d(t(q^1_l))^{n^1_l},\;\ldots,\;\overrightarrow{\prod_{l=1\ldots
n_M}} d(s(q^M_l))^{m^M_l}\,h^M_l\,d(t(q^M_l))^{n^M_l}\right)
\end{eqnarray}

\noindent where the $h_l^k$ are the $h_m$ corresponding to the
decomposition $p_k=q^k_1\circ\ldots\circ q^k_{n_k}$ given by
(\ref{Gl:ActionOfSOnAPath}). Since the integration in
(\ref{Gl:IntegrapOverThetaTransformedFunction}) uses the
bi-invariant Haar measure on $G^{M'}$, we can perform a shift of
integration variables
\begin{eqnarray}
h^k_l\;\longrightarrow\;d(s(q^k_l))^{-m^k_l}\;h^k_l\;d(t(q^k_l))^{-n^k_l}
\end{eqnarray}

\noindent which gives us
\begin{eqnarray}\nonumber
\int_{\AL}d\m_{AL}(A)\;\Theta_{S,d}f(A)\;&=&\;\int_{G^{M'}}d\m_H(h_1,\ldots,h_{M'})\;F\big(h_1^1\cdots
h_{n_1}^1\,,\,\ldots\,,\,h_1^M\cdots h^M_{n_M}\big)\\[5pt]
&=&\;\int_{G^{M}}d\m_H(h_1,\ldots,h_{M})\;F\big(h_1\,,\,\ldots\,,\,h_M\big)\\[5pt]\nonumber
&=&\;\int_{AL}d\m_{AL}(A)\;f(A),
\end{eqnarray}

\noindent since $\m\leq\m'$ is an inclusion of primitive subgraphs,
and the integral of $f$ over $\AL$ does not depend on the choice of
primitive metagraph over which it is cylindrical. Thus, we have
\begin{eqnarray}\label{Gl:ThetaActsMeasurePreserving}
\int_{\AL}d\m_{AL}(A)\;\Theta_{S,d}f(A)\;&=&\int_{AL}d\m_{AL}(A)\;f(A)
\end{eqnarray}

\noindent for all cylindrical functions $f$, which form a dense
subset in $C(\AL)$, and since $\Theta_{S,d}$ acts continuously on
$\AL$, we have that (\ref{Gl:ThetaActsMeasurePreserving}) leaves
$\m_{AL}$ invariant.
\end{itemize}

\noindent The action of $\Theta_{S, d}$ can naturally be pulled back
to $\Hkin \,=\,L^2(\AL,d\m_{AL})$ by, where they act as unitary
operators $\hat U(S,d)$. These operators generalize the
exponentiated fluxes that one can build from surfaces and smooth
fields $E_I^a$ on $\Sig$. In fact, one can recover generalized
fluxes $E(S)$ from the $\Theta_{S, d}$.

\begin{Lemma}\label{Lem:CommutingDs}
Let $\Sig$ be a manifold, $G$ be a compact Lie group, $S$ be a
generalized oriented surface, and $d_1,\,d_2\,:\Sig\to G$ two
pointwise commuting maps, i.e. $d_1(x)d_2(x)=d_2(x)d_1(x)$ for all
$x\in\Sig$. Then
\begin{eqnarray}
\Theta_{S, d_1}\,\circ\,\Theta_{S, d_2}\;=\;\Theta_{S, d_1d_2}
\end{eqnarray}
\end{Lemma}

\noindent\textbf{Proof:} By direct calculation. Let $p$ be a path in
$\Sig$, then by the definition of $S$, there are $S$-trivial paths
$p_1,\ldots,\,p_N$ with $p=p_1\circ\cdots\circ p_N$ and numbers
$m_1,\ldots,\,m_N$, $n_1,\ldots,\,n_N$ such that
\begin{eqnarray}
S(p_k)\;=\;\left[\begin{array}{c}p_k\\m_k\\n_k\end{array}\right]
\end{eqnarray}

\noindent Since $S$ is a functor, we have

\begin{eqnarray}
S(p)\;=\;\left[\begin{array}{ccc}p_1,\,&\ldots,\,&p_N\\m_1,\,&\ldots,\,&m_N\\n_1,\,&\ldots,\,&n_N\end{array}\right]
\end{eqnarray}

\noindent Thus we see that one can choose a representant of $S(p)$
contains $S$-trivial paths only. We now have

\begin{eqnarray}
\Theta_{S,d_1}\circ\Theta_{S,d_2}A\,(p)\;&=&\;\overrightarrow{\prod_{k=1\ldots
N}}
d_1(s(p_k))^{m_k}\;\Big(\Theta_{S,d_2}A\Big)(p_k)\;d_1(t(p_k))^{n_k}\\[5pt]\nonumber
&=&\;\overrightarrow{\prod_{k=1\ldots N}}
d_1(s(p_k))^{m_k}\;\Big(d_2(s(p_k))^{m_k}A(p_k)d_2(t(p_k))^{n_k}\Big)\;d_1(t(p_k))^{n_k}\\[5pt]\nonumber
&=&\;\overrightarrow{\prod_{k=1\ldots N}}
(d_1d_2)(s(p_k))^{m_k}\;A(p_k)\;(d_2d_1)(t(p_k))^{n_k}\\[5pt]\nonumber
&=&\;\Theta_{S,d_1d_2}A(p)
\end{eqnarray}

\noindent since $d_1$ and $d_2$ commute pointwise. Here the property
that all $p_k$ are $S$-trivial was essential for the proof.\\

\noindent We now show that a generalized flux can be obtained by
differentiating the action of the exponentiated fluxes.

\begin{Lemma}\label{Lem:ExistenceOfFluxes}
Let $\Sig$ be a manifold, $G$ be a compact Lie group, $S$ a
generalized oriented surface and $k:\Sig\to\mathfrak{g}$ be a map
from $\Sig$ to the Lie algebra of $G$. Then the operator

\begin{eqnarray}
E_k(S)\;:=\;-i\frac{d}{dt}\hat U\big(S,\,e^{itk}\big)_{\big|_{t=0}}
\end{eqnarray}

\noindent defines a self-adjoint operator on $\Hkin$, the domain of
definition of which contains the cylindric functions $Cyl$ on $\AL$.
\end{Lemma}

\noindent\textbf{Proof:} Lemma \ref{Lem:CommutingDs} shows that
\begin{eqnarray}\label{Gl:DefinitionOfGeneralizedFluxes}
\Theta_{S, e^{itk}}\circ\Theta_{S, e^{isk}}\;=\;\Theta_{S,
e^{i(t+s)k}},
\end{eqnarray}

\noindent i.e. the map $t\,\longmapsto\,\hat U(S, e^{itk})$ defines
a one-parameter family of unitarities on $\Hkin$.  Thus, the
Stone-von-Neumann theorem guarantees the existence of
(\ref{Gl:DefinitionOfGeneralizedFluxes}), if the limits of matrix
elements

\begin{eqnarray}
\lim_{t\to 0}\,\langle f|\,\hat U(S,e^{itk})\,|f'\rangle\;=\;\langle
f\,|\,f'\rangle
\end{eqnarray}

\noindent exist for all $f,\,f'$ being in a dense set
$D\subset\Hkin$. Let $f,f'\in Cyl$ be two smooth cylindrical
functions over a metagraph $\m$, which can, without loss of
generality be chosen to be the same for $f$ and $f'$. Then there is,
by definition of $S$, a metagraph $\m'\geq \m$ consisting of
$S$-trivial morphisms $q_1,\ldots,\,q_M$. The functions $f$ and $f'$
are then obviously also cylindrical over $\m'$, i.e.  there are
smooth functions $F, F':G\,\longrightarrow\C$ such that

\begin{eqnarray*}
f(A)\;&=&\;F(A(q_1),\ldots,\,A(q_M))\\[5pt]
f'(A)\;&=&\;F'(A(q_1),\ldots,\,A(q_M))
\end{eqnarray*}

\noindent Hence, since
$S(q_l)=\left[\begin{array}{c}q_l\\m_l\\n_l\end{array}\right]$, we
have

\begin{eqnarray*}
\Big(U(S,e^{itk})f'\Big)(A)\;&=&\;f'\big(\Theta_{S,e^{itk}}A\big)\\[5pt]
&=&\;F'\Big(e^{itm_1k^s_1}A(q_1)e^{itn_1k^t_1},\;\ldots,\;e^{itm_Mk^s_M}A(q_M)e^{itn_Mk^t_M}\Big)
\end{eqnarray*}

\noindent with $k^s_l:=k(s(q_l))\in\mathfrak{g}$ and
$k^t_l=k(t(q_l))\in\mathfrak{g}$. From this it follows that

\begin{eqnarray*}
\langle\psi|\,U(S,e^{itk})\,|\phi\rangle\;&=&\;\int_{G^M}d\m_{H}(h_1,\ldots,h_M)\overline{F(h_1,\ldots,h_M)}\,F'\Big(e^{itm_1k^s_{1}}h_1e^{itn_1k^t_1},\;
\ldots,\;e^{itm_Mk^s_M}h_Me^{itn_Mk^t_M}\Big).
\end{eqnarray*}

\noindent Since $F,\,F'$ are smooth and $G$ is compact, the
expression is smooth in $t$. Since the smooth cylindrical functions
$Cyl$ over metagraphs are dense in $\Hkin$, the claim follows from
the Stone-von-Neumann theorem.

\subsection{Algebra relations}

\noindent The smooth cylindrical functions $f\in Cyl$ and the maps
$\Theta_{S, d}$ generate an algebra, which can be completed to a
$C^*$-algebra $\A_{\text{\rm cat}}$, similar to the Weyl algebra of
quantum geometry \cite{FLEISCHHACK2}. $Cyl$, which is an abelean
subalgebra of $\A_{\text{\rm cat}}$ is generated by functions
cylindrical over just one morphism. Let $f$ be a function
cylindrical over $p$, i.e. there is a smooth function $F:G\to \C$
with $f(A)=F(A(p))$. Let furthermore
\begin{eqnarray}
S(p)\;=\;\left[\begin{array}{ccc}p_1,\,&\ldots,\,&p_N\\m_1,\,&\ldots,\,&m_N\\n_1,\,&\ldots,\,&n_N\end{array}\right]
\end{eqnarray}

\noindent The function $f$ acts via multiplication on, say,
$C(\AL)$. Then, for any $g\in C(\AL)$, we have

\begin{eqnarray}
\Theta_{S, d}\,f\,\Theta_{S,d}^{-1}\,g\;(A)\;&=&\;\big(f\,\Theta_{S,
d}^{-1}\,g\big)(\Theta_{S, d}A)\\[5pt]\nonumber
&=&\;f(\Theta_{S, d}A)\;\cdot\;\Theta_{S, d}^{-1}g\,(\Theta_{S,
d}A)\\[5pt]\nonumber
&=&\;f(\Theta_{S, d}A)\;\cdot\;g(A)
\end{eqnarray}

\noindent which gives

\begin{eqnarray}\label{Gl:WeylAlgebraicRelation}
\Theta_{S, d}\,f\,\Theta_{S,d}^{-1}\;=\;f\circ\Theta_{S, d},
\end{eqnarray}

\noindent which is the usual algebraic relation which also
determines the Weyl algebra of quantum geometry. To cast
(\ref{Gl:WeylAlgebraicRelation}) into a more frequently used form,
we go over to the (generalized) fluxes  $E_f(S)$. As operators on
$\Hkin$, (\ref{Gl:WeylAlgebraicRelation}) reads

\begin{eqnarray}
\hat U(S, d)\,\hat f\,\hat U(S,
d)^{-1}\;=\;\widehat{f\circ\Theta_{S, d}}
\end{eqnarray}

\noindent where $\hat f$ is the operator corresponding to
multiplication with $f\in Cyl$ on $\Hkin$. For $k:\Sig\to
\mathfrak{g}$ any map, we get, by lemma \ref{Lem:ExistenceOfFluxes}

\begin{eqnarray}
-i\frac{d}{dt}\Big[\hat U(S, e^{itk})\,\hat f\,\hat U(S,
e^{itk})^{-1}\Big]_{\big|_{t=0}}\;&=&\;\big[\hat E_k(S),\,\hat
f\big]
\end{eqnarray}

\noindent On the other hand, the derivative of the function
$f\circ\Theta_{S, e^{itk}}$ with respect to $t$ reads

\begin{eqnarray}\nonumber
-i\frac{d}{dt}\Big[f(\Theta_{S,
e^{itk}}A)\Big]_{\big|_{t=0}}\;&=&\;-i\frac{d}{dt}\left[\overrightarrow{\prod_{l=1,\ldots,N}}e^{itm_lk_l^s}\,A(p_l)\,e^{itn_lk^t_l}\right]_{\Big|_{t=0}}\\[5pt]
&=&\;\tilde F\big(A(p_1),\,\ldots,\,A(p_N)\big)
\end{eqnarray}

\noindent where
$k^s_l=k(s(p_l)),\,k^t_l=k(t(p_l))\,\in\,\mathfrak{g}$. The function
$\tilde F:G^N\to\C$ is given by

\begin{eqnarray}
\tilde
F(h_1,\,\ldots,\,h_N)\;=\;\sum_{l=0}^N(n_l+m_{l+1})\,F\,\big(h_1\cdot\ldots\cdot
k_l\,k^t_{l}\,h_{l+1}\cdot\ldots\cdot h_N\big).
\end{eqnarray}

\noindent In this notation, $k^t_0:=k^s_1=k(s(p))$. Furthermore
$n_0:=m_{N+1}:=0$, also $h_1\cdot\ldots\cdot h_0:=\mathbbm{1}$, as
well as $h_{N+1}\cdot\ldots\cdot h_N:=\mathbbm{1}$. We thus get

\begin{eqnarray}\label{Gl:CommutatioinRelationsGeneralized}
\big[\hat E_k(S),\,\hat
f\big](A)\;=\;\sum_{l=0}^N\e(S,p_l)\;F\,\big(A(p_1)\cdot\ldots\cdot
A(p_l)\,k^t_{l}\,A(p_{l+1})\cdot\ldots\cdot A(p_N)\big)
\end{eqnarray}

\noindent Here $\e(S, p_l):=n_l+m_{l+1}$. If $S$ is an analytic
surface and $p$ is an edge, this formula reduces to the one employed
in the literature (modulo prefactors as $l_p^2$ and the Immirzi
parameter, which can be absorbed into the definition of $E_k(S)$).
We have

\begin{eqnarray}
S(p)\;=\;\left[\begin{array}{ccc}p_1,\,&\ldots,\,&p_N\\m_1,\,&\ldots,\,&m_N\\n_1,\,&\ldots,\,&n_N\end{array}\right],
\end{eqnarray}

\noindent so $S$ cuts the edge $p$ into the pieces $p_1,\ldots,p_N$,
and thus the $k^t_l,\,l=0,\ldots,N$ are the values of the smearing
function $k$ at the endpoints of $p$ and the points where $S$
intersects $p$. At these points, the values $k^t_l\in\mathfrak{g}$
are inserted in the arguments of the function $F$, and the sum is
taken over all these points, with prefactors $(n_l+m_{l+1})$.  If
one restricts the values of the intersection functions $m_l,\,n_l$
to $\{0,\,\pm1\}$, then the factor $\e(S, p_l)$ keeps track of the
way in which the segment $p_l$ touches $S$ at $s(p_l)\in\Sig$.

Note that, since $S(p)$ is an equivalence class, so strictly
speaking $\tilde F$ depends on the choice of the representant. With
the precise definition of $S(p)$ however, one can easily see that
the function $\tilde f(A)\,=\,\tilde
F\big(A(p_1),\,\ldots,\,A(p_N)\big)$ does not.\\

This shows that (\ref{Gl:CommutatioinRelationsGeneralized})
generalizes the canonical commutation relations for quantum gravity
to a $*$-algebra obtained from $\A_{\text{\rm cat}}$, which is
generated by smooth cylindrical functions $f$ over metagraphs and
generalized electric fluxes $E_k(S)$.\\


\begin{thebibliography}{}

\bibitem{ALLMT} Ashtekar, A., Lewandowski, J., Marolf, D., Mourao, J.,
Thiemann, T. \emph{Quantization of diffeomorphism invariant theories
of connections with local degrees of freedom} 1995 J. Math. Phys.
{\bf 36} 6456 [arXiv:gr-qc/9504018]

\bibitem{ROVELLI} Rovelli, C.: \emph{Quantum
Gravity} (Cambridge Monographs on Mathematical Physics) Cambridge
University Press 2004

\bibitem{SMOLIN} Smolin, L.: \emph{An invitation to Loop
Quantum Gravity} [arXiv:hep-th/0408048]

\bibitem{INTRO} Thiemann, T.: \emph{Modern Canonical Quantum General Relativity} (Cambridge Monographs on Mathematical
Physics) Cambridge University Press 2007

\bibitem{ROVELLIFAIBAIRN} Fairbairn, W., Rovelli, C.: \emph{Separable Hilbert space in Loop Quantum
Gravity} 2004 J.Math.Phys. {\bf 45}  2802-2814 [arXiv:gr-qc/0403047]

\bibitem{KOSL} Koslowski, T.: \emph{Physical Diffeomorphisms in Loop Quantum
Gravity} [arXiv:gr-qc/0610017]

\bibitem{ALSTATUS} Ashtekar, A., Lewandowski, J.: \emph{Background Independent Quantum Gravity: A Status
Report} 2004 Class.Quant.Grav. {\bf 21}  R53 [arXiv:gr-qc/0404018]

\bibitem{LOST} Lewandowski, J., Okolow, A., Sahlmann, H., Thiemann,
T. \emph{Uniqueness of diffeomorphism invariant states on
holonomy-flux algebras} 2006 Commun. Math. Phys. {\bf 267} No. 3,
703 [arXiv:gr-qc/0504147]

\bibitem{ZAP2} Zapata, J.A::
\emph{Combinatorial space from Loop Quantum Gravity}, 1998
Gen.Rel.Grav. {\bf 30}  1229-1245 [arXiv:gr-qc/9703038]

\bibitem{FLEISCHHACK2} Fleischhack, C.: \emph{Representations of the
Weyl algebra in quantum geometry} [arXiv:math-ph/0407006]

\bibitem{BAEZCAT} Baez, J.: \emph{Spin Networks in Gauge Theory}  1996 Adv. Math. {\bf 117}, 253-272

\bibitem{VELH2} Velhinho, J. M.: \emph{Functorial aspects of the
space of generalized connections} 2005 Mod.Phys.Lett. \textbf{A20}
1299 [arXiv:math-ph/0411073]

\bibitem{ZAP1} Zapata, J.A.: \emph{A combinatorial approach to
diffeomorphism invariant quantum gauge theories} 1997 J. Math. Phys.
\textbf{38} 5663 [arXiv:gr-qc/9703037]

\bibitem{TINA1} Giesel, K., Thiemann, T.\emph{Algebraic Quantum
Gravity (AQG) I. Conceptual Setup}, 2007 Class. Quant. Grav. {\bf
24}, 2465 [gr-qc/0607099], \emph{Algebraic Quantum Gravity (AQG) II.
Semiclassical Analysis}, 2007 Class. Quant. Grav. {\bf 24}, 2499
[gr-qc/0607100], \emph{Algebraic Quantum Gravity (AQG) III.
Semiclassical Perturbation Theory}, 2007 Class. Quant. Grav. {\bf
24}, 2565 [gr-qc/0607101]

\bibitem{ALGRAPHS} Ashtekar, A., Lewandowski, J.: \emph{Differential Geometry on the Space of Connections via Graphs and Projective
Limits} 1995 J.Geom.Phys. {\bf 17} 191-230 [arXiv:hep-th/9412073]

\bibitem{FLEISCHHACK1} Fleischhack, C.: \emph{Hyphs and the
Ashtekar-Lewandowski measure} 2003 Jour. Geom. Phys. \textbf{45} 231
[MIS-preprint 3/2000]

\bibitem{BRATT} Bratteli, O., Robinson, J.: \emph{Operator algebras
and Quantum Statistical Mechanics I, 2nd edition} 1987 Springer
Verlag Heidelberg

\bibitem{HANNO} Shalmann, H.: \emph{Exploring the diffeomorphism invariant Hilbert space of a scalar
field} 2007 Classical and Quantum Gravity, \textbf{24}, 4601-4615
[arXiv:gr-qc/0609032]

\bibitem{FREI} Freidel, L., Livine, E. \emph{Spin Networks for Non-Compact Groups} 2003 J.Math.Phys. {\bf 44}
1322 [arXiv:hep-th/0205268]

\bibitem{ROVELLI1} Rovelli, C.: \emph{Partial observables} 2002 Phys.Rev. D \textbf{65}  124013
[arXiv:gr-qc/0110035], Dittrich, B.: \emph{Partial and Complete
Observables for Hamiltonian Constrained Systems} 2006
Class.Quant.Grav. \textbf{23} 6155-6184 [arXiv:gr-qc/0507106]

\bibitem{QSD8} Thiemann, T. \emph{The Phoenix Project: Master Constraint Programme for Loop Quantum Gravity} 2006 Class. Quant. Grav.
\textbf{23}, 2211  [arXiv:gr-qc/0305080], Thiemann, T. \emph{Quantum
Spin Dynamics VIII. The Master Constraint} 2006 Class. Quant. Grav.
{\bf 23},  2249 [arXiv:gr-qc/0510011]

\bibitem{BEAZSF} Baez, J.: \emph{Spin Foam Models} 1998 Class.Quant.Grav. {\bf 15}
1827-1858 [arXiv:gr-qc/9709052]

\bibitem{HYKGIR} Pfeiffer, H,: \emph{Higher gauge theory and a non-Abelian generalization of 2-form
electrodynamics} 2003 Annals Phys. {\bf 308} 447-477
[arXiv:hep-th/0304074], Girelli, F., Pfeiffer, H.: \emph{Higher
gauge theory -- differential versus integral formulation},
J.Math.Phys. 2004 {\bf 45} 3949-3971 [arXiv:hep-th/0309173]

\bibitem{MAAK} Mackaay, M.: \emph{Spherical 2-categories and 4-manifold
invariants} Adv. Math. 1999 {\bf 143}, 288-348 [arXiv:math/9805030]

\bibitem{CAT} Ad\'{a}mek, Ji\v{r}\'{\i}, Herrlich, Horst,  Strecker, George E.  \emph{Abstract and concrete categories} (1990) John Wiley and Sons

\bibitem{COMB} Magnus, W.,
Karrass, A., Solitar, D.: \emph{Combinatorial Group theory:
Presentations of Groups in Terms of Generators and Relations} 1976
Dover Publications, Inc. (New York)

\bibitem{FLEISCHHACK3} Fleischhack, C.:
\emph{Construction of Generalized Connections}
[arXiv:math-ph/0601005]



%

%
%
%
%
%
%
%
%
%
%
%
%
%
%
%








\end{thebibliography}
\end{document}